\documentclass[aps,prx,twocolumn,notitlepage,showkeys,showpacs,superscriptaddress,longbibliography,floatfix]{revtex4-1}
\usepackage{epsf}
\usepackage{bm}
\usepackage{times}
\usepackage{amsfonts}
\usepackage{amssymb}
\usepackage{amsmath,mathtools}
\usepackage{array}
\usepackage{enumerate,dsfont,here,url}
\usepackage{dcolumn,multirow}
\usepackage{slashbox}
\usepackage{bbding}
\usepackage{tabularx}
\usepackage[colorlinks=true,citecolor=blue,linkcolor=blue]{hyperref}
\usepackage[normalem]{ulem}
\usepackage{makecell}
\usepackage{color,xcolor}
\newcommand{\noi}[1]{\noindent (#1)}

\begin{document}
\title{Complete classification of band nodal structures}
\author{Feng Tang}\email{fengtang@nju.edu.cn}
\author{Xiangang Wan}\email{xgwan@nju.edu.cn}
\affiliation{National Laboratory of Solid State Microstructures and School of Physics, Nanjing University, Nanjing 210093, China and Collaborative Innovation Center of Advanced Microstructures, Nanjing University, Nanjing 210093, China}
\begin{abstract}
Nodal structures (NSrs) where energy bands meet to be degenerate in the Brillouin zone (BZ) in the form of point, line or surface, received immense research interest in the past decade. However, the nearly NSrs with negligible gaps, can also own many exotic quantum responses and nontrivial topological properties and deserve a systematic investigation.  Here, we provide a complete list of all symmetry-diagonalizable strictly and nearly NSrs in the 1651 magnetic space groups (MSGs) and 528 magnetic layer groups (MLGs) in both spinful and spinless settings.  We first apply compatibility relations (CRs) which encode how  bands split from a band node (BN) (located at a symmetric point of BZ), to obtain all NSrs (emanating from the BN). The NSrs by CRs are definitely strict and are exhaustively enumerated based on all irreducible representations of the little groups for all BNs.  We then construct $k\cdot p$ models around BNs and prove that there is a cutoff $k\cdot p$ order (6 at most) for any BN which is sufficient to predict the corresponding strictly NSrs and including higher-order $k\cdot p$ terms cannot gap the NSrs. We provide all $k\cdot p$ models up to the cutoff orders around all the BNs, and comprehensively explore all the nearly NSrs by lowering the $k\cdot p$ order. Our results reveal that the same as BNs, the NSrs, especially the nearly NSrs, are also ubiquitous. For example, while strictly nodal surface can only exist in some nonsymmorphic MSGs,  nearly nodal surface can occur even in many low symmetric MSGs. Moreover, we also find that for some BNs, the method of $k\cdot p$ modeling could give more strictly NSrs than CRs.  With our complete list, one can conveniently obtain all possible NSrs for any two/three-dimensional and nonmagnetic/magnetic material, and given a target NSr, one can also  design/search for  materials realizations based on the MSGs/MLGs. Finally, we predict hundreds of ever-synthesized magnetic materials with essential and extended nodal line or surface.
\end{abstract}
\date{\today}
\maketitle
The nodal structures (NSrs) in energy bands, classified by degeneracy, dimensionality (point, line or surface), topological charge etc., have attracted  extensive attention  \cite{PRB-Wan, TaAs-PRX, TaAs-NC, ChenFang-largeChern, new-fermions,CoSi-Tang,CoSi-Zhang, QuanShengWu-S, SMYoung, Na3Bi, Cd3As2,Nagaosa-NC,Du-1,Du-2,new-fermions,APL,zxl,DoubleDSM,triple-Weng, nodal-line-balents,nodal-line-fangchen,cupdn-1,cupdn-2,hourglass-n,nodal-chain-n,hopf-link-1,hopf-link-2,hopf-link-3,hopf-link-4,hopf-link-5,hopf-link-6,hopf-link-7,wwk,n-sym-enf} since they can underlie  fascinating physical properties, like surface Fermi arcs \cite{PRB-Wan}, chiral anomaly \cite{DTson}, drumhead surface states \cite{nodal-line-balents}, Klein tunneling \cite{klein} and so on. In most cases, the stability of an NSr is guaranteed by space group (SG) symmetry \cite{bradley,bilbao}. In addition to the novel quantum excitations and responses \cite{AV-RMP}, breaking the protecting symmetry may also result in nontrivial insulating phases typical examples being quantum spin Hall effect \cite{TI-RMP-1, TI-RMP-2} and valleytronics \cite{berryphase, valley} proposed in graphene. The SG symmetry also benefits efficient identification of band node (BN), a degenerate point in the Brillouin zone (BZ) \cite{Chiu-RMP, AV-RMP, LBQ-RMP, Ando, Bala, tqc, si, slager-prx,n-1,n-2,n-3,n-4}. The previous systematic studies of BNs mostly focused  on specific types of BNs: For example, the location of a BN is at a high-symmetry point (HSP) \cite{new-fermions, APL, Manes, SMYoung, DoubleDSM, KramersWeyl, Nagaosa-NC}, the degeneracy is larger than 2 \cite{new-fermions, APL, SMYoung, DoubleDSM,Nagaosa-NC} and the low energy dispersion should be linear \cite{Manes, SMYoung}, etc. Very recently, considering all strictly degenerate BNs in HSPs, high-symmetry lines (HSLs) and high-symmetry planes (HSPLs)  in the 230 SGs with additional time-reversal ($T$) symmetry by $k\cdot p$ modeling, Ref. \cite{yao} constructed a very useful encyclopedia. They find that all possible degeneracies of BNs in crystals are $2,3,4,6$ and $8$ and the largest topological charge is 4 \cite{yao}. Furthermore, it is found that when $T$ symmetry is considered, the charge-4 Weyl point (WP) only exists in the spinless setting \cite{yao,TT}, thus dubbed as spinless charge-4 WP here. However, it is known that the symmetry groups are far from 230 SGs when magnetic order is formed so that there are another 1421 magnetic SGs (MSGs) describing magnetic three-dimensional (3D) systems \cite{bradley}. Besides, the 528 magnetic layer groups (MLGs) are also interesting for they describe symmetries of two-dimensional (2D) magnetic or nonmagnetic materials like graphene \cite{graphene-S}, as well as the boundaries between different insulating phases \cite{wallpaper}.

While the BNs have been well-investigated \cite{yao,new-fermions, APL, Manes, KramersWeyl, Nagaosa-NC}, a complete classification of all NSrs is still lacking. Though the NSrs  can be deduced by constructing $k\cdot p$ model around a BN \cite{new-fermions, APL, AV-RMP, LBQ-RMP, yao,zxl}, it can only predict local NSrs around the BN.  Moreover, the $k\cdot p$ model is commonly only expanded to a fixed order in $\mathbf{q}$ (the momentum measured from the BN)  \cite{new-fermions, APL, AV-RMP, LBQ-RMP, yao,zxl}, thus whether including symmetry-allowed higher $k\cdot p$ terms will gap the predicted NSrs is not known.  Notably, the same as the strictly NSrs, the nearly  NSrs could also bring about novel properties and have attracted much attention. For example, spin-orbit coupling (SOC) can gap the $PT$ ($P$: inversion) protected nodal lines (NLs)  \cite{nodal-line-balents,nodal-line-fangchen,cupdn-1,cupdn-2}, thus those NLs actually only exist nearly since the SOC always exists. Despite the tiny band gaps induced by SOC, these nearly NLs and the associated drumhead surface states have been verified in realistic materials and have attracted enormous attentions \cite{NL-exp-2, NL-exp-1, NL-exp-3}. Similarly, though the  Dirac point in graphene at $K$ is a nearly nodal point (NP), which can also be gapped by SOC,  the  Klein tunneling of Dirac fermions in graphene \cite{klein} has been  experimentally confirmed \cite{klein-exp}.  All in all, a complete  list of all possible BNs and strictly and nearly NSrs  in the 1651 MSGs and 528 MLGs and in both spinful and spinless settings is not only important in revealing new types of BNs and NSrs,  and it is also useful for efficient materials predictions.

In this work, we combine compatibility relations (CRs) and $k\cdot p$ models  to obtain such complete list.  Noticing that any symmetry-protected NSr owns at least one point (namely, the BN mentioned above) pinned in symmetric $k$ point,  we can exhaust all possible symmetry-diagonalizable  NSrs by systematically analyzing all BNs located at HSP/HSL/HSPL.  Starting from any BN, whose participating irreducible representation(s) (irrep(s)) or co-irrep(s) can be known from its little group, we exploit CRs  to obtain how  bands split in the vicinity of the BN \cite{bradley, bilbao}. This is because CRs relate the (co-)irrep of the (energy level of) BN to those of any neighboring $k$ point \cite{bradley, bilbao}.  From the obtained CR-required band splitting pattern (CR-BSP) along all HSLs and HSPLs passing the BN, we obtain all NSrs (definitely strictly) emanating from the BN. These NSrs required by CR-BSPs always exist as long as the BN is formed and can be extended in the Brillouin zone (BZ) when they are HSLs/HSPLs. It is worth mentioning that these extended NSrs  have a great chance of crossing the Fermi surfaces, significant in designing devices with large excitation and exotic responses \cite{n-sym-enf}.

In order to obtain all nearly NSrs, we then construct $k\cdot p$ models around all the BNs.  By comparing the NSrs by CR-BSPs and those by $k\cdot p$ models, we prove that there exists a cutoff $k\cdot p$ order for any BN:  For most BNs, we can find a least order to which the $k\cdot p$ model  can give exactly the same NSrs as those by CR-BSPs and such order can be chosen to be the cutoff order. For the rest BNs, we can also find a least cutoff $k\cdot p$ order which can not only reproduce all the NSrs predicted by CR-BSPs but also provide additional strictly NSrs. These additional NSrs cannot be captured simply using CRs though.  An important usage of  $k\cdot p$ models  is that  the $k\cdot p$ models up to the cutoff orders facilitate  us to explore all nearly NSrs by lowering the $k\cdot p$ orders.  We find that the maximal cutoff order is 6, thus to obtain strictly NSrs,  $k\cdot p$ model up to 6th order is sufficient.  The cutoff order quantitatively characterizes order of magnitude of the gap opened in the nearly NSrs by considering higher-order $k\cdot p$ terms.

Our exhaustive study indicate that the nearly NSrs enrich the NSrs significantly: Among 98970 BNs in total in MSGs, around 22.4\% can have nearly NSrs. For example, strictly nodal surface (NS) can only occur in very restricted symmetry conditions \cite{yao}: In total, 10636 BNs are found to lie in strictly NSs for some nonsymmorphic MSGs. However, nearly NSs can occur for another 19520 BNs. Concretely, there are only 254 (486) MSGs allowing strictly NS while there are 789 (1316) MSGs allowing nearly NSs, in the spinful (spinless) setting. Such enrichment on NSrs is expected to be important from an experimentalist's perspective: The nearly NSr in some material near the Fermi level can induce large responses even though the material cannot allow strictly NSr.   Based on the explicit $k\cdot p$ models, we also find that the charge-4 WP,  believed to only exist in the spinless setting \cite{yao, TT}, can exist in magnetic materials with significant SOC. The  material realization of charge-4 WP in SrCuTe$_2$O$_6$ proposed in this work is verified experimentally \cite{private}. We also list hundreds of realistic magnetic materials with symmetry-enforced essential and extended NLs/NSs.

Our comprehensive results are gathered in \textbf{Supplementary Information I-IV} in both top-down and bottom-up forms:  Simply searching for the MSG  of any crystalline material in Sec. \ref{page} of \textbf{Appendix}, and then knowing the pages for the corresponding results in \textbf{Supplementary Information III}, one can quickly retrieve all possible CR-BSPs and the strictly NSrs by CR-BSPs. Then combined with \textbf{Supplementary Information I}, the corresponding  $k\cdot p$ models  and the nearly NSrs can also be found.  Inversely, using \textbf{Supplementary Information II} and \textbf{IV}, one can also know concretely in which MSG/MLG to  realize any specific  $k\cdot p$ model or CR-BSP, and then finding materials realizations based on MSG/MLG becomes much more convenient.

\section{work flow and results}
\begin{figure*}[!htb]
  \includegraphics[width=1.0\textwidth]{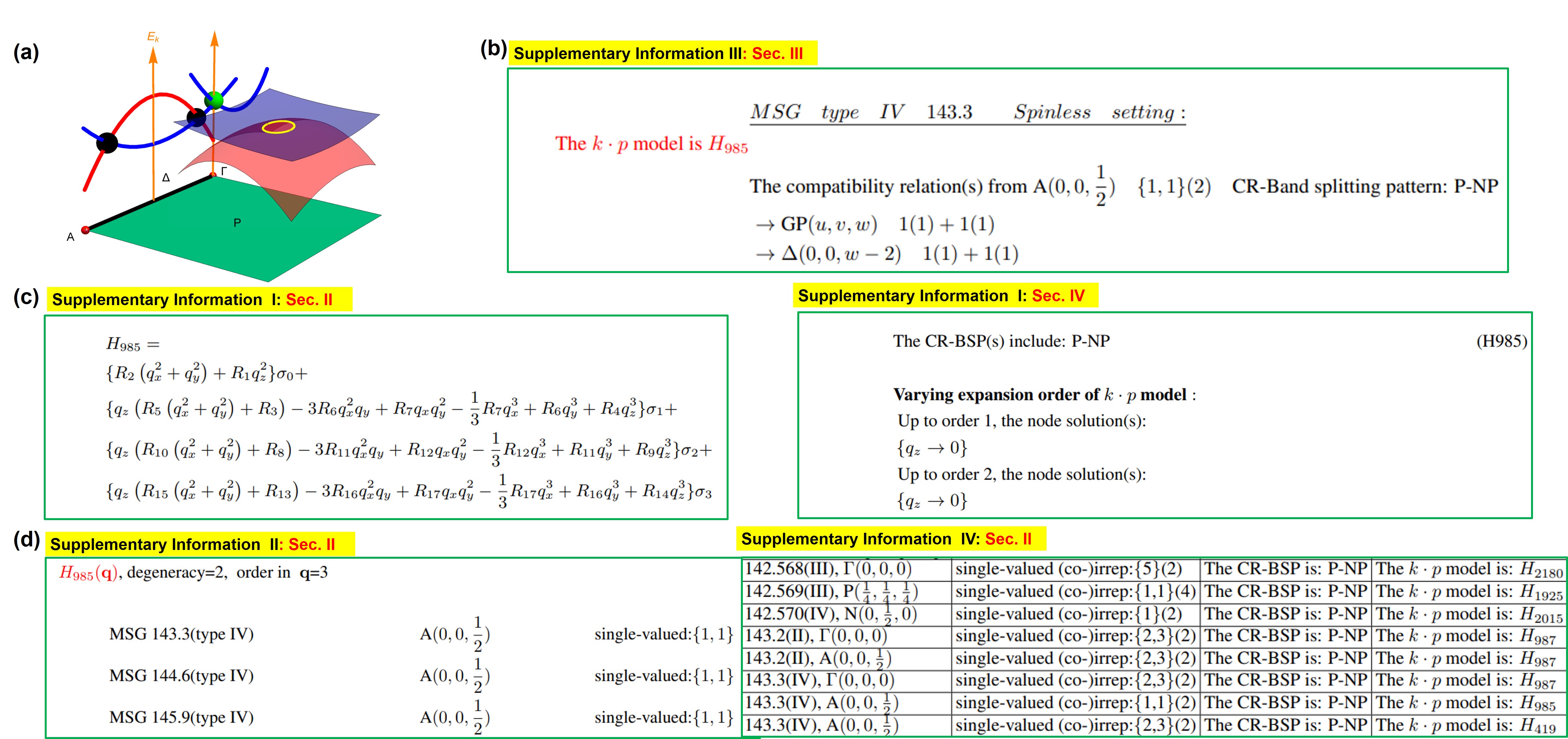}\\
  \caption{\textbf{All symmetry-diagonalizable  BNs (a) and an example of using our results (b-d).}
  (a) The HSL $\Delta$ connects two HSPs $\Gamma$ and $A$, and they all lie within an HSPL $P$ (green plane). Along $\Delta$, two bands belonging to different (co-)irreps (denoted by red and blue colors) cross each other, resulting in two BNs (black spheres, which cannot be predicted using the symmetry information of $\Gamma$ and $A$ since they satisfy the CRs). Such BNs could own CR-BSP L-NP or L-NL.  The green sphere denotes BN at HSP, whose CR-BSP can be P-NP, P-NL, P-NS or P-NSL. The yellow loop, the intersection of two energy surfaces with two different (co-)irreps, is an NL lying in the HSPL. Every point in the NL is a BN with CR-BSP PL-NL. Besides, each point in the HSPL can be a BN with CR-BSP PL-NS. Every point in the HSL may also be a BN with CR-BSP L-sNL or L-NS. (b-d) An example and demonstration of using our results are shown. (b) \textbf{Supplementary Information III} contains results of top-down type, namely, given a BN, the corresponding $k\cdot p$ model and CR-BSP are given along with all related CRs: Here, for the BN at HSP A of MSG 143.3 with  single-valued (co-)irrep being \{1,1\}, the CR-BSP is P-NP and the $k\cdot p$ model is $H_{985}$. (c) \textbf{Supplementary Information I} lists all $k\cdot p$ models realized around all BNs up to the cutoff orders.  Here we show the $k\cdot p$ model, $H_{985}$ whose $k\cdot p$ cutoff order is 3, as the example. Its solutions of all NSrs are given in Sec. IV of \textbf{Supplementary Information I} as shown here: The CR-BSP for this $k\cdot p$ model can be P-NP thus to the cutoff order 3, the NSr is simply an NP at the BN. Up to order 1 and 2, the NSrs by solving the corresponding $k\cdot p$ models are both $q_z=0$ (namely, an NS coinciding with $q_x$-$q_y$ plane). (d) \textbf{Supplementary Information II} and \textbf{IV} contain results of bottom-up type, namely, stating from a given $k\cdot p$ model (\textbf{Supplementary Information II}) and a given CR-BSP (\textbf{Supplementary Information IV}), we show all corresponding BNs: For example, given a $k\cdot p$ model, $H_{985}$, the corresponding BNs can be realized in HSP A of MSG 143.3 with  single-valued (co-)irrep being \{1,1\}; Given CR-BSP P-NP, the corresponding BNs can be realized at HSP A of MSG 143.3 with  single-valued (co-)irrep being \{1,1\}.}\label{figure-1}
\end{figure*}

\subsection{CR-BSPs}\label{CR}
Firstly, we describe the classification work flow adopted in the work. We are concerned with the BNs located in HSP/HSL/HSPL and thus they can be efficiently identified by the irreps of energy bands \cite{yao,tang-kp}. Each BN may contain one degenerate (co-)irrep \cite{remark-3} when it is in HSP/HSL/HSPL, or two different (co-)irreps when it is in HSL or HSPL. A schematic diagram for various BNs  is shown in Fig. \ref{figure-1}(a).   Based on all the (co-)irrep matrices in Ref. \cite{tang-kp} and the calculated CRs, we finally obtain 98970 and 9869 BNs for MSGs and MLGs, respectively. Here the counting includes both spinless and spinful settings.  Knowing the participating (co-)irrep(s) in a BN, CRs can encode the band splitting pattern away from the BN since CRs describe the reduction from the (co-)irrep(s) of the BN to the (co-)irreps of $k$ points in the vicinity around the BN.  Next, we briefly describe the formation condition for all possible CR-BSPs and the examples are given in Sec. \ref{condition} of \textbf{Appendix} where  the detailed conditions $C_{i}$ as shown in Table \ref{table-1} can also be found.  Though there are infinite number of $k$ points in the vicinity of the BN, they can be classified into HSL, HSPL and generic point (GP). Note that from the BN to any generic direction, the bands should split by definition otherwise the $k\cdot p$ model is simply a constant term.  Hence we check the CRs between the BN and all possible HSLs and HSPLs which contains the BN (hereafter they are called neighboring HSLs and HSPLs of the BN),  and we can be encountered with the following two cases. In total there are 10 types of CR-BSPs, also   listed in Table \ref{table-1}, named by X-Y where X (P, L or PL) denotes the location (HSP, HSL or HSPL) of the BN and Y denotes the type of NSr (emanating from the BN).

\begin{table*}[!htb]
  \centering
  \begin{tabular}{c|c|c|c|c|c|c|c|c|c|c}
    \hline\hline
    The location of BN & \multicolumn{4}{c|}{HSP} & \multicolumn{4}{c|}{HSL} & \multicolumn{2}{c}{HSPL} \\\hline
    CR-BSP& P-NP& P-NL& P-NS& P-NSL & L-sNL& L-NP& L-NL& L-NS & PL-NS&PL-NL \\\hline
    Condition&$C_1$&$C_2$&$C_3$&$C_4$&$C_2$&$C_1$&$C_3$&$C_3$&$C_3$&$C_3$\\\hline
    NSr by CR-BSP& -& \makecell{coinciding\\ with\\ neighboring \\HSL}& \makecell{coinciding\\ with\\ neighboring\\ HSPL}& \makecell{coinciding\\ with\\ neighboring\\ HSL\\ and HSPL}& \makecell{coinciding\\ with\\ the HSL}&-&\makecell{lying in\\ neighboring\\ HSPL}& \makecell{coinciding\\ with\\ neighboring\\ HSPL}&\makecell{coinciding\\ with\\ the HSPL}&\makecell{lying in\\ the HSPL}
    \\
    \hline
  \end{tabular}
  \caption{\textbf{The 10 types of CR-BSPs and the resulting NSrs.} The name of CR-BSP is in the form of ``X-Y'' where X denotes the location of the BN while Y denotes the type of the NSr. For example, P-NP means that the BN is at an HSP and it is required to be an NP by CR-BSPs. Note that when X=L, namely, the BN occurs in some HSL (namely, X=L), we use Y=``sNL'' and ``NL'' both  denoting NL.  But ``s'' in ``sNL'' implies that the NL is straight coinciding with the HSL (namely, the NL is simply the HSL), while ``NL'' represents NL lying in neighboring HSPL containing the HSL. P-NSL means that from a BN at an HSP, there exist NSrs coinciding with neighboring HSL and HSPL (containing the HSP).  The conditions for these CR-BSPs and examples are given in Sec. \ref{condition} of \textbf{Appendix}. In the fourth row, we show the NSrs required by CR-BSPs, which are printed in red in \textbf{Supplementary Information III}, except those for CR-BSP L-sNL and PL-NS since the NSrs for them are simply the HSL and HSPL hosting the BN, respectively.}\label{table-1}
\end{table*}

\subsubsection{Case 1: The BN contains only one (co-)irrep}
In this case, the BN is symmetry protected and constituted by one (co-)irrep whose dimension is larger than one. When CRs requires that the (co-)irrep in the BN is reduced to several (co-)irreps away from the BN in some direction, the bands should split in this direction.  Concretely, the BN in this case can be located at an HSP or lie in an HSL or HSPL. When the BN is at an HSP,  the resulting CR-BSPs can be P-NL, P-NS and P-NSL:  P-NL (P-NS) means that there exist at least one neighboring HSL (HSPL) being an NL (NS). P-NSL means that there exist both neighboring HSL and HSPL  corresponding to an NL and NS, respectively.  If there exist no neighboring HSL or HSPL being an NL or NS,   the BN is simply an NP and the CR-BSP is named by P-NP.  When the BN lies in  an HSL,  the resulting CR-BSPs can be L-sNL and L-NS.  L-sNL means that the HSL where the BN is located, is itself an NL. L-NS means that there exists neighboring HSPL being an NS. When the BN lies in  an HSPL,  the resulting CR-BSP can only be PL-NS, meaning that the HSPL where the BN is located, is itself an NS.

\subsubsection{Case 2: The BN contains two (co-)irreps}
In this case the BN can lie in an HSL or HSPL, formed by two branches of bands with two different (co-)irreps. The formation of such BN is due to band inversion, and is stable against any  symmetry-preserved perturbations such as the Dirac points in Na$_3$Bi \cite{Na3Bi} and Cd$_3$As$_2$ \cite{Cd3As2}.  Denote that two band branches as $\epsilon_1(\mathbf{k})$ and $\epsilon_2(\mathbf{k})$, where $\epsilon_1$ and $\epsilon_2$ carry two different (co-)irreps and $\mathbf{k}$ is restricted in the HSL or HSPL.  In general,  a BN lying in the HSL or HSPL can be formed by requiring that $\epsilon_1(\mathbf{k})=\epsilon_2(\mathbf{k})$ at the BN.  When the BN lies in an HSPL, it is easy to know that there is an NL (passing the BN) lying in the HSPL and the corresponding CR-BSP is named by PL-NL.   When the BN lies in an HSL, if CRs require that the two different (co-)irreps in the BN keep to be two different (co-)irreps in some neighboring HSPL, the BN then lies in an NL lying in this neighboring  HSPL (the CR-BSP is denoted as L-NL), otherwise the BN is simply an NP (the CR-BSP is L-NP). The two kinds of NLs for L-NL and PL-NL in this case both lie in an HSPL, and can be a loop, namely, not straight as the NL coinciding with an HSL.

Based on the above classification of CR-BSPs, it's clear that CRs can require NSrs (emanating from the BN) along some HSL or HSPL.  All CR-BSPs for all BNs are listed successively in \textbf{Supplementary Information III}, where the degeneracy of the BN, CR-BSP, NSr, related CRs and the $k\cdot p$ model (up to the cutoff order) are all shown.   One can know all possible NSrs  for any 3D/2D materials simply by looking up \textbf{Supplementary Information III}.  The NSr may coincide with  an HSL or HSPL (case 1) or lie in an HSPL (case 2) and the corresponding HSL or HSPL is printed in red as we show in the following by examples.

We first take the example as shown in Fig. \ref{figure-1}(b) to illustrate how to use our results.  If one is interested in a compound crystallizing in MSG 143.3 (type IV) in the Belov-Neronova-Smirnova (BNS) \cite{bns} notation. By searching 143.3 in Sec. \ref{page} of \textbf{Appendix}, one can find that all possible BNs for MSG 143.3 are given in pages from 4880-4882, 15049-15052, 23370-23379 and 35547-35555 in \textbf{Supplementary Information III}. For example, as shown in page 15050 in \textbf{Supplementary Information III} (a snapshot is given in Fig. \ref{figure-1}(b)),   A is an HSP of MSG 143.3  which allows the single-valued co-irrep, denoted by \{1,1\}  (the notation of (co-)irreps follows those in Ref. \cite{tang-kp}). The dimension of this co-irrep is 2 (the number in the parentheses following the name of co-irrep), thus the BN formed by this co-irrep is of degeneracy 2. The CR-BSP can be found to P-NP, which means that the BN is simply an NP and no NL or NS is emanating from the BN, namely, energy bands split in all directions away from the BN.  For readers' convenience, we also show the $k\cdot p$ model (up to the cutoff order) around the BN in \textbf{Supplementary Information III}, and here it's given by $H_{985}$ whose concrete expression can be found in \textbf{Supplementary Information I} as clarified in next section.

We then show another two examples whose NSrs by CR-BSPs are NLs. First, consider a BN  at $\Gamma$ of MSG 147.14 (type II) in the spinless setting (see the snapshot in \textbf{Supplementary Information III} in Fig. \ref{C2}(a) of \textbf{Appendix}). The concrete page numbers for this MSG in \textbf{Supplementary Information III} can be found in Sec. \ref{page} of \textbf{Appendix}.   The co-irrep of the BN is \{2,6\} and the CR-BSP is P-NL for which the NSr by CR-BSPs is along an HSL. The HSL is found to be $\Delta$ which is printed in red as shown in Fig. \ref{C2}(a) of \textbf{Appendix}.  The $k\cdot p$ model is $H_{2373}$ for this case, also shown in Fig. \ref{C2}(a) of \textbf{Appendix}.  Second,  see Fig. \ref{C3-2}(c) of \textbf{Appendix}.  The BN is formed by two different co-irreps \{3\} and \{4\} in HSL $\Delta$ of MSG 183.185 and the CR-BSP is L-NL. The corresponding  NSrs by CR-BSPs are also NLs but lying in neighboring HSPLs  (all these HSPLs contain the HSL $\Delta$ where the BN is located). These HSPLs are all printed in red as shown in Fig. \ref{C3-2}(c) of \textbf{Appendix}.  The corresponding $k\cdot p$ model  is given by $H_{2469}$.

To summarize, \textbf{Supplementary Information III} can be thought of as a top-down list: Given an MSG or MLG (the MLG is expressed by MSG, see Sec. \ref{mlg} of \textbf{Appendix}), when a BN is formed in some $k$ point with (co-)irrep(s) known, \textbf{Supplementary Information III} can be exploited to obtain the corresponding CR-BSP and $k\cdot p$ model quickly.  However, one may be encountered with an inverse problem: How to realize a given  CR-BSP in concrete materials. Hence, we also show all BNs corresponding to a given CR-BSP in \textbf{Supplementary Information IV}. For example, as shown in the right inset of Fig. \ref{figure-1}(d), regarding CR-BSP which is P-NP, we can find that the corresponding BNs can be realized at $\Gamma$ of MSG 143.3 with the single-valued co-irrep \{2,3\}, at A of MSG 143.3 with the single-valued co-irreps \{1,1\} or \{2,3\}, and so on. The $k\cdot p$ models are also given, and here as shown in Fig. \ref{figure-1}(d) they are $H_{987}$, $H_{985}$ and $H_{419}$, successively.

\subsection{$k\cdot p$ models}\label{kp}
We also construct explicit $k\cdot p$ models around the BNs to obtain all associated NSrs. The method of  $k\cdot p$ modeling, has been extensively applied to study the BNs in the field of topological semimetals  and the obtained $k\cdot p$ models can also be used to directly calculate the topological characters \cite{new-fermions, APL, yao}.

Any symmetry-allowed $k\cdot p$ models automatically satisfies the symmetry constraints from the little group of the corresponding BN and all related CRs \cite{tang-kp, bradley}. Thus by solving the $k\cdot p$ model, one can definitely obtain the NSrs which include those predicted by CRs.   Three questions naturally arise: (1) up to which order can the $k\cdot p$ model reproduce all the NSrs by CR-BSP? (2) Can a $k\cdot p$ model predict additional strictly NSrs which cannot be captured by CR-BSP? (3) Can a $k\cdot p$ model give nearly NSrs?  We address these questions in the following.

To tackle the above questions, we construct $k\cdot p$ model starting with $k\cdot p$ order=1  following the method in Ref. \cite{tang-kp}. Then we gradually increase the $k\cdot p$ order and obtain the corresponding $k\cdot p$ model. Using the explicit $k\cdot p$ model up to each order, we can obtain the corresponding NSrs. By definition, the NSr is the region in the BZ where  the $k\cdot p$ model is reduced to a constant term (see equations in Sec. \ref{node} of \textbf{Appendix}). Besides, we require that the NSr should pass the BN. Note that there could exist NSr from a $k\cdot p$ model that is isolated from the BN, such NSr can be not physical since the $k\cdot p$ model is only meaningful locally. In fact, such NSr is not excluded from our analysis since it passes another BN. We find that a $k\cdot p$ order exists for each BN, so that the NSrs by the corresponding $k\cdot p$ model are all strict. Besides, increasing the $k\cdot p$ order further couldn't reduce the number of NSrs any more, thus we are only concerned with a least order, denoted as cutoff order $o$. Hence, any $k\cdot p$ model up to the order  which is smaller than $o$,  definitely gives nearly NSrs.

For most BNs, we find that we can obtain a least order  for which the NSrs are exactly the same as those by CR-BSPs, and then the order is chosen as the cutoff order.  For the rest BNs, we can obtain a least order for which the NSrs contain two parts: One includes exactly the NSrs by CR-BSPs while the other includes strictly NSrs not captured by CR-BSPs. In this case, such order is chosen as the cutoff order. We also prove that the nearly NSrs can be obtained by solving the $k\cdot p$ model with the $k\cdot p$ order smaller than the cutoff order. We call these nearly NSrs not predicted by CR-BSPs as $k\cdot p$ order enriched NSrs.

Interestingly, the  cutoff $k\cdot p$ order is found to be 6 at most, for both the MSGs and MLGs, which means that one can obtain all possible strictly and nearly NSrs by a $k\cdot p$ model up to the sixth order.  Here, up to the cutoff orders, we construct all the $k\cdot p$ models for all the 98970 and 9869 BNs in MSGs and MLGs, respectively. Since many of the corresponding 98970 and 9869 $k\cdot p$ models share the same form, we finally collect 3026 and 601 $k\cdot p$ models expressed by $q_x, q_y, q_z$, free real parameters $R_i$ and matrix basis sets (see Sec. \ref{matrixbasis} of \textbf{Appendix}) for MSGs and MLGs, respectively. Here $(q_x, q_y, q_z)$ is the Cartesian coordinate of $\mathbf{q}$ and the Cartesian coordinate system adopts that in Ref. \cite{bradley}.  These $k\cdot p$ models are complete to describe all possible BNs and can be exploited to calculate both exactly and nearly NSrs. They are given explicitly in Secs. I and II of \textbf{Supplementary Information I}, denoted by $H_i (i=1,2,\ldots,3026)$ and $h_i (i=1,2,\ldots,601)$ for MSGs and MLGs, respectively.  An example $H_{985}$ is shown in Fig. \ref{figure-1}(c) for the HSP A in the BZ of MSG 143.3.  In \textbf{Supplementary Information I}, we show explicit expressions of NSrs by CR-BSPs  and the $k\cdot p$ order enriched NSrs for MSGs. For MLGs, we directly show the solutions of NSrs. Also, we find that MLGs have no any $k\cdot p$ enriched NSrs, since for all the BNs in MLGs, the $k\cdot p$ models up to order $o-1$ are always  constant terms (see $h_{426}$ in page 466 of \textbf{Supplementary Information I}, which is $\frac{R_5q_xq_y(-10q_x^2q_y^2+3q_x^4+3q_y^4)}{3}\sigma_z$ ($\sigma_{x/y/z}$ is a Pauli matrix), the constant term omitted. For this $k\cdot p$ model, the NSrs occur in lines obtained by  operating successively rotation by $\frac{\pi}{6}$ around $z$ axis on the line expressed by $q_x=0$, all along HSLs, by CR-BSPs).

As shown in Fig. \ref{figure-1}(c), for the $k\cdot p$ model, $H_{985}$, there is no NSr since it describes an NP (the CR-BSP is P-NP). However, lowering $k\cdot p$ order to 1 and 2 would result in an NS along $q_z=0$ ($q_x$-$q_y$ plane). Hence, in a relatively large vicinity of the BN, the bands would keep to be very close to each other, thus possibly giving rise to large responses. As a matter of fact, the $k\cdot p$ model $H_{985}$ is found to describe a charge-3 WP, discussed below.
\subsubsection{Large Berry curvatures associated with nearly NSr}\label{large}
We take the parameters as shown in Sec. \ref{peak} of \textbf{Appendix} in the $k\cdot p$ model $H_{985}$ whose realization can be found in Figs. \ref{figure-1}(b) and (d) and calculate the distribution of Berry curvatures in a unit sphere around the BN. With respect to the unit of $\mathbf{q}$, it is chosen to be $q_c$ and $q_c$ is defined as follows. The scope around the BN where  $k\cdot p$ modeling is suitable has a boundary and $q_c$ is the least distance from the boundary to the BN. From Fig. \ref{figure-4} of \textbf{Appendix}, the Berry curvatures peak around the nearly NS $q_z=0$ plane (namely, $\theta=\frac{\pi}{2}$ in Fig. \ref{figure-4} of \textbf{Appendix}), as expected. The large Berry curvatures indicate a large Chern number probably and the integral of the Berry curvatures shows that the $k\cdot p$ model $H_{985}$ actually describes a charge-3 WP.

When the corresponding BN is realized in realistic material, the predicted charge-3 WP tells us that Fermi arcs \cite{PRB-Wan} can be observed in the surface. However, the large Berry curvatures around the nearly NS provide more information and have prominent consequences on the electronic transport behaviors \cite{berryphase}. Besides, the nearly NS could also result in interesting surface states \cite{wwk} and  also implies that when we numerically study the associated topological characters of the BN, we should be cautious on the $k$ points very close to the nearly NS.
\subsubsection{Example: Enriched nodal structures by $k\cdot p$ order}\label{order}
Then we show an example of BN whose cutoff order of $k\cdot p$ model is 6 to show nearly NSrs by lowering $k\cdot p$ order. The BN is formed by two single-valued irreps 3 and 4 in the HSL $\Delta$ of MSG 183.185.  As shown in Fig. \ref{C3-2}(c) of \textbf{Appendix},  the CR-BSP for this BN is L-NL, namely, no matter how large the $k\cdot p$ order is, the CRs require that there should exist NLs  (red lines in Fig. \ref{figure-3}(c)) lying in HSPLs (six gray planes in Fig. \ref{figure-3}(c)) for this BN. The $k\cdot p$ model is  $H_{2469}$, which is also a $2\times 2$ matrix, whose explicit expression can be found in \textbf{Supplementary Information I}. As the cutoff order for this BN is 6, thus $H_{2469}$ has the order of 6 in $\mathbf{q}$.  As shown in Fig. \ref{figure-3}(c), the yellow surface represents the node solution for the $k\cdot p$ model with order 2, while the cyan surface represents the node solution of the $k\cdot p$ model with order 1.  The yellow curved surface in Fig. \ref{figure-3}(c) is nearly but an almost strictly NS, while the strictly NS stabilized by MSG symmetries can only coincide with an HSPL \cite{yao}. In total, there are 796 out of the 3026 $k\cdot p$ models $H_i^,$s for MSGs  found to show nearly NSrs, corresponding to 22159 BNs in total, which means that nearly $\frac{1}{4}$ of BNs can demonstrate $k\cdot p$ order enriched NSrs. Concretely, among 98970 BNs in total for all MSGs,  we combine CRs and $k\cdot p$ models to reveal that 10636 can lie in strictly NS (the CR-BSP is P-NS, P-NSL, P-NSL or PL-NS) and 60200 can lie in strictly NLs (the CR-BSP is P-NL, P-NSL, L-NL, L-sNL or PL-NL), while 19520 can lie in nearly NS (the CR-BSP is not P-NS, P-NSL, P-NSL or PL-NS) and 57 can lie in nearly NLs ((the CR-BSP is not P-NL, P-NSL, L-NL, L-sNL or PL-NL)). Roughly, considering nearly NSrs, the number of BNs with NSrs are doubled.  We list all MSGs realizing strictly or/and nearly NSs in spinless or/and spinful setting in Sec. \ref{enrichns} of  \textbf{Appendix}. Overall, majority of MSGs allow NSs, while only nonsymmorphic MSGs can allow the formation of strictly NS.
\begin{table}[!htb]
  \begin{tabular}{c|c|c|c|c|c}
    \hline\hline
    Type &IV&  I & IV & I & IV \\\hline
    MSG & 198.11 & 212.59 & 212.62 & 213.63 & 213.66 \\\hline
    $H_i$& $H_{1415}$&$H_{1508}$&$H_{1508}$&$H_{1508}$&$H_{1508}$\\
    \hline
  \end{tabular}
  \caption{\textbf{MSGs allowing spinful charge-4 WP.} For all MSGs here, the charge-4 WP is pinned at HSP $(\frac{1}{2},\frac{1}{2},\frac{1}{2})$.  All 2D (co-)irreps for this HSP in these MSGs correspond to spinful charge-4 WPs. Other than spinful charge-4 WPs, this HSP can also host spin-1 NP \cite{new-fermions}. $H_{1415}$ and $H_{1508}$ are the $k\cdot p$ models which can be found in \textbf{Supplementary Information I}: $H_{1415}=[R_2 (q_z^2-q_x^2)+R_3(q_y^2-q_x^2)]\sigma_x+\frac{R_3(q_x^2+q_y^2-2q_z^2)+R_2(2q_y^2-q_x^2-q_z^2)}{\sqrt3}\sigma_y+ R_4q_xq_yq_z\sigma_z$ and $H_{1508}=R_2 (q_z^2-\frac{q_x^2+q_y^2}{2}) \sigma_x+\frac{\sqrt3 R_2(q_y^2-q_z^2)}{2}\sigma_y+ R_3q_xq_yq_z\sigma_z$  where the constant terms are omitted.}\label{table-2}
\end{table}

Inversely, for each $k\cdot p$ model, $H_i$ or $h_i$, all corresponding BNs are given in  \textbf{Supplementary Information II}, which could be useful in inversely searching for materials realizations given a $k\cdot p$ model. As shown in Fig. \ref{figure-1}(d), given $H_{985}$, the corresponding BNs can be realized in HSP A of MSGs 143.3, 144.6, 145.9 and so on. In the examples shown below in Secs. \ref{PT}, \ref{order} and \ref{large}, we use $k\cdot p$ models $H_{2026}$, $H_{2373}$, $H_{2469}$ and $H_{985}$, but remember that the concrete realizations in MSGs can be found in \textbf{Supplementary Information II} simply by searching ``H2026'', ``H2373'', ``H2469'' and ``H985'' respectively. Starting from these $k\cdot p$ models, we are then ready to study the topological characters. In this work, we numerically calculate all possible Chern numbers for all NPs, and find that spinful charge-4 WP can be realized in some type-I and IV MSGs as listed in Table \ref{table-2}. SrCuTe$_2$O$_6$ \cite{magndata} is found to be crystallized in  MSG 213.63 and could host the spinful charge-4 WP, which has never been proposed to be realized in electronic systems \cite{yao,TT}.

\subsection{Strictly NLs not captured by CRs}
As stated above, we find that there are several exceptions for which  there exist more additional strictly NSrs by $k\cdot p$ models  than those by CR-BSPs. The CR-BSPs for these exceptions are: P-NP, L-NP and P-NL, in which cases there could exist additional NLs by $k\cdot p$ models,  protected to be stable against any high order $k\cdot p$ corrections (see Tables. \ref{NSOC-mirror-NL}, \ref{SOC-mirror-NL}, \ref{NSOC-PT-NL}, \ref{NSOC-PT-NL-P-NL}, \ref{NSOC-mirror-NL-P-NL} and \ref{SOC-mirror-NL-P-NL} of \textbf{Appendix} for concrete $k\cdot p$ models). The corresponding BNs can be found in \textbf{Supplementary Information II} given $H_i$ in these tables. We find that these NLs are protected by $PT$/mirror/glide symmetry, through checking the little groups for the BNs for these $k\cdot p$ models:  For CR-BSP P-NP, when the BN owns a mirror/glide symmetry, we find that for all BNs corresponding the Tables. \ref{NSOC-mirror-NL} and \ref{SOC-mirror-NL} of \textbf{Appendix}, the eigenvalues of the mirror/glide operation are inverse. Since the BN is pinned at an HSP, there is no reason that the BN lies in an NL within an HSPL invariant under the mirror/glide operation. However, we use the explicit $k\cdot p$ models to obtain  that strictly NL for the BN indeed can be formed. Similarly, for CR-BSP P-NL in Tables \ref{NSOC-mirror-NL-P-NL} and \ref{SOC-mirror-NL-P-NL} of \textbf{Appendix}, there exist more strictly NLs than those by CR-BSPs, also protected by some mirror/glide symmetry. For Tables. \ref{NSOC-PT-NL} and \ref{NSOC-PT-NL-P-NL} of \textbf{Appendix}, the CR-BSP can be P-NP, L-NP or P-NL, there exist more strictly NLs protected by $PT$ symmetry in the spinless setting: Examples  are shown in Sec. \ref{PT} using the $k\cdot p$ models $H_{2026}$ and $H_{2373}$. Interestingly, such $PT$ protected  NLs  are fascinating since the points in these NLs are all flexible in the BZ other than one point pinned at the BN (thus can be diagnosed efficiently).  To our best knowledge, this type of NL, has not been reported before.

\subsubsection{Examples: $PT$ protected NLs}\label{PT}
\begin{figure*}[!htb]
  \includegraphics[width=1\textwidth]{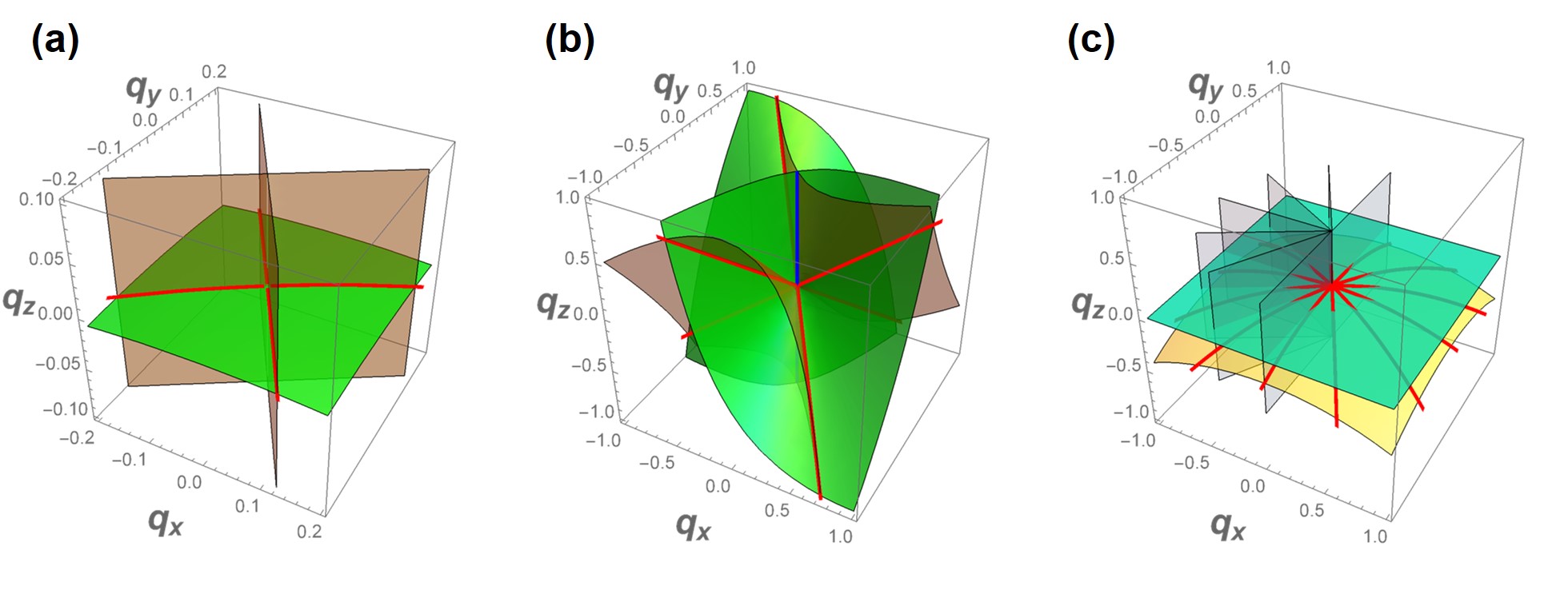}\\
  \caption{\textbf{$PT$ protected NLs in (a) and (b), $k\cdot p$ order enriched nodal structures in (c).} In (a), for the $k\cdot p$ model $H_{2026}=[R_4(q_y^2-q_x^2)+R_5q_xq_y]\sigma_x+[R_6q_z+R_7q_z^2+R_8(q_x^2+q_y^2)]\sigma_z$, whose CR-BSP can only be L-NP, we choose the parameters $R_4=R_5=0.1R_6=R_7=R_8=0.1$. The NLs for this $k\cdot p$ model can be visualized the intersection of two energy surfaces: The orange surface is $0.1(q_y^2-q_x^2)+0.1 q_x q_y=0$ while the green surface is $(q_z+0.1q_z^2)+0.1(q_x^2+q_y^2)=0$ and their intersection contains two NLs (in red) which are  pinned at the BN located at the origin. In (b), for the $k\cdot p$ model $H_{2373}=[R_4(q_y^2-q_x^2)+R_5q_xq_z+R_3q_yq_z+R_6q_xq_y]\sigma_x+(-R_3q_zq_x+2R_4q_xq_y+R_5q_yq_z+R_6\frac{q_x^2-q_y^2}{2})\sigma_y$, whose CR-BSP can be P-NL, we choose all the parameters to be 1.  The orange surface is $q_y^2-q_x^2+q_xq_z+q_yq_z+q_xq_y=0$ while the green surface is  $-q_zq_x+2q_xq_y+q_yq_z+\frac{q_x^2-q_y^2}{2}=0$. Their intersection contains one NL (in blue) along the $q_z$ axis which is required by CR-BSP and the other three NLs (in red) not required by CR-BSP. (c) demonstrates the $k\cdot p$ order enriched nodal structures using the $k\cdot p$ model $H_{2469}$: The  six NLs in red, lying in six HSPLs (in gray)  are required by CR-BSP (the CR-BSP is L-NL) but they only appear when the $k\cdot p$ order is 6 or larger. The cyan plane is along $q_x$-$q_y$ plane which is the solution for nodes when the $k\cdot p$ order takes $1$. The yellow surface is the solution for nodes when the $k\cdot p$ order is 2. When the $k\cdot p$ order is 3, 4 or 5, the yellow surface is almost still the node solution. The parameters we choose for the demonstration are shown in Sec. \ref{para} of \textbf{Appendix}.}\label{figure-3}
\end{figure*}

We give an example of a BN formed by the crossing of two bands with two different co-irreps: single-valued co-irreps \{1\} and \{3\},  within HSL $\Lambda$ or V of MSG 83.44 (type II). According to \textbf{Supplementary Information III}, the CR-BSP is L-NP and the $k\cdot p$ model is $H_{2026}$. From \textbf{Supplementary Information I},  the $k\cdot p$ model $H_{2026}$ is $[R_4(q_y^2-q_x^2)+R_5q_xq_y]\sigma_x+[R_6q_z+R_7q_z^2+R_8(q_x^2+q_y^2)]\sigma_z$ to the cutoff order 2,  where the constant term is omitted.   Though CRs require that the $k\cdot p$ model $H_{2026}$ should describe an NP by CR-BSP L-NP, $H_{2026}$ allows NL solutions, which are then found to be protected by $PT$ symmetry. Fig. \ref{figure-3}(a) demonstrates the $PT$ protected NLs where the concrete parameters are also given.

We then show another example, a BN formed at HSP $\Gamma$ of MSGs 147.14, 147.16, 148.18 or 148.20, with a single-valued co-irrep being \{2,6\}. According to \textbf{Supplementary Information III}, its CR-BSP is P-NL, as shown in Fig. \ref{C2}(a) of \textbf{Appendix}, where the NSr can be found along the HSL $\Delta$.  The $k\cdot p$ model is $H_{2373}=[R_4(q_y^2-q_x^2)+R_5q_xq_z+R_3q_yq_z+R_6q_xq_y]\sigma_x+(-R_3q_zq_x+2R_4q_xq_y+R_5q_yq_z+R_6\frac{q_x^2-q_y^2}{2})\sigma_y$ where the constant term is omitted, from \textbf{Supplementary Information I}.  Obviously, $q_x=0, q_y=0$ is the node solution of this model, namely, the corresponding NL is along $z$ direction (parallel to the HSL $\Delta$). However, there exist another three solutions corresponding to three NLs protected by $PT$ symmetry, as demonstrated in Fig. \ref{figure-3}(b) where we take all the parameters to be unity.

Note that though $PT$ symmetry can protect the formation of NL which can occur in any GP in principle \cite{nodal-line-balents,nodal-line-fangchen,cupdn-1,cupdn-2},  previously such NL \cite{cupdn-1,cupdn-2} was shown to be constrained to coincide with an HSL or lie within an HSPL, respectively, which is required by CR-BSP.   Here we show another possibility: We can also identify the $PT$ protected NL which is only pinned at BN,  simply by the irreps of bands in HSPs for all the BNs (all are doubly-degenerate) corresponding to Table. \ref{NSOC-PT-NL}  of \textbf{Appendix}.  The corresponding NLs cannot be diagnosed by CR-BSPs. For example, CR-BSP could require the BN to be simply an NP, and the $k\cdot p$ model restricted by the little group of the BN can  take the form of $(q_x^2+q_y^2+q_z^2)\sigma_x$ (thus the solution for NSr is simply $\mathbf{q}=(0,0,0)$).  As a matter of fact, the exhaustive study here shows  that  all BNs corresponding to Table. \ref{NSOC-PT-NL} of \textbf{Appendix} can give NLs by $k\cdot p$ models, even their CR-BSPs require the BNs are NPs. The NLs are only pinned at the BNs and the other parts can be tuned to be flexible in the BZ. The corresponding example has been shown using the $k\cdot p$ model $H_{2026}$.  For the BNs corresponding to Table \ref{NSOC-PT-NL-P-NL} of \textbf{Appendix}, whose CR-BSPs are all P-NL, other than the NLs (coinciding with some HSLs) by CR-BSPs, there exist other NLs only pinned at the HSP with the other parts flexible in the BZ. The corresponding example has been shown using the $k\cdot p$ model $H_{2373}$.  Moreover, the above two  $k\cdot p$ models $H_{2026}$ and $H_{2373}$ can be realized in type-II MSGs, for nonmagnetic materials. The newly discovered properties of NSrs in these systems are unveiled only by combining CRs and $k\cdot p$ models.

\section{Magnetic materials with symmetry-enforced extended NL/NS}\label{mat}
As an application, apply our results in the search for magnetic topological semimetals.  Compared with the great advancements in the nonmagnetic topological materials \cite{TI-RMP-3,franz}, the field of topological  magnetic materials \cite{tokura} is still in its infancy  possibly due to the scarcity of realistic magnetic materials with experimentally identified magnetic structures. Another intrinsic difficulty in predicting magnetic topological materials is originated in the electronic correlation, which should be considered very carefully in the first-principles calculations. Here we obtain a complete list of MSGs  with essentially NSs and NLs, corresponding to the CR-BSP PL-NS and L-sNL, respectively, as listed in detail in \textbf{Supplementary Information V}. Here ``essentially'' means that any (co-)irrep for these HSPLs/HSLs can constitute a BN lying in NS or NL. The information on essentially NLs/NSs in \textbf{Supplementary Information V} are thus useful for materials predictions, for both nonmagnetic and magnetic materials, and also applicable for bosonic excitations.

To date, the Bilbao MAGNDATA database \cite{magndata} has more than 1500 magnetic materials whose magnetic structures have already been measured experimentally.  Here, simply by matching MSGs of the magnetic materials  in this materials database with the MSGs in the spinful setting shown in  \textbf{Supplementary Information V}, we find hundreds of magnetic materials with commensurate magnetic structures  also listed in \textbf{Supplementary Information V}. In Table \ref{table-3} of the main text, we list the stoichiometric magnetic materials whose chemical formulas exclude O and haloid elements. The Fermi surfaces are expected to cross the extended nodal structures in these materials and lead to large responses and exotic properties \cite{n-sym-enf}. The present materials predictions are purely from symmetry and expected to be robust against weak electronic correlation \cite{filling-L,m-si,filling-np,filling-prb}, thus they are promising to be verified by experiments in near future. Note that  that these NLs and NSs can be judged by the information at HSPs according to the CR-BSPs P-NL, P-NS, P-NSL,  all possible patterns of them  being schematically shown in Sec. \ref{pattern} of \textbf{Appendix}. Interestingly, for P-NL, the maximal number of the extended NLs emanating from the HSP, is found to be 13. We expect this exotic nodal structures, diagnosed simply using the symmetry information at HSP, could show fascinating quantum responses when electrons shuttle around these NLs.

{\begin{table*}[!t]\footnotesize
  \begin{tabular}{l|l|l|l|l|l|l|l}
    \hline\hline
    {\color{black}1.136: AgCrS$_2$}&{\color{black}1.120: BaFe$_2$Se$_3$}&{\color{blue}0.449: Tb$_2$Pt}&{\color{cyan}0.481: HoNi}&{\color{cyan}0.664: Mn$_3$Sn$_2$}&{\color{black}0.176: Mn$_3$Ti$_2$Te$_6$}& {\color{black}0.226: NiCo$_2$}& {\color{black}0.397: Mn$_3$Si$_2$Te$_6$}\\\hline
    {\color{black}0.690: NdPt}&{\color{black}2.10: HoP}&{\color{black}2.70: GdMg}&{\color{blue}1.349: CoNb$_3$S$_6$}&{\color{blue}0.710: MnNb$_3$S$_6$}&{\color{cyan}1.98: DyFe$_4$Ge$_2$}&{\color{black}1.335: Nd$_2$Pd$_2$In}&{\color{cyan}1.0.48: MnSe$_2$}\\\hline
    {\color{cyan}1.18: MnS$_2$}&{\color{black}2.51: EuMnBi$_2$}&{\color{black}0.768: SrMnSb$_2$}&{\color{cyan}1.444: Er$_2$Pt}&{\color{cyan}1.86: GeV$_4$S$_8$}&{\color{cyan}1.33: ErAuGe}&{\color{cyan}1.34: HoAuGe}&
    {\color{cyan}1.504: GdCuSn}\\\hline
    {\color{cyan}1.505: GdAgSn}& {\color{cyan}1.506: GdAuSn}&{\color{black}1.267: Dy$_2$Co$_3$Al$_9$}&{\color{black}1.67: TmPtIn}&{\color{black}0.342: Tb$_3$Ge$_5$}&{\color{black}0.345: Tb$_2$C$_3$}&{\color{black}1.632: ErFe$_6$Ge$_6$}&{\color{black}1.633: YFe$_6$Sn$_6$}\\\hline
    {\color{black}2.12: TbMg}&{\color{black}2.11: TbMg}&{\color{black}1.356: Ho$_3$Ge$_4$}&{\color{black}1.628: PrMnSi$_2$}&{\color{black}1.634: YFe$_6$Ge$_6$}&{\color{black}1.457: NdNiMg$_{15}$}&{\color{black}2.30: CeRh$_2$Si$_2$}&{\color{black}1.22: DyB$_4$}\\\hline
    {\color{black}0.468: ErB$_4$}&{\color{black}1.368: Tb$_2$Ni$_3$Si$_5$}&{\color{black}2.68: FeGe$_2$}&{\color{black}0.563: Ce$_2$Ni$_3$Ge$_5$}&{\color{black}1.362: Er$_3$Ge$_4$}&{\color{black}1.8: CeRu$_2$Al$_{10}$}&{\color{blue}2.80: ErFe$_6$Ge$_6$}&{\color{cyan}2.57: TbMn$_2$Si$_2$}\\\hline
    {\color{cyan}2.60: NbMn$_2$Si$_2$}&{\color{black}1.252: CaCo$_2$P$_2$}&{\color{black}1.305: Mn$_5$Si$_3$}&{\color{black}1.460: PrCuSi}&{\color{black}2.67: FeSn$_2$}&{\color{black}0.562: Ce$_2$Ni$_3$Ge$_5$}&
    {\color{black}1.450: Pr$_6$Fe$_{13}$Sn}&{\color{black}1.556: FeSn$_2$}\\\hline{\color{black}1.557: FeGe$_2$}&{\color{black} 2.31: Mn$_3$ZnN}&{\color{black}1.0.47: MnSe$_2$}&{\color{black}2.1: EuFe$_2$As$_2$}&{\color{cyan}0.185: Nd$_5$Ge$_4$}&{\color{cyan}0.408: PrSi}&{\color{cyan}0.437: Ho$_3$NiGe$_2$}&{\color{cyan}0.438: Pr$_3$CoGe$_2$}\\\hline{\color{cyan}0.480: HoNi}&
    {\color{cyan}0.685: ErPt}&{\color{cyan}0.686: HoPt}&{\color{cyan}0.688: TmPt}&{\color{cyan}0.407: NdSi}&{\color{cyan}0.409: TmNi}&
    {\color{cyan}0.445: MnCoGe}&{\color{cyan}0.684: TbPt}\\\hline{\color{cyan}0.687: DyPt}&{\color{cyan}0.439: Tb$_3$NiGe$_2$}&{\color{cyan}0.662: Mn$_3$Sn$_2$}&{\color{cyan}0.767: SrMnSb$_2$}&{\color{black}1.130: Cr$_2$As}&{\color{black}1.131: Fe$_2$As}&{\color{black}1.132: Mn$_2$As}&{\color{black}1.28: CrN}\\\hline{\color{black}1.507: NdPd$_5$Al$_2$}&{\color{black}2.36: TbGe$_3$}&{\color{black}1.334: Pr$_2$Pd$_2$In}&{\color{black}1.475: DyNiAl$_4$}&{\color{black}1.401: Nd$_5$Pb$_3$}&{\color{blue}0.199: Mn$_3$Sn}&
    {\color{blue}0.279: Mn$_3$As}&{\color{blue}0.709: MnNb$_4$S$_8$}\\\hline{\color{blue}0.711: MnTa$_4$S$_8$}&{\color{blue}0.200: Mn$_3$Sn}&{\color{blue}0.280: Mn$_3$As}&{\color{blue}0.377: Mn$_3$Ge}&{\color{blue}0.706: Tb$_2$Ir$_3$Ga$_9$}&{\color{blue}0.771: PrMnSi$_2$}&
    {\color{blue}0.774: NdMnSi$_2$}&{\color{blue}0.776: CeMnSi$_2$}\\\hline{\color{blue}0.778: LaMnSi$_2$}&{\color{blue}3.3: Ho$_2$RhIn$_8$}&
    {\color{black}0.149: Nd$_3$Ru$_4$Al$_{12}$}&{\color{black}0.173: Pr$_3$Ru$_4$Al$_{12}$}&{\color{black}0.395: MnPtGa}&{\color{blue}0.561: NdNiGe$_2$}&{\color{black}0.689: PrPt}&{\color{black}1.195: Er$_2$Ni$_2$In}\\\hline{\color{black}1.415: Tb$_2$Pd$_2$In}&{\color{black}1.446: CeCoAl$_4$}&{\color{black}3.13: CeB$_6$}&{\color{black}1.392: KCuMnS$_2$}&{\color{black}1.142: CeMgPb}&{\color{black}0.403: NdCo$_2$}&
    {\color{black}2.28: NpNiGa$_5$}&{\color{blue}0.184: Nd$_5$Si$_4$}\\\hline{\color{cyan}1.85: $\alpha$-Mn}&{\color{black}1.152: Ce$_3$NIn}&{\color{cyan}1.574: NdBiPt}&{\color{black}0.448: Ce$_4$Ge$_3$}&{\color{black}0.681: Ce$_4$Sb$_3$}&{\color{black}2.44: KCuMnS$_2$}&
    {\color{black}1.451: Nd$_6$Fe$_{13}$Sn}&{\color{black}2.14: NdMg}\\\hline{\color{black}1.253: CeCo$_2$P$_2$}&{\color{black}1.425: UGeTe}&{\color{black}1.453: EuMn$_2$Si$_2$}&{\color{black}1.468: TbMn$_2$Si$_2$}&{\color{black}1.469: YMn$_2$Si$_2$}&{\color{black}1.488: CeMn$_2$Si$_2$}&{\color{black}1.491: PrMn$_2$Si$_2$}&{\color{black}1.493: NdMn$_2$Si$_2$}\\\hline{\color{black}1.496: YMn$_2$Ge$_2$}&{\color{black}0.320: U$_2$Pd$_2$In}&{\color{black}0.321: U$_2$Pd$_2$Sn}&{\color{black}1.102: U$_2$Ni$_2$In}&{\color{blue}0.593: UPSe}&{\color{blue}0.594: UAsS}&
    {\color{black}1.215: UP$_2$}&{\color{black}1.271: CeSbTe}\\\hline{\color{black}1.384: USb$_2$}&{\color{black}1.426 UGeS}&{\color{black}1.535 UPd$_2$Ge$_2$}&{\color{black}1.143: Mn$_3$Pt}&{\color{black}2.13: UP}&{\color{black}2.20: UAs}&{\color{black}1.103: U$_2$Rh$_2$Sn}&{\color{black}1.479: U$_2$Ni$_2$Sn}\\\hline{\color{black}1.539: KMnP}&{\color{black} 1.541: RbMnP}&{\color{black}1.543: RbMnAs}&{\color{black}1.545: RbMnBi}&{\color{black}1.546: CsMnBi}&{\color{black}1.547: CsMnP}&
    {\color{black}1.550: LiMnAs}&{\color{black}1.553: KMnAs}\\\hline{\color{black}0.236: CaFe$_4$Al$_8$}&{\color{black}0.324: CdYb$_2$S$_4$}&{\color{black}0.325: CdYb$_2$Se$_4$}&{\color{black}0.227: NdCo$_2$}&{\color{black}0.613: FeCr$_2$S$_4$}&{\color{black}0.169: U$_3$As$_4$}&
    {\color{black}0.170: U$_3$P$_4$}&{\color{black}2.32: Dy$_3$Ru$_4$Al$_{12}$}\\\hline{\color{black}0.286: Mn$_5$Ge$_3$}&{\color{black}0.673: MnFe$_4$Si$_3$}&{\color{blue}0.528: CrSb}&{\color{black}1.412: Au$_{72}$Al$_{14}$Tb$_{14}$}&{\color{black}0.150: NiS$_2$}&{\color{black}0.20: MnTe$_2$}&
    {\color{black}3.6: DyCu}&{\color{red}0.440: SrCuTe$_2$O$_6$}
    \\\hline
  \end{tabular}
  \caption{\textbf{Magnetic materials realizing charge-4 WP, symmetry-enforced essential and extended NL (L-sNL)  and NS (PL-NS).} The materials are given by the entry label in Bilbao MAGNDATA database  \cite{magndata} and the chemical formula. The red/black/blue colors are for materials with charge-4 WP/essential NL/essential NS, respectively, while the cyan color is for materials with coexisting essential NL and NS.}\label{table-3}
\end{table*}}

\section{Conclusion and Discussion}
Though there have been many comprehensive studies on BNs, rare attention has been paid to a thorough investigation on all NSrs, especially NLs and NSs. An exhaustive study on all possible MSGs and MLGs for BNs is also lacking.  Also, nearly NSrs should be placed on the same footing as strictly ones, as we show in this work.  To summarize, we  obtained a comprehensive classification and calculation of all BNs and their NSrs  in all the 1651 MSGs and 528 MLGs for both spinful and spinless settings: In \textbf{Supplementary Information I}, we list all $k\cdot p$ models in $H_i$ ($i=1,2,\ldots,3026$) and $h_i$ ($i=1,2,\ldots,601$) which are complete to describe all low energy models around all BNs in both MSGs and MLGs and in both spinful and spinless settings. We also show solutions of strictly and nearly NSrs in  \textbf{Supplementary Information I}. In \textbf{Supplementary Information II}, for each $k\cdot p$ model,  $H_i$ or $h_i$, all possible BNs (the MSG/MLG, spinful/sinless setting, $k$ point and (co-)irrep(s)) are given.  For detailed CRs from the BNs, the reader can find the results in \textbf{Supplementary Information III}, given in the order of MSGs/MLGs. In \textbf{Supplementary Information IV}, we show all BNs (the MSG/MLG, spinful/spinless setting, $k$ point, (co-)irrep(s) and the $k\cdot p$ model in the form of $H_i$ or $h_i$) for a given CR-BSP.

The results for type-I, III and IV MSGs/MLGs can be applied in magnetic materials while those for type-II MSGs/MLGs are applicable in nonmagnetic materials. Notably, considering nearly NSrs leads to more fruitful band NSrs. The cutoff $k\cdot p$ order $o$ can quantitatively characterize the order of magnitude of gap between the bands (required to be separated by MSG/MLG symmetries).  The nearly NSrs are  meaningful experimentally since they could be applied to interpret the observed phenomena better than simply the strictly NSrs, whose possibilities are restricted by the (co-)irreps in crystalline materials.

The new types of BNs uncovered in this work:  Weyl NLs pinned at HSPs/HSLs and $k\cdot p$ order enriched curved NS can be realized in concrete materials. The magnetic materials shown in Table \ref{table-3} for extended NLs/NSs are expected to be verified experimentally in near future. Also, with the development of determining magnetic structures by neutron scattering or other techniques, more magnetic materials crystallizing in MSGs as listed in Table \ref{table-2} can be synthesized to realize the spinful charge-4 WP.

Though the BNs in this work are all symmetry-enforced or diagonalizable by symmetry eigenvalues, the BNs in GPs are easily to analyzed: When a BN is formed at a GP in the spinless setting \cite{nodal-line-fangchen, nodal-line-balents} and the BN is only subject to the $PT$ symmetry, the $k\cdot p$ model is $\mathbf{v}_1\cdot\mathbf{q}\sigma_x+\mathbf{v}_2\cdot\mathbf{q}\sigma_z$ ($\mathbf{v}_i$ is a real vector), thus the NSr is an NL: $\mathbf{v}_1\cdot\mathbf{q}=0$ and $\mathbf{v}_2\cdot\mathbf{q}=0$. For the other cases, including WP at GP \cite{PRB-Wan, TaAs-PRX, TaAs-NC} in 3D systems, WP protected by $PT$  symmetry in 2D systems in the spinless setting, or WP protected by $c_2T$ ($c_2$ is a two-fold rotation, possibly followed by a fractional translation) symmetry in 2D systems ($(c_2T)^2=1$), CRs predict that these BNs are simply an NP, the same as the predictions by the $k\cdot p$ model.

The NSrs in this work are all emanating from the BNs.  For Hopf-link nodal loops \cite{hopf-link-1,hopf-link-2,hopf-link-3,hopf-link-4,hopf-link-5,hopf-link-6,hopf-link-7}, whose NSrs are nested, our results can be applied to identify each nodal loop and whether the obtained nodal loops are nested need a case-by-case check. Finally, our methodology of classifying NSrs combining CRs and $k\cdot p$ models might be applied to superconductors with particle-hole symmetry \cite{sc-1,sc-2} or other situations with symmetries that escape standard SG analysis \cite{dual} in future works.
\section{Acknowledgement}
F.T. was supported by National  Natural  Science Foundation of China (NSFC) under Grants No. 12104215. F.T. and X.W. were supported by the National Key R\&D Program of China (Grants No. 2018YFA0305704), NSFC Grants No. 12188101, No. 11834006, No. 51721001, and No. 11790311, and the excellent program at Nanjing University. X.W. also acknowledges the support from the Tencent Foundation through the XPLORER PRIZE.
\clearpage
\appendix
\begin{widetext}
\renewcommand{\thefigure}{A\arabic{figure}}
\renewcommand{\thetable}{A\arabic{table}}
\renewcommand{\thesection}{A\arabic{section}}
\renewcommand{\theequation}{A\arabic{equation}}
Organization of the Appendix:\\
\noi{\ref{condition}} We describe the formation conditions for ten CR-BSPs and the concrete examples are also given.\\
\noi{\ref{pattern}} We show all possible patterns for CR-BSP P-NL, P-NS and P-NSL.\\
\noi{\ref{mlg}} We show how to obtain MLG from an MSG.\\
\noi{\ref{matrixbasis}} We give explicitly all bases to express the $k\cdot p$ models around all BNs.\\
\noi{\ref{node}} We show all equations for NSrs of a given $k\cdot p$ model, expressed by the matrix basis in Sec. \ref{matrixbasis}.\\
\noi{\ref{more}} We give all $k\cdot p$ models, $H_i^,$s, whose BNs have additional NLs, not captured by CR-BSPs.\\
\noi{\ref{para}} We give the parameters used in $H_{2469}$ to plot Fig. \ref{figure-3}(c) in the main text.\\
\noi{\ref{peak}} We show the distribution of Berry curvature in the sphere around the NP, whose $k\cdot p$ model is $H_{985}$, and the parameters are also provided.\\
\noi{\ref{enrichns}} We print all the 1651 MSGs in 10 different colors and each color denotes which type of NS can be realized in the MSG. The meaning of each color is illustrated in Table \ref{color}.\\
\noi{\ref{page}} We give all page numbers in \textbf{Supplementary Information III} corresponding to the results of any given MSG.\\

\section{The four conditions for forming ten CR-BSPs}\label{condition}
In this section of \textbf{Appendix}, we describe in detail the formation conditions $C_1$-$C_4$ for all possible ten CR-BSPs. As shown below, we use $u$ to describe the representation in the BN. It might contain only one degenerate (co-)irrep when the BN is located at HSP/HSL/HSPL. It might also contain two different (co-)irreps when the BN is located at HSL/HSPL.  We also call $u$ is stuck in $k'$ where $k'$ may be an HSL/HSPL or even GP, the neighborhood of the BN, if:

1. the CR is $u\rightarrow u'$ ($u'$ is some (co-)irrep of $k'$), when $u$ contains only one (co-)irrep, or,

2. the CR is $u\rightarrow u'+u''$ ($u'$ and $u''$ are two different (co-)irreps of $k'$), when $u$ contains two different (co-)irreps.

For a BN to occur, $u$ should not be stuck in all directions from the BN. However, in some special direction, $u$ might be stuck, thus this direction is possible to become an NSr. Note that when $u$ contains two different (co-)irreps, it must be a BN, since it definitely becomes more than one (co-)irreps in GP.
\subsection{$C_1:$  $u$ is not stuck in the whole neighborhood.}
This is the condition for the formation of NP coinciding with the BN. The BN can be pinned at an HSP, owning one degenerate (co-)irrep  or lie in an HSL owing two different (co-)irreps.  The resulting CR-BSP is P-NP or L-NP, respectively. Fig. \ref{C1} shows the examples. In Fig. \ref{C1}(a), the BN occurs at the HSP D for type IV MSG 13.74. The co-irrep of the BN is \{1,1\} \cite{tang-kp} whose degeneracy is 4, in the spinless setting as shown in the figure. The $k\cdot p$ model is $H_{1788}$. The CR-BSP is P-NP. From this co-irrep, all related CRs are shown indicated by $\rightarrow$. The related directions are GP, HSL V and HSPL P1. All the CRs imply that the bands away from the BN should  split: along the direction for GP, the bands should split to four nondegenerate bands; along the direction for V and P1, the bands should split to two doubly-degenerate bands.

In Fig. \ref{C1}(b), we take type III  MSG 84.54 as the example of CR-BSP L-NP. The BN is formed of two different co-irreps \{1\} and \{3\} of HSL V, whose dimensions are both 1. Hence, the degeneracy of the BN is 2. All possible CRs are listed: the CRs to GP, P4 and P5. Obviously, the CRs require that, in the directions for P4 and P5,  the two irreps of the BN become the same co-irrep: \{1\}. Thus, the bands cannot be stuck in these directions. For the direction for GP, the bands definitely split.

\begin{figure}[!htb]
  \includegraphics[width=1\textwidth]{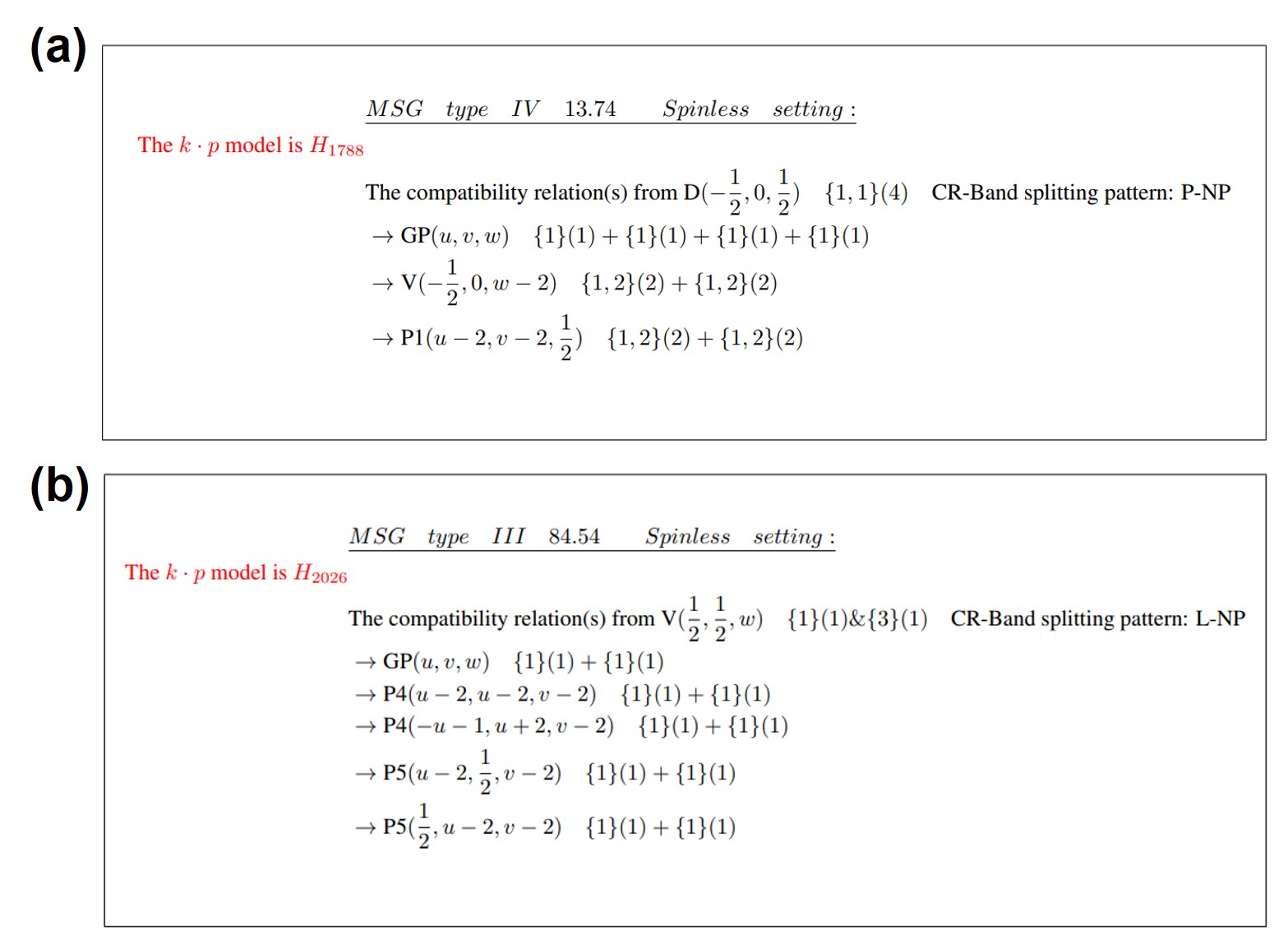}\\
  \caption{\textbf{Examples of $C_1$ to form CR-BSP P-NP (a) or L-NP (b).}}\label{C1}
\end{figure}

%With respect to BN at HSP,  we need to check the CRs for all neighboring independent HSLs and HSPLs. Obviously, once some neighboring HSL lies within a neighboring HSPL, we only need to require that the CR for the HSL should not be stuck. With respect to BN in HSL, along the HSL, $u$ definitely splits. Hence, we only need to check neighboring HSPLs containing the HSL: The two (co-)irreps in the BN should not keep to two different ones in any neighboring HSPL.
\begin{figure}[!htb]
  \includegraphics[width=1\textwidth]{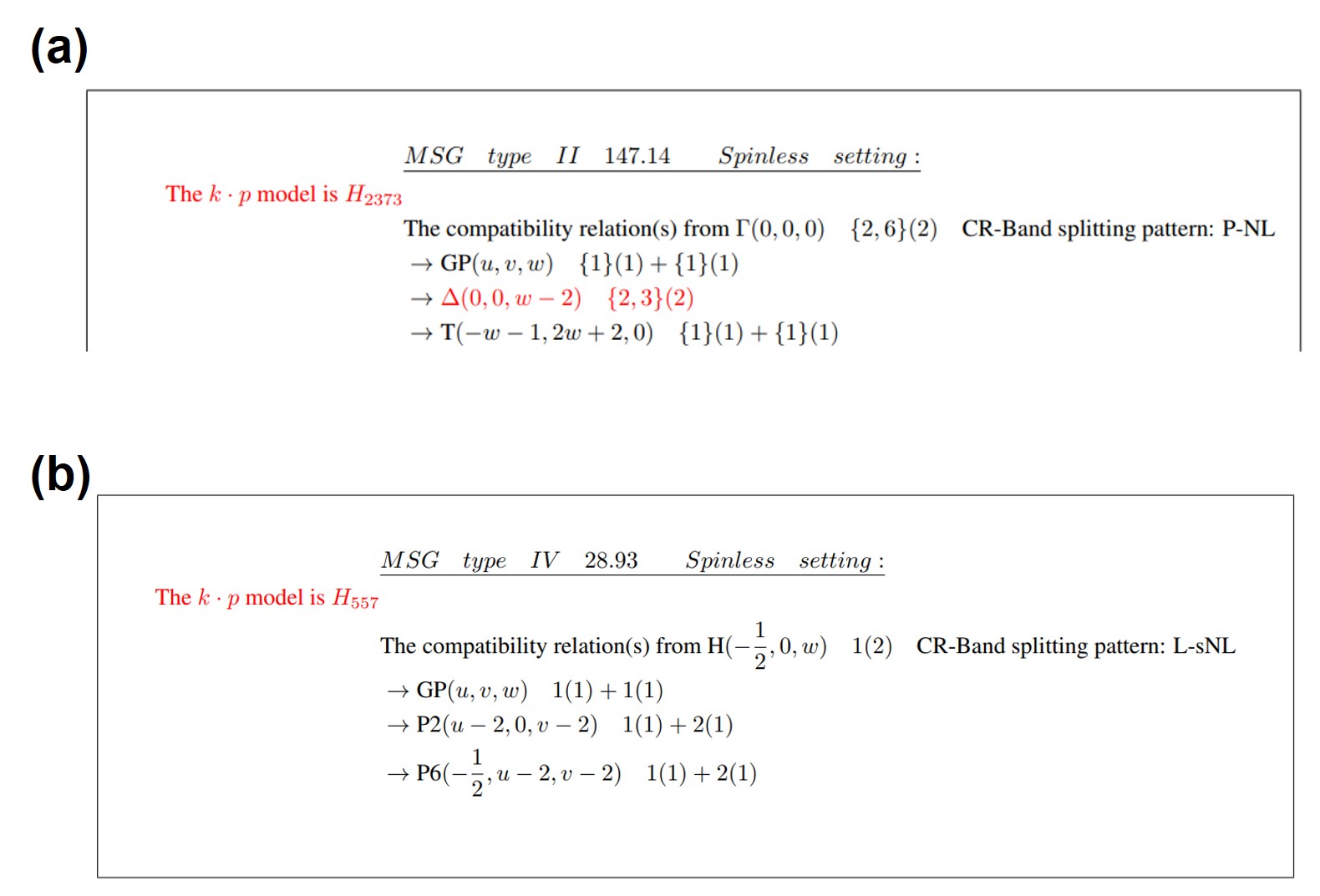}\\
  \caption{\textbf{Examples of $C_2$ to form CR-BSP P-NL (a) or L-sNL (b).}}\label{C2}
\end{figure}

\subsection{$C_2:$  $u$ is not stuck in  the neighborhood other  than some HSL(s).}
This is the condition for the formation of NL coinciding with an HSL or several HSLs, thus the NL should be straight. The BN can be pinned at an HSP or within an HSL, both containing one degenerate (co-)irrep, corresponding CR-BSP P-NL or L-sNL, respectively. When the BN is within an HSL, the NL simply coincides with the HSL.

Figs. \ref{C2}(a) and (b) demonstrate the examples of CR-BSP P-NL and L-sNL, respectively. For P-NL, we take type II MSG 147.14 as the example. The $k\cdot p$ model is $H_{2373}$. The co-irrep in the BN is \{2,6\} of the HSP $\Gamma$ and the degeneracy is 2. The bands split in all directions other than that for the HSL $\Delta$, as printed in red in Fig. \ref{C2}(a) (Note CRs are not shown completely here).  In Fig. \ref{C2}(b), the HSL H of type IV MSG 28.93 is an NL when the participating irrep is 1. The CR-BSP for any BN is this HSL is L-sNL.
\begin{figure}[!htb]
  \includegraphics[width=1\textwidth]{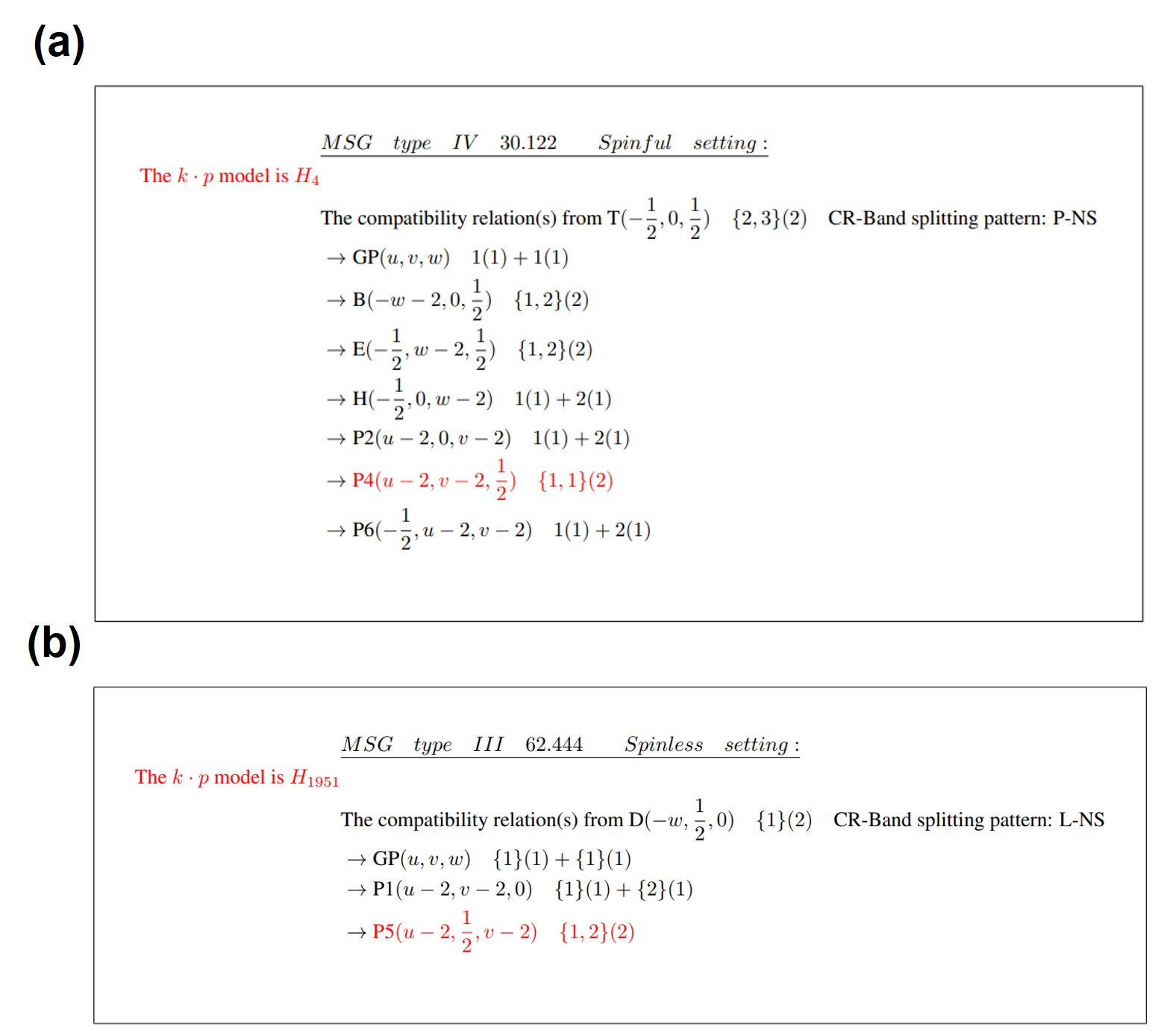}\\
  \caption{\textbf{Examples of $C_3$ to form CR-BSP P-NS (a) or L-NS (b).}}\label{C3-1}
\end{figure}

\begin{figure}[!htb]
  \includegraphics[width=1\textwidth]{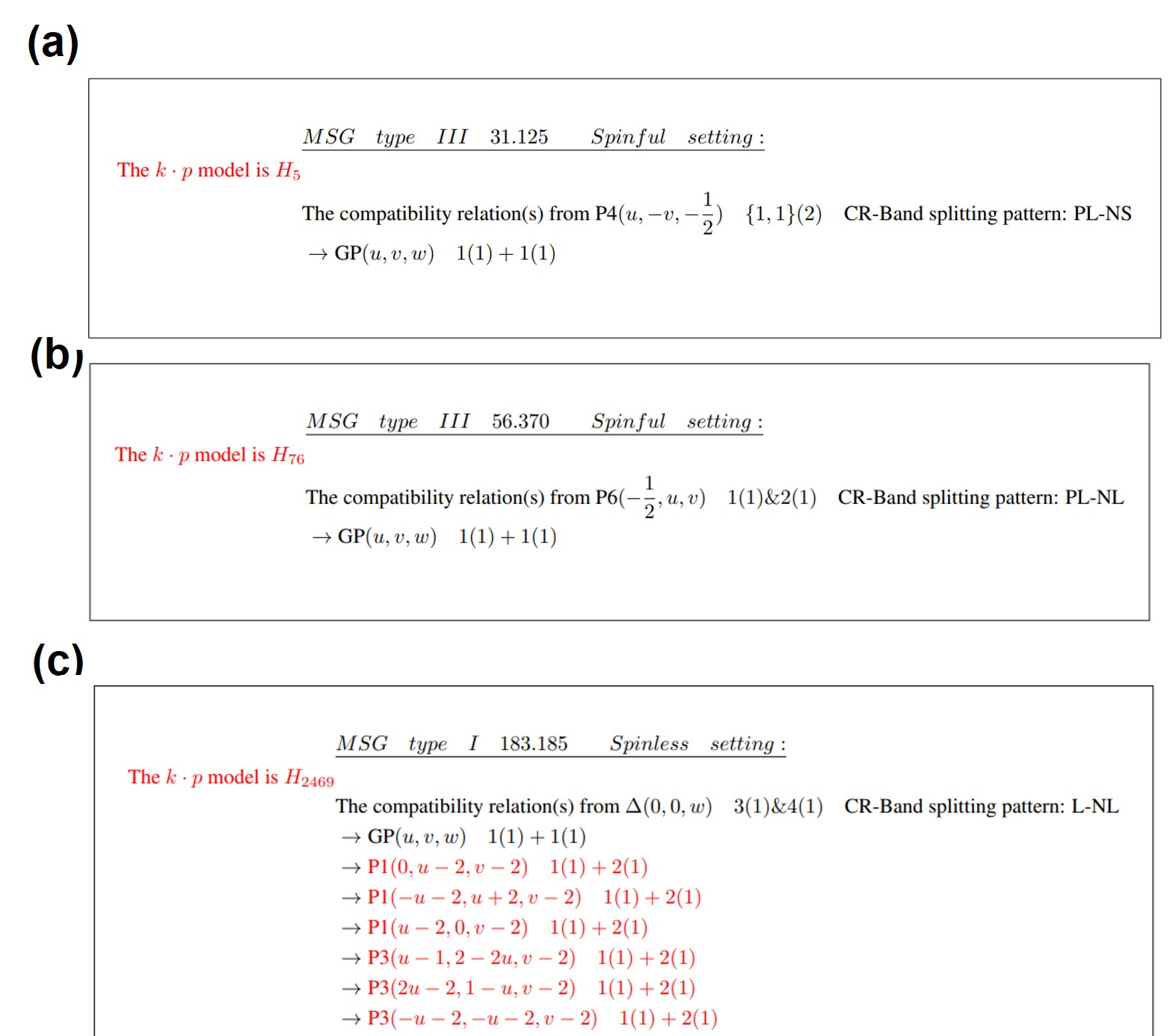}\\
  \caption{\textbf{Examples of $C_3$ to form CR-BSP  PL-NS (a), PL-NL (b) or L-NL (c).}}\label{C3-2}
\end{figure}

\subsection{$C_3:$  $u$ is not stuck in  the neighborhood other than some HSPL(s).}
This is the condition for the formation of NS or NL   coinciding with or lying in an HSPL, respectively. For NS, the BN can be pinned at an HSP, within an HSL or HSPL, all containing one degenerate (co-)irrep.  The corresponding CR-BSP can be P-NS, L-NS or PL-NS, respectively, and the NS simply coincides with an  HSPL. When the BN is pinned at an HSP or within a HSL, the NS coincides with a neighboring HSPL. When the BN is in  an HSPL, the NS simply coincides with this HSPL.  For NL, the BN can be in an HSL or HSPL, both containing two different (co-)irreps. The CR-BSP can be L-NL or PL-NL, respectively. When the BN is in an HSL, the HSPL(s) containing the HSL host the NL(s), thus a nodal chain structure could emerge if more than one HSPLs exist. When the BN is in an HSPL, this HSPL hosts the NL.

We show the examples of CR-BSP P-NS, L-NS and PL-NS in Figs. \ref{C3-1} (a), (b) and Fig. \ref{C3-2}(a). For P-NS, the MSG is 30.122, of type IV. The BN occurs in the HSP T with co-irrep \{2,3\}. From the CRs related with GP, B, E, H, P2, P4 and P6, we can see that only along B, E and P4 should the bands keep to be degenerate. However, the HSLs B and E are not NL since they both lie in the HSPL P4. Actually, P4 is an NS, thus B and E are parts of this NS. For L-NS, the MSG is 62.444 of type III. The BN is at the HSL D. Away from the BN and along GP, P1 and P5, all the related CRs indicate that the bands split other than P5. Along P5, the bands keep degenerate, which is actually an NS. In Fig. \ref{C3-2}(a), the HSPL P4 for MSG 31.125 is an NS. Actually, the MSG has no $PT$ symmetry otherwise, the HSPL is not an NS since GP would be always doubly-degenerate (in the spinful setting).

Figs. \ref{C3-2}(b) and (c) both show a BN whose NSr(s) are NL(s) and containing two different irreps. In Fig. \ref{C3-2}(b), the BN is in an HSPL (P6), containing two different irreps, 1 and 2. They are definitely two different irreps in the HSPL. Then the BN lies in an NL within in the HSPL. The corresponding CR-BSP is PL-NL.   In Fig. \ref{C3-2}(c), the BN is in an HSL ($\Delta$) and contains two different irreps 3 and 4, which are found to keep different in neighboring HSPLs P1 and P3 printed in red. Hence, the HSPLs P1 and P3 are NSs. The CR-BSP is L-NL.

\subsection{$C_4:$  $u$ is not stuck in the neighborhood other  than some HSL(s) and HSPL(s).}
This is only for BN pinned at HSP and the NSrs contain both NL and NS, coinciding with HSL and HSPL, respectively. As shown in Fig. \ref{C4}, the bands away from the BN at the HSP Z should be stuck in the directions along HSLs U, $\Lambda$, S and HSPL P2. P2 is an NS, which contains U and S while $\Lambda$ is an NL. Finally, the NSrs contain both NL and NS. The pattern is shown in Fig. \ref{figure-2}(b).
\begin{figure}[!htb]
  \includegraphics[width=1\textwidth]{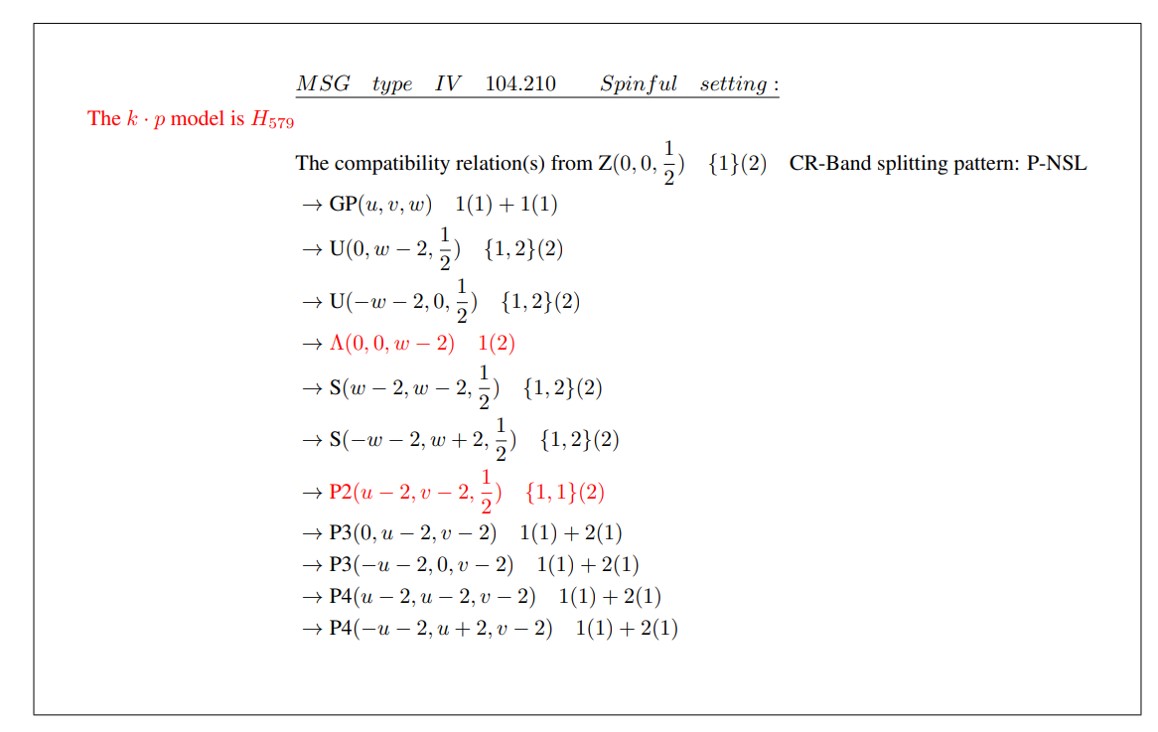}\\
  \caption{\textbf{Example of $C_4$ to form CR-BSP P-NSL.}}\label{C4}
\end{figure}

\section{All possible patterns of CR-BSPs P-NL, P-NS and P-NSL}\label{pattern}
Figure \ref{figure-2} shows all possible patterns of CR-BSPs P-NL, P-NS and P-NSL. We emphasize  the CR-BSP P-NL with the maximal number (13) of NLs coinciding with HSLs:  (g)+$c_3$(g)+$c_3^{-1}$(g)+(j)  where $c_3$ denotes the rotation of $2\pi/3$ around $(1,1,1)$ direction. The diagnosis of NLs need no calculations of Berry phase, but simply from the symmetry information at an HSP. The large number of NLs emanating from the HSP  and their extended nature in the BZ makes this kind of CR-BSP promising to result in exotic responses to external fields.
\begin{figure*}[!htbp]
  \includegraphics[width=1\textwidth]{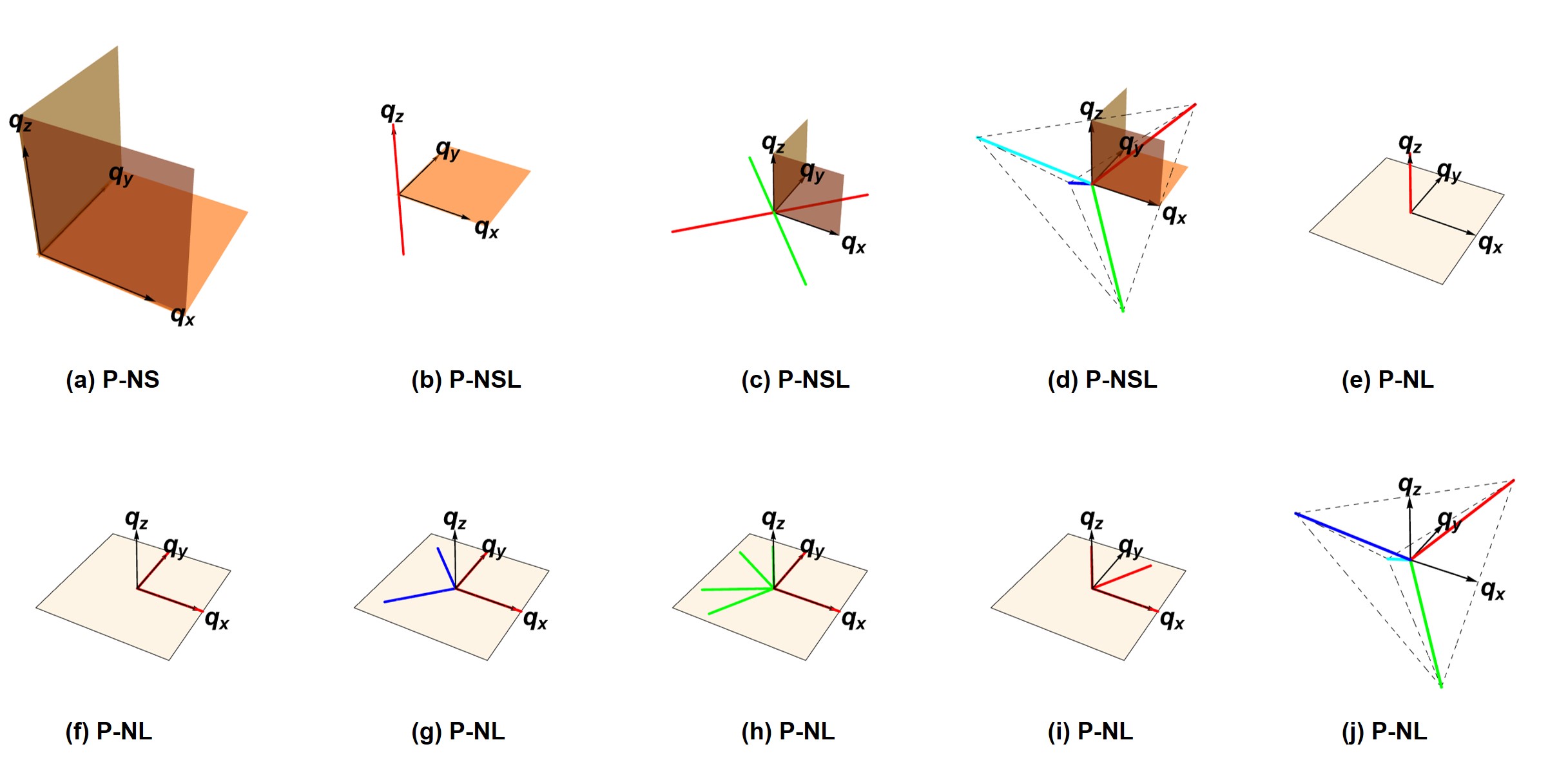}
  \caption{\textbf{All possible patterns of CR-BSPs P-NL, P-NS and P-NSL.} The BN at an HSP is set as the origin. In (a-d), orange plane denotes the NS emanating from the HSP and coinciding with some HSPL. Red, green, cyan and blue thick lines denote the NL emanating from the HSP and coinciding with some HSLs.  In (a), we demonstrate one pattern of P-NS with three planes being NSs.  Note that there are another two patterns of P-NS, not shown here, with only one plane  or two planes as NS(s), respectively. In (b-d), we show all three different patterns of P-NSL. In (b), the NL is along $q_z$ axis while the NS is along $q_x$-$q_y$ plane. In (c), two NSs are along $q_z$-$q_x$ and $q_y$-$q_z$ planes while two NLs, perpendicular to each other, lie in the $q_x$-$q_y$ plane but have an angle $\pi/4$ with respect to the $q_x$ axis. In (d), three NSs are along $q_x$-$q_y$, $q_y$-$q_z$ and $q_z$-$q_x$ planes. Four NLs are along directions $(1,1,1),(1,-1,-1),(-1,1,-1),(-1,-1,1)$, respectively. In (e-j), we show 6 (of 13 in total) patterns of P-NL while the other 7 patterns can be obtained by adding or rotating them. In (e), there is only one NL. In (f), two NLs are perpendicular to each other. In (g), the NLs in blue or red are perpendicular to each other but the blue NLs have an angle $\pi/4$ with respect to the red ones. All these NLs lie in one plane. In (h), there are 6 NLs lying in one plane. The red NLs are along $q_x$ or $q_y$ axes.  The green NLs have an angle $\pi/6$ with respect to the $q_x$ or $q_y$ axis. In (i), all three NLs lie in one plane and are along $q_x$ axis, directions $(1,\sqrt3,0)$ and $(-1,\sqrt 3,0)$, respectively. In (j), the four NLs are the same as those in (d). The rest 7 patterns of P-NL are: (e)+(f), (e)+(g), (e)+(h), (e)+(i), (e)+(f)+(j), (g)+$c_3$(g)+$c_3^{-1}$(g),  (g)+$c_3$(g)+$c_3^{-1}$(g)+(j).  Here $c_3$ denotes the rotation of $2\pi/3$ around direction $(1,1,1)$. }\label{figure-2}
\end{figure*}
\section{From MSGs to MLGs}\label{mlg}
In this work, we express MLGs based on parent MSGs. Only the following Bravais lattices in Table \ref{layer} could allow layer structures. Then to see whether an MSG could allow a layer structure, one crystal axis $\mathbf{a}, \mathbf{b}$ or $\mathbf{c}$ is chosen to be the direction along which the 2D layers are stacking to form the 3D structure. In practical first-principles calculation, the 2D layer structure is treated as a 3D structure in such way but  its length along the stacking direction should be set to be very large. This is why we express the MLG in terms of an MSG.  Consider the point group operations of the MSG corresponding to an MLG. They should let the stacking direction to be itself or the inverse. Note that for  monoclinic base-centered Bravais lattice, the stacking direction can only take $\mathbf{b}$, while   for  orthorhombic base-centered Bravais lattice, the stacking direction can only take $\mathbf{c}$. Through the construction in this way, we obtain 597 MLGs in total (which can be reduced to 528 different MLGs). For each MLG, we show the parent MSG and the stacking direction. All $k$ points and the (co-)irreps of their little groups still follow those for the parent MSGs \cite{bradley, tang-kp}.
\begin{table*}[!h]
  \centering
  \begin{tabular}{c|c|c|c|c|c|c}
    \hline
    Triclinic & Monoclinic & Monoclinic base-centered & Orthorhombic & Orthorhombic base-centered & Tetrahedral& Hexagonal\\
    \hline
  \end{tabular}
  \caption{\textbf{The Bravais lattices allowing layer structures}}\label{layer}
\end{table*}
\section{Matrix basis sets to express $k\cdot p$ models}\label{matrixbasis}
To express the $k\cdot p$ model for the BN, we  need to specify the matrix basis sets for $2\times 2$ matrix, $3\times 3$ matrix and $4\times 4$ matrix.  The degeneracy of BN can take 2, 3, 4, 6, 8. For $6\times 6$ $k\cdot p$ model $H_j$, we first divide it into a block form as $\left(\begin{array}{cc}\{H_j\}_{1\sim 3,1\sim3}&\{H_j\}_{1\sim 3,4\sim6}\\\{H_j\}_{1\sim 3,4\sim6}^\dag&\{H_j\}_{4\sim 6,4\sim6}\end{array}\right)$ so that each block can be expanded on matrix basis set for $3\times 3$ matrix. Similarly, for $8\times 8$ $k\cdot p$ model $H_j$, we first divide it into a block form as $\left(\begin{array}{cc}\{H_j\}_{1\sim 4,1\sim4}&\{H_j\}_{1\sim 4,5\sim8}\\\{H_j\}_{1\sim 4,5\sim8}^\dag&\{H_j\}_{5\sim 8,5\sim8}\end{array}\right)$ so that each block can be expanded on matrix basis set for $4\times 4$ matrix.\\

With respect to the matrix basis set for $2\times2$ matrix, we use four matrices: $\sigma_{\mu}, \mu=0,1,2,3$. $\sigma_0$ is $2\times 2$ identity matrix. $\sigma_1$, $\sigma_2$ and $\sigma_3$ are Pauli matrices. For $4\times4$ $k\cdot p$ model, we use 16 $\Gamma$ matrices defined by $\Gamma_{\mu\nu}=\sigma_\mu\otimes\sigma_\nu$.  For $3\times3$ $k\cdot p$ model, we use 9 matrices to expand it as defined in the following:
$L_x=\left(
\begin{array}{ccc}
 0 & 0 & 0 \\
 0 & 0 & -i \\
 0 & i & 0 \\
\end{array}
\right)$,$L_y=\left(
\begin{array}{ccc}
 0 & 0 & i \\
 0 & 0 & 0 \\
 -i & 0 & 0 \\
\end{array}
\right)$,$L_z=\left(
\begin{array}{ccc}
 0 & -i & 0 \\
 i & 0 & 0 \\
 0 & 0 & 0 \\
\end{array}
\right)$,$S_1=\left(
\begin{array}{ccc}
 0 & 1 & 0 \\
 1 & 0 & 0 \\
 0 & 0 & 0 \\
\end{array}
\right)$,$S_2=\left(
\begin{array}{ccc}
 0 & 0 & 1 \\
 0 & 0 & 0 \\
 1 & 0 & 0 \\
\end{array}
\right)$,$S_3=\left(
\begin{array}{ccc}
 0 & 0 & 0 \\
 0 & 0 & 1 \\
 0 & 1 & 0 \\
\end{array}
\right)$,$I_0=\left(
\begin{array}{ccc}
 \sqrt{\frac{2}{3}} & 0 & 0 \\
 0 & \sqrt{\frac{2}{3}} & 0 \\
 0 & 0 & \sqrt{\frac{2}{3}} \\
\end{array}
\right)$,$I_1=\left(
\begin{array}{ccc}
 \frac{1}{\sqrt{3}} & 0 & 0 \\
 0 & \frac{1}{\sqrt{3}} & 0 \\
 0 & 0 & -\frac{2}{\sqrt{3}} \\
\end{array}
\right)$ and $I_2=\left(
\begin{array}{ccc}
 1 & 0 & 0 \\
 0 & -1 & 0 \\
 0 & 0 & 0 \\
\end{array}
\right)$.

\section{Solve for NSrs of a given $k\cdot p$ model} \label{node}
Using the above matrix set, it is then convenient to obtain a series of equations on $q_x, q_y, q_z$ whose solutions are those for BNs (among which we are only interested on the emanating ones, namely, the solutions with $\mathbf{q}=(0,0,0)$ being a special case). In the following, we explicitly list these equations for all situations that could be encountered.\\
For $2\times 2$ $k\cdot p$ model $H=\sum_{\mu=0}^3f_\mu\sigma_\mu$, the equations for BNs are:
\[\left\{\begin{array}{c}
f_1=0,\\
f_2=0,\\
f_3=0
\end{array}\right.\]
For $4\times 4$ $k\cdot p$ model $H=\sum_{\mu,\nu=0}^3f_{\mu\nu}\Gamma_{\mu\nu}$, the equations for BNs are:
\[\left\{\begin{array}{cccc}
f_{01}=0,&f_{02}=0,&f_{03}=0,&\\
f_{10}=0,&f_{11}=0,&f_{12}=0,&f_{13}=0,\\
f_{20}=0,&f_{21}=0,&f_{22}=0,&f_{23}=0,\\
f_{30}=0,&f_{31}=0,&f_{32}=0,&f_{33}=0
\end{array}\right.\]
For $3\times 3$ $k\cdot p$ model $H=f_1L_x+f_2L_y+f_3L_z+f_4S_1+f_5S_2+f_6S_3+f_7I_0+f_8I_1+f_9I_2$, the equations for BNs are:
\[\left\{\begin{array}{ccc}
f_1=0,&f_2=0,&f_3=0,\\
f_4=0,&f_5=0,&f_6=0,\\
f_8=0,&f_9=0&
\end{array}\right.\]

For $6\times 6$ $k\cdot p$ model $H$ with $\{H\}_{1\sim3,1\sim3}=f_1L_x+f_2L_y+f_3L_z+f_4S_1+f_5S_2+f_6S_3+f_7I_0+f_8I_1+f_9I_2$, $\{H\}_{4\sim6,4\sim6}=f'_1L_x+f'_2L_y+f'_3L_z+f'_4S_1+f'_5S_2+f'_6S_3+f'_7I_0+f'_8I_1+f'_9I_2$, and $\{H\}_{1\sim3,4\sim6}=g_1L_x+g_2L_y+g_3L_z+g_4S_1+g_5S_2+g_6S_3+g_7I_0+g_8I_1+g_9I_2$,  the equations for BNs are:
\[\left\{\begin{array}{cccccc}
f_1=0,&f_2=0,&f_3=0,&f_4=0,&f_5=0,&f_6=0,\\
f_8=0,&f_9=0,&f'_1=0,&f'_2=0,&f'_3=0&f'_4=0,\\
f'_5=0,&f'_6=0,&f'_8=0,&f'_9=0,&f_7=f'_7,&g_1=0,\\
g_2=0,&g_3=0,&g_4=0,&g_5=0,&g_6=0,&g_7=0,\\
g_8=0,&g_9=0&&&&
\end{array}\right.\]

For $8\times 8$ $k\cdot p$ model $H$ with $\{H\}_{1\sim4,1\sim4}=\sum_{\mu,\nu=0}f_{\mu\nu}\Gamma_{\mu\nu}$, $\{H\}_{5\sim8,5\sim8}=\sum_{\mu,\nu=0}f'_{\mu\nu}\Gamma_{\mu\nu}$, and $\{H\}_{1\sim4,5\sim8}=\sum_{\mu,\nu=0}g_{\mu\nu}\Gamma_{\mu\nu}$,  the equations for BNs are:
\[\left\{\begin{array}{cccccccc}
f_{01}=0,&f_{02}=0,&f_{03}=0,&f_{10}=0,&f_{11}=0,&f_{12}=0,&f_{13}=0,&f_{20}=0,\\
f_{21}=0,&f_{22}=0,&f_{23}=0,&f_{30}=0,&f_{31}=0,&f_{32}=0,&f_{33}=0,&f'_{01}=0,\\
f'_{02}=0,&f'_{03}=0,&f'_{10}=0,&f'_{11}=0,&f'_{12}=0,&f'_{13}=0,&f'_{20}=0,&f'_{21}=0,\\
f'_{22}=0,&f'_{23}=0,&f'_{30}=0,&f'_{31}=0,&f'_{32}=0,&f'_{33}=0,&f_{00}=f'_{00},&g_{00}=0,\\
g_{01}=0,&g_{02}=0,&g_{03}=0,&g_{10}=0,&g_{11}=0,&g_{12}=0,&g_{13}=0,&g_{20}=0,\\
g_{21}=0,&g_{22}=0,&g_{23}=0,&g_{30}=0,&g_{31}=0,&g_{32}=0,&g_{33}=0&\\
\end{array}\right.\]
Note that in the above equations, the coefficients as $f$ or $f'$ are all real numbers while those as $g$ are all complex numbers. Thanks to the symmetry constraints, the above equations can have solutions other than $\mathbf{q}=(0,0,0)$ since the number of actual independent equations might be very small. We present all explicit solutions for $k\cdot p$ models of all BNs in MLGs in Sec. IV of \textbf{Supplementary Information I}. We also present all explicit solutions for $k\cdot p$ models with lowered orders of all BNs in MSGs in Sec. III of \textbf{Supplementary Information I}, where the CR required NSrs are also shown.
\section{Mirror/glide/$PT$ protected NLs in MSGs}\label{more}
\subsection{$H_i$ allowing mirror/glide protected NL for CR-BSP P-NP}
In the following, Table \ref{NSOC-mirror-NL} lists the values $i$ in $H_i$ for the spinless setting, which correspond to CR-BSP P-NP, but allow additional NL(s) protected by mirror or glide symmetry. Table \ref{SOC-mirror-NL} lists the values $i$ in $H_i$ for the spinful setting, which correspond to CR-BSP P-NP, but allow additional NL(s) protected by mirror or glide symmetry.\\
\begin{table}[!htb]
\begin{tabular}{ccccccccccccccccccccccc}
  \hline
  8 & 14 & 19 & 133 & 134 & 408 & 435 & 436 & 615 & 699 &981 & 982 & 1011 & 1020 & 1033 & 1078 & 1090 & 1558 & 1738 & 1784&
  1787 & 2009 & 2010 \\ 2011 & 2014 & 2015 & 2016 & 2017 & 2035 & 2036& 2039 & 2068 & 2105 & 2121 & 2130 & 2138 & 2150 & 2174 & 2175 & 2179&
  2180 & 2231 & 2278 & 2316 & 2329 & 2347 \\ 2349 & 2350 & 2351 & 2370 & 2404 & 2412 & 2433 & 2440 & 2444 & 2447 & 2458 & 2724 & 2751 & 2811&
  2932 & 2934 & 2937& 2952 & 2953 & 2962 & 2963 &  &  \\\hline
\end{tabular}\\\caption{\textbf{The values of $i$ in $H_i$ which allows mirror or glide protected NL(s) for CR-BSP P-NP in the spinless setting.}}\label{NSOC-mirror-NL}
\end{table}

\begin{table}[!htb]
\begin{tabular}{ccccccccccccccccc}
  \hline
  8 & 14 & 15& 17& 19& 22& 23& 133& 134& 407&408&435&436&439&441&615&698\\699&981&982&
  1011&1020&1033&1045&1058&1068&1078&1090&1091&1092&1558&1718&1735&1738\\\hline
\end{tabular}\\\caption{\textbf{The values of $i$ in $H_i$ which allows mirror or glide protected NL(s) for CR-BSP P-NP in the spinful setting.}}\label{SOC-mirror-NL}
\end{table}
\subsection{$H_i$ allowing $PT$ protected NL for CR-BSP P-NP or L-NP in the spinless setting}
In the following, Table \ref{NSOC-PT-NL} lists the values $i$ in $H_i$ which correspond to CR-BSP P-NP or L-NP, but allow additional NL(s) protected by $PT$ symmetry in the spinless setting.\\
\begin{table}[!htb]
\begin{tabular}{ccccccccccccccccccc}
  \hline
  1777&1778&1814&1815&2026&2032&2181&2346&2379&2402&2420&2428&2441&2450&2665&2836&2838&2935&2939\\\hline
\end{tabular}\\\caption{\textbf{The values of $i$ in $H_i$ which allows $PT$ protected NL(s) for CR-BSP P-NP or L-NP in the spinless setting.}}\label{NSOC-PT-NL}
\end{table}
\subsection{$H_i$ allowing $PT$ protected nodal line for CR-BSP P-NL}
The emanating BNs of $H_{2373}$ and $H_{2374}$ are found to include $PT$ protected NL(s) in the spinless setting other than CR required NL(s) coinciding with HSL(s), which are not pinned within any HSPL, as listed in Table \ref{NSOC-PT-NL-P-NL}.
\begin{table}[!htb]
\begin{tabular}{cc}
  \hline
 2373& 2374\\\hline
\end{tabular}\\\caption{\textbf{The values of $i$ in $H_i$ which allows $PT$ protected NL(s) for CR-BSP P-NL in the spinless setting.}}\label{NSOC-PT-NL-P-NL}
\end{table}
\subsection{$H_i$ allowing mirror/glide protected NL for CR-BSP P-NL}
 In the following, Table \ref{NSOC-mirror-NL} lists the values $i$ in $H_i$ in the spinless setting, which correspond to CR-BSP P-NL, but allow additional NL(s) protected by mirror or glide symmetry. Table \ref{SOC-mirror-NL} lists the values $i$ in $H_i$ in the spinful setting, which correspond to CR-BSP P-NL, but allow additional NL(s) protected by mirror or glide symmetry.\\
\begin{table}[!htb]
\begin{tabular}{cccccccccccccccccccc}
  \hline
9&135&136&622&1012&1021&1027&1032&1034&1079&1562&1599&1747&1785&1786&2012&2013&2034&2037&2040\\
2089&2204&2205&2206&2230&2250&2267&2314&2321&2381&2406&2414&2435&2445&2864&2912&2933&2948&2955&2956\\\hline
\end{tabular}\\\caption{\textbf{The values of $i$ in $H_i$ which allows mirror/glide protected NL(s) for CR-BSP P-NL in the spinless setting.}}\label{NSOC-mirror-NL-P-NL}
\end{table}

\begin{table}[!htb]
\begin{tabular}{cccccccccccccccc}
  \hline
9&135&136&431&432&622&1012&1021&1027&1032&1034&1077&1079&1562&1599&1747\\\hline
\end{tabular}\\\caption{\textbf{The values of $i$ in $H_i$ which allows mirror/glide protected NL(s) for CR-BSP P-NL in the spinful setting.}}\label{SOC-mirror-NL-P-NL}
\end{table}
\section{Parameters used in $H_{2469}$ to show $k\cdot p$ order enriched nodal structures in the main text}\label{para}
Table \ref{para-1} shows the concrete parameters used to plot the demonstration of $k\cdot p$ order enriched nodal structures in Fig. \ref{figure-3}(c).
\begin{table}[!htb]
  \centering
  \begin{tabular}{c|c|c|c|c|c|c|c|c|c|c|c|c|c|c|c|c}

    \hline\hline
    $R_1$ & $R_2$ &  $R_3$ & $R_4$ & $R_5$ & $R_6$ & $R_7$ & $R_8$ & $R_9$ & $R_{10}$ &$R_{11}$ & $R_{12}$ & $R_{13}$ & $R_{14}$ & $R_{15}$ & $R_{16}$&$R_{17}$ \\\hline
    1 & 0.1 & 0.01 & 0.001 & 0.0001 & 0.00001 & 0.12 & 0.02 & 0.001 & 0.0002 & 0.00002 & 0.002 & 0.0003 & 0.000025 & 0.00003 & 0.000015 & 0.000012 \\\hline
   $R_{18}$ & $R_{19}$ & $R_{20}$ & $R_{21}$ & $R_{22}$ & $R_{23}$ & $R_{24}$ & $R_{25}$ & $R_{26}$ & $R_{27}$ & $R_{28}$ &$R_{29}$ & $R_{30}$ & $R_{31}$ & $R_{32}$ & $R_{33}$ & $R_{34}$ \\\hline
    0.000022 & 1 & 0.15 & 0.02 & 0.003 & 0.00025 & 0.000011 & 0.2 & 0.022 & 0.0015 & 0.00015 & 0.00001 & 0.0025 & 0.00012 & 0.00001 & 0.00001 & 0.00001 \\\hline
  \end{tabular}
  \caption{\textbf{34 parameters in $H_{2469}$.}}\label{para-1}
\end{table}
\section{Peak of Berry curvatures by nearly  NSr of  $H_{985}$}\label{peak}
\begin{figure}[!htb]
  \includegraphics[width=0.8\textwidth]{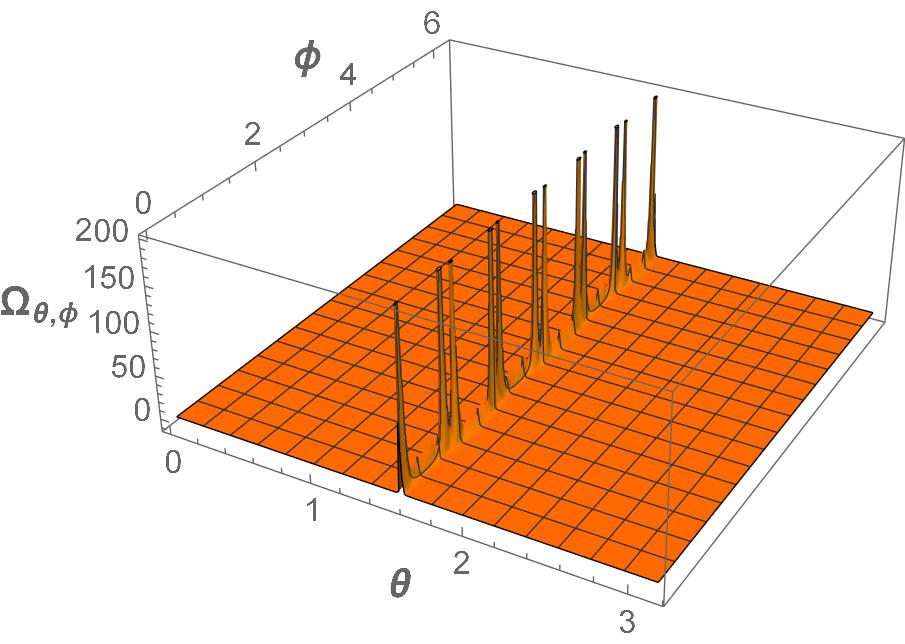}\\
  \caption{\textbf{Peak of Berry curvatures.} Using the $k\cdot p$ model $H_{985}$ with parameters shown in Table \ref{para-2-1} in the following, we show the Berry curvatures $\Omega_{\theta\phi}$ in a unit sphere around the BN at the origin. Here $\theta$ and $\phi$ are azimuthal and polar angles, respectively.}\label{figure-4}
\end{figure}

Table \ref{para-2-1} in the following shows the concrete parameters used to plot the demonstration of Berry curvatures $\Omega_{\theta\phi}$ in Fig. \ref{figure-4}.
\begin{table}[!htb]
  \centering
  \begin{tabular}{c|c|c|c|c|c|c|c|c|c|c|c|c|c|c|c|c}

  \hline\hline
    $R_1$ & $R_2$ &  $R_3$ & $R_4$ & $R_5$ & $R_6$ & $R_7$ & $R_8$ & $R_9$ & $R_{10}$ &$R_{11}$ & $R_{12}$ & $R_{13}$ & $R_{14}$ & $R_{15}$ & $R_{16}$&$R_{17}$ \\\hline
     0.1& 0.11 & 1 & 0.01 & 0.02 & 0.03 & 0.011 & 1.1 & 0.013 & 0.03 & 0.012 & 0.005 & 1.2 & 0.012 & 0.007 & 0.02 & 0.002 \\\hline
  \end{tabular}
  \caption{\textbf{17 parameters in $H_{985}$.}}\label{para-2-1}
\end{table}

\section{$k\cdot p$ order enriched NS}\label{enrichns}
In the following table, we show the enrichment by $k\cdot p$ order on NS by printing each MSG in a specific color. The meaning of each color can be found in Table \ref{color}. It is easy to find that majority of MSGs can allow NS (these MSGs are printed in color other than black). \\

\begin{table}[!htb]
  \centering
  \begin{tabular}{|c|c|c|c|c|c|c|c|c|c|c|}
    \hline
    NS&{\bf \color{black}MSG} & {\bf \color{red}MSG}  & {\bf \color{blue}MSG}  & {\bf \color{magenta}MSG}  & {\bf \color{green}MSG}  & {\bf \color{cyan}MSG}  & {\bf \color{yellow}MSG} &{\bf \color{orange}MSG}  & {\bf \color{brown}MSG}  & {\bf \color{gray}MSG}  \\\hline
    spinless strictly NS&$\times$& $\checkmark$& $\checkmark$ & $\checkmark$ & $\times$ & $\times$ & $\times$ & $\checkmark$ & $\checkmark$ & $\checkmark$ \\\hline
     spinless nearly NS&$\times$& $\checkmark$& $\checkmark$ & $\times$ & $\checkmark$ & $\checkmark$ & $\times$ & $\times$ & $\checkmark$ & $\checkmark$ \\\hline
      spinful strictly NS&$\times$& $\checkmark$& $\checkmark$ & $\checkmark$ & $\times$ & $\times$ & $\times$ & $\times$ & $\times$ & $\times$ \\\hline
       spinful nearly NS&$\times$& $\checkmark$& $\times$ & $\times$ & $\checkmark$ & $\times$ & $\checkmark$ & $\times$ & $\times$ & $\checkmark$ \\\hline
  \end{tabular}
  \caption{The meaning of color to indicate the enrichment of NS by $k\cdot p$ order. The first column indicates the NS is nearly or strictly and occurs in the spinless or spinful setting. ``MSG'' is printed in ten colors and for each color, an NS (e.g. spinless nearly NS) can occur ($\checkmark$) or cannot occur ($\times$) .}\label{color}
\end{table}

{\bf \centering
\begin{tabular}{|c|c|c|c|c|c|c|c|c|c|c|c|c|c|c|}\hline
1.1&
1.2&
1.3&
2.4&
2.5&
2.6&
2.7&
3.1&
3.2&
3.3&
3.4&
{\color{magenta}3.5}&
{\color{magenta}3.6}&
4.7&
{\color{magenta}4.8}
\\\hline
{\color{magenta}4.9}&
{\color{magenta}4.10}&
4.11&
4.12&
5.13&
5.14&
5.15&
5.16&
5.17&
6.18&
6.19&
6.20&
{\color{yellow}6.21}&
6.22&
{\color{green}6.23}
\\\hline
7.24&
{\color{cyan}7.25}&
7.26&
{\color{green}7.27}&
7.28&
{\color{green}7.29}&
{\color{green}7.30}&
7.31&
8.32&
8.33&
8.34&
{\color{yellow}8.35}&
8.36&
9.37&
{\color{cyan}9.38}
\\\hline
9.39&
9.40&
{\color{cyan}9.41}&
10.42&
10.43&
10.44&
10.45&
10.46&
10.47&
{\color{orange}10.48}&
{\color{orange}10.49}&
11.50&
{\color{orange}11.51}&
{\color{orange}11.52}&
11.53
\\\hline
{\color{magenta}11.54}&
{\color{orange}11.55}&
11.56&
{\color{cyan}11.57}&
12.58&
12.59&
12.60&
12.61&
12.62&
12.63&
12.64&
13.65&
{\color{cyan}13.66}&
13.67&
13.68
\\\hline
13.69&
{\color{cyan}13.70}&
{\color{brown}13.71}&
13.72&
{\color{orange}13.73}&
{\color{brown}13.74}&
14.75&
{\color{brown}14.76}&
{\color{orange}14.77}&
14.78&
{\color{magenta}14.79}&
{\color{brown}14.80}&
{\color{cyan}14.81}&
{\color{orange}14.82}&
14.83
\\\hline
{\color{cyan}14.84}&
15.85&
{\color{cyan}15.86}&
15.87&
15.88&
15.89&
15.90&
{\color{cyan}15.91}&
16.1&
16.2&
16.3&
{\color{magenta}16.4}&
{\color{magenta}16.5}&
{\color{magenta}16.6}&
17.7
\\\hline
{\color{magenta}17.8}&
17.9&
{\color{magenta}17.10}&
{\color{magenta}17.11}&
17.12&
{\color{magenta}17.13}&
{\color{magenta}17.14}&
{\color{magenta}17.15}&
18.16&
{\color{magenta}18.17}&
{\color{magenta}18.18}&
{\color{magenta}18.19}&
{\color{magenta}18.20}&
{\color{magenta}18.21}&
{\color{magenta}18.22}
\\\hline
18.23&
{\color{magenta}18.24}&
19.25&
{\color{magenta}19.26}&
{\color{magenta}19.27}&
{\color{magenta}19.28}&
{\color{magenta}19.29}&
19.30&
20.31&
{\color{magenta}20.32}&
20.33&
{\color{magenta}20.34}&
20.35&
{\color{magenta}20.36}&
20.37
\\\hline
21.38&
21.39&
21.40&
21.41&
{\color{magenta}21.42}&
21.43&
{\color{magenta}21.44}&
22.45&
22.46&
22.47&
22.48&
23.49&
23.50&
23.51&
23.52
\\\hline
24.53&
24.54&
24.55&
24.56&
{\color{cyan}25.57}&
{\color{cyan}25.58}&
25.59&
25.60&
{\color{blue}25.61}&
{\color{green}25.62}&
{\color{green}25.63}&
{\color{red}25.64}&
{\color{red}25.65}&
{\color{cyan}26.66}&
{\color{blue}26.67}
\\\hline
{\color{magenta}26.68}&
{\color{magenta}26.69}&
26.70&
{\color{red}26.71}&
{\color{red}26.72}&
{\color{green}26.73}&
{\color{green}26.74}&
{\color{green}26.75}&
{\color{red}26.76}&
{\color{green}26.77}&
{\color{cyan}27.78}&
{\color{cyan}27.79}&
27.80&
{\color{green}27.81}&
{\color{blue}27.82}
\\\hline
{\color{green}27.83}&
{\color{green}27.84}&
{\color{red}27.85}&
{\color{red}27.86}&
{\color{green}28.87}&
{\color{green}28.88}&
28.89&
28.90&
28.91&
{\color{green}28.92}&
{\color{green}28.93}&
{\color{red}28.94}&
{\color{red}28.95}&
{\color{red}28.96}&
{\color{green}28.97}
\\\hline
{\color{red}28.98}&
{\color{green}29.99}&
{\color{red}29.100}&
{\color{magenta}29.101}&
{\color{magenta}29.102}&
{\color{green}29.103}&
{\color{red}29.104}&
{\color{red}29.105}&
{\color{green}29.106}&
{\color{green}29.107}&
{\color{green}29.108}&
{\color{red}29.109}&
{\color{green}29.110}&
{\color{green}30.111}&
{\color{green}30.112}
\\\hline
30.113&
30.114&
{\color{green}30.115}&
{\color{green}30.116}&
{\color{green}30.117}&
{\color{red}30.118}&
{\color{red}30.119}&
{\color{red}30.120}&
{\color{green}30.121}&
{\color{red}30.122}&
{\color{green}31.123}&
{\color{red}31.124}&
{\color{magenta}31.125}&
{\color{magenta}31.126}&
31.127
\\\hline
{\color{red}31.128}&
{\color{red}31.129}&
{\color{green}31.130}&
{\color{green}31.131}&
{\color{green}31.132}&
{\color{red}31.133}&
{\color{green}31.134}&
{\color{green}32.135}&
{\color{green}32.136}&
32.137&
{\color{green}32.138}&
{\color{red}32.139}&
{\color{green}32.140}&
{\color{green}32.141}&
{\color{red}32.142}
\\\hline
{\color{red}32.143}&
{\color{green}33.144}&
{\color{red}33.145}&
{\color{magenta}33.146}&
{\color{magenta}33.147}&
{\color{green}33.148}&
{\color{red}33.149}&
{\color{red}33.150}&
{\color{green}33.151}&
{\color{green}33.152}&
{\color{green}33.153}&
{\color{red}33.154}&
{\color{green}33.155}&
{\color{green}34.156}&
{\color{green}34.157}
\\\hline
34.158&
{\color{green}34.159}&
{\color{green}34.160}&
{\color{red}34.161}&
{\color{red}34.162}&
{\color{green}34.163}&
{\color{red}34.164}&
{\color{cyan}35.165}&
{\color{cyan}35.166}&
35.167&
35.168&
{\color{blue}35.169}&
{\color{cyan}35.170}&
{\color{blue}35.171}&
{\color{cyan}36.172}
\\\hline
{\color{blue}36.173}&
{\color{magenta}36.174}&
{\color{magenta}36.175}&
36.176&
{\color{green}36.177}&
{\color{blue}36.178}&
{\color{green}36.179}&
{\color{cyan}37.180}&
{\color{cyan}37.181}&
37.182&
{\color{green}37.183}&
{\color{blue}37.184}&
{\color{cyan}37.185}&
{\color{blue}37.186}&
{\color{cyan}38.187}
\\\hline
{\color{cyan}38.188}&
38.189&
38.190&
38.191&
{\color{green}38.192}&
{\color{green}38.193}&
{\color{green}38.194}&
{\color{cyan}39.195}&
{\color{cyan}39.196}&
39.197&
39.198&
39.199&
{\color{cyan}39.200}&
{\color{green}39.201}&
{\color{green}39.202}
\\\hline
{\color{green}40.203}&
{\color{green}40.204}&
40.205&
40.206&
40.207&
{\color{green}40.208}&
{\color{green}40.209}&
{\color{green}40.210}&
{\color{green}41.211}&
{\color{green}41.212}&
41.213&
41.214&
41.215&
{\color{green}41.216}&
{\color{green}41.217}
\\\hline
{\color{green}41.218}&
{\color{cyan}42.219}&
{\color{cyan}42.220}&
42.221&
42.222&
{\color{cyan}42.223}&
{\color{green}43.224}&
{\color{green}43.225}&
43.226&
{\color{green}43.227}&
{\color{green}43.228}&
{\color{cyan}44.229}&
{\color{cyan}44.230}&
44.231&
44.232
\\\hline
{\color{green}44.233}&
{\color{green}44.234}&
{\color{cyan}45.235}&
{\color{green}45.236}&
45.237&
{\color{green}45.238}&
{\color{cyan}45.239}&
{\color{cyan}45.240}&
{\color{cyan}46.241}&
{\color{cyan}46.242}&
46.243&
46.244&
46.245&
{\color{green}46.246}&
{\color{green}46.247}
\\\hline
{\color{cyan}46.248}&
{\color{cyan}47.249}&
{\color{cyan}47.250}&
{\color{cyan}47.251}&
47.252&
47.253&
{\color{gray}47.254}&
{\color{gray}47.255}&
{\color{gray}47.256}&
{\color{green}48.257}&
{\color{cyan}48.258}&
{\color{cyan}48.259}&
{\color{green}48.260}&
{\color{cyan}48.261}&
{\color{gray}48.262}
\\\hline
{\color{brown}48.263}&
{\color{brown}48.264}&
{\color{green}49.265}&
{\color{cyan}49.266}&
{\color{cyan}49.267}&
{\color{cyan}49.268}&
{\color{green}49.269}&
49.270&
{\color{cyan}49.271}&
{\color{gray}49.272}&
{\color{brown}49.273}&
{\color{gray}49.274}&
{\color{gray}49.275}&
{\color{gray}49.276}&
{\color{green}50.277}
\\\hline
{\color{cyan}50.278}&
{\color{cyan}50.279}&
{\color{cyan}50.280}&
{\color{green}50.281}&
{\color{green}50.282}&
{\color{cyan}50.283}&
{\color{brown}50.284}&
{\color{gray}50.285}&
{\color{gray}50.286}&
{\color{brown}50.287}&
{\color{gray}50.288}&
{\color{green}51.289}&
{\color{brown}51.290}&
{\color{brown}51.291}&
{\color{brown}51.292}
\\\hline
{\color{cyan}51.293}&
51.294&
{\color{magenta}51.295}&
{\color{magenta}51.296}&
51.297&
{\color{gray}51.298}&
{\color{gray}51.299}&
{\color{cyan}51.300}&
{\color{gray}51.301}&
{\color{gray}51.302}&
{\color{gray}51.303}&
{\color{gray}51.304}&
{\color{green}52.305}&
{\color{gray}52.306}&
{\color{brown}52.307}
\\\hline
{\color{brown}52.308}&
{\color{cyan}52.309}&
{\color{green}52.310}&
{\color{red}52.311}&
{\color{red}52.312}&
{\color{cyan}52.313}&
{\color{gray}52.314}&
{\color{gray}52.315}&
{\color{cyan}52.316}&
{\color{brown}52.317}&
{\color{gray}52.318}&
{\color{gray}52.319}&
{\color{gray}52.320}&
{\color{green}53.321}&
{\color{brown}53.322}
\\\hline
{\color{brown}53.323}&
{\color{brown}53.324}&
{\color{cyan}53.325}&
53.326&
{\color{magenta}53.327}&
{\color{red}53.328}&
{\color{cyan}53.329}&
{\color{gray}53.330}&
{\color{gray}53.331}&
{\color{cyan}53.332}&
{\color{brown}53.333}&
{\color{gray}53.334}&
{\color{gray}53.335}&
{\color{gray}53.336}&
{\color{green}54.337}
\\\hline
{\color{gray}54.338}&
{\color{brown}54.339}&
{\color{brown}54.340}&
{\color{cyan}54.341}&
{\color{green}54.342}&
{\color{red}54.343}&
{\color{magenta}54.344}&
{\color{cyan}54.345}&
{\color{gray}54.346}&
{\color{gray}54.347}&
{\color{cyan}54.348}&
{\color{gray}54.349}&
{\color{brown}54.350}&
{\color{gray}54.351}&
{\color{gray}54.352}
\\\hline
{\color{green}55.353}&
{\color{brown}55.354}&
{\color{brown}55.355}&
{\color{brown}55.356}&
{\color{magenta}55.357}&
{\color{magenta}55.358}&
{\color{cyan}55.359}&
{\color{gray}55.360}&
{\color{gray}55.361}&
{\color{gray}55.362}&
{\color{cyan}55.363}&
{\color{gray}55.364}&
{\color{green}56.365}&
{\color{gray}56.366}&
{\color{brown}56.367}
\\\hline
{\color{brown}56.368}&
{\color{red}56.369}&
{\color{red}56.370}&
{\color{cyan}56.371}&
{\color{gray}56.372}&
{\color{gray}56.373}&
{\color{gray}56.374}&
{\color{cyan}56.375}&
{\color{gray}56.376}&
{\color{green}57.377}&
{\color{gray}57.378}&
{\color{brown}57.379}&
{\color{brown}57.380}&
{\color{gray}57.381}&
{\color{magenta}57.382}
\\\hline
{\color{magenta}57.383}&
{\color{magenta}57.384}&
{\color{cyan}57.385}&
{\color{brown}57.386}&
{\color{brown}57.387}&
{\color{gray}57.388}&
{\color{gray}57.389}&
{\color{gray}57.390}&
{\color{cyan}57.391}&
{\color{gray}57.392}&
{\color{green}58.393}&
{\color{brown}58.394}&
{\color{brown}58.395}&
{\color{brown}58.396}&
{\color{red}58.397}
\\\hline
{\color{magenta}58.398}&
{\color{cyan}58.399}&
{\color{gray}58.400}&
{\color{gray}58.401}&
{\color{gray}58.402}&
{\color{cyan}58.403}&
{\color{brown}58.404}&
{\color{green}59.405}&
{\color{brown}59.406}&
{\color{brown}59.407}&
{\color{brown}59.408}&
{\color{magenta}59.409}&
{\color{magenta}59.410}&
59.411&
{\color{brown}59.412}
\\\hline
{\color{gray}59.413}&
{\color{gray}59.414}&
{\color{cyan}59.415}&
{\color{gray}59.416}&
{\color{green}60.417}&
{\color{gray}60.418}&
{\color{brown}60.419}&
{\color{brown}60.420}&
{\color{gray}60.421}&
{\color{red}60.422}&
{\color{magenta}60.423}&
{\color{red}60.424}&
{\color{cyan}60.425}&
{\color{gray}60.426}&
{\color{brown}60.427}
\\\hline
{\color{gray}60.428}&
{\color{gray}60.429}&
{\color{gray}60.430}&
{\color{cyan}60.431}&
{\color{brown}60.432}&
{\color{green}61.433}&
{\color{gray}61.434}&
{\color{gray}61.435}&
{\color{magenta}61.436}&
{\color{cyan}61.437}&
{\color{gray}61.438}&
{\color{brown}61.439}&
{\color{cyan}61.440}&
{\color{green}62.441}&
{\color{gray}62.442}
\\\hline
{\color{brown}62.443}&
{\color{brown}62.444}&
{\color{gray}62.445}&
{\color{magenta}62.446}&
{\color{magenta}62.447}&
{\color{magenta}62.448}&
{\color{cyan}62.449}&
{\color{gray}62.450}&
{\color{gray}62.451}&
{\color{gray}62.452}&
{\color{gray}62.453}&
{\color{brown}62.454}&
{\color{gray}62.455}&
{\color{cyan}62.456}&
{\color{green}63.457}
\\\hline
{\color{brown}63.458}&
{\color{brown}63.459}&
{\color{brown}63.460}&
{\color{cyan}63.461}&
63.462&
{\color{magenta}63.463}&
{\color{magenta}63.464}&
63.465&
{\color{cyan}63.466}&
{\color{brown}63.467}&
{\color{cyan}63.468}&
{\color{green}64.469}&
{\color{brown}64.470}&
{\color{brown}64.471}&
{\color{brown}64.472}
\\\hline
{\color{cyan}64.473}&
64.474&
{\color{magenta}64.475}&
{\color{magenta}64.476}&
64.477&
{\color{cyan}64.478}&
{\color{brown}64.479}&
{\color{cyan}64.480}&
{\color{cyan}65.481}&
{\color{cyan}65.482}&
{\color{cyan}65.483}&
{\color{cyan}65.484}&
65.485&
65.486&
65.487
\\\hline
{\color{gray}65.488}&
{\color{cyan}65.489}&
{\color{gray}65.490}&
{\color{green}66.491}&
{\color{cyan}66.492}&
{\color{cyan}66.493}&
{\color{cyan}66.494}&
{\color{green}66.495}&
66.496&
{\color{cyan}66.497}&
{\color{brown}66.498}&
{\color{cyan}66.499}&
{\color{brown}66.500}&
{\color{cyan}67.501}&
{\color{cyan}67.502}
\\\hline
{\color{cyan}67.503}&
{\color{cyan}67.504}&
67.505&
67.506&
67.507&
{\color{gray}67.508}&
{\color{cyan}67.509}&
{\color{gray}67.510}&
{\color{green}68.511}&
{\color{cyan}68.512}&
{\color{cyan}68.513}&
{\color{cyan}68.514}&
{\color{green}68.515}&
68.516&
{\color{cyan}68.517}
\\\hline
{\color{brown}68.518}&
{\color{cyan}68.519}&
{\color{brown}68.520}&
{\color{cyan}69.521}&
{\color{cyan}69.522}&
{\color{cyan}69.523}&
69.524&
69.525&
{\color{cyan}69.526}&
{\color{green}70.527}&
{\color{cyan}70.528}&
{\color{cyan}70.529}&
{\color{green}70.530}&
{\color{cyan}70.531}&
{\color{cyan}70.532}
\\\hline
{\color{cyan}71.533}&
{\color{cyan}71.534}&
{\color{cyan}71.535}&
71.536&
71.537&
{\color{cyan}71.538}&
{\color{cyan}72.539}&
{\color{cyan}72.540}&
{\color{cyan}72.541}&
{\color{cyan}72.542}&
{\color{green}72.543}&
72.544&
{\color{cyan}72.545}&
{\color{cyan}72.546}&
{\color{cyan}72.547}
\\\hline
{\color{cyan}73.548}&
{\color{cyan}73.549}&
{\color{cyan}73.550}&
{\color{green}73.551}&
73.552&
{\color{cyan}73.553}&
{\color{cyan}74.554}&
{\color{cyan}74.555}&
{\color{cyan}74.556}&
{\color{cyan}74.557}&
74.558&
74.559&
74.560&
{\color{cyan}74.561}&
{\color{cyan}74.562}
\\\hline
{\color{green}75.1}&
{\color{green}75.2}&
{\color{cyan}75.3}&
{\color{red}75.4}&
{\color{green}75.5}&
{\color{red}75.6}&
{\color{green}76.7}&
{\color{red}76.8}&
{\color{green}76.9}&
{\color{green}76.10}&
{\color{red}76.11}&
{\color{green}76.12}&
{\color{green}77.13}&
{\color{green}77.14}&
{\color{cyan}77.15}
\\\hline
{\color{red}77.16}&
{\color{green}77.17}&
{\color{red}77.18}&
{\color{green}78.19}&
{\color{red}78.20}&
{\color{green}78.21}&
{\color{green}78.22}&
{\color{red}78.23}&
{\color{green}78.24}&
{\color{green}79.25}&
{\color{green}79.26}&
{\color{cyan}79.27}&
{\color{green}79.28}&
{\color{green}80.29}&
{\color{green}80.30}
\\\hline
{\color{green}80.31}&
{\color{green}80.32}&
81.33&
81.34&
81.35&
{\color{magenta}81.36}&
81.37&
{\color{magenta}81.38}&
82.39&
82.40&
82.41&
82.42&
{\color{green}83.43}&
{\color{cyan}83.44}&
83.45
\\\hline
{\color{cyan}83.46}&
{\color{cyan}83.47}&
{\color{brown}83.48}&
{\color{cyan}83.49}&
{\color{brown}83.50}&
{\color{green}84.51}&
{\color{cyan}84.52}&
{\color{cyan}84.53}&
{\color{cyan}84.54}&
{\color{cyan}84.55}&
{\color{gray}84.56}&
{\color{cyan}84.57}&
{\color{gray}84.58}&
{\color{green}85.59}&
{\color{cyan}85.60}
\\\hline
{\color{cyan}85.61}&
{\color{cyan}85.62}&
{\color{cyan}85.63}&
{\color{gray}85.64}&
{\color{cyan}85.65}&
{\color{gray}85.66}&
{\color{green}86.67}&
{\color{cyan}86.68}&
{\color{cyan}86.69}&
{\color{cyan}86.70}&
{\color{cyan}86.71}&
{\color{gray}86.72}&
{\color{cyan}86.73}&
{\color{gray}86.74}&
{\color{green}87.75}
\\\hline
{\color{cyan}87.76}&
{\color{cyan}87.77}&
{\color{cyan}87.78}&
{\color{cyan}87.79}&
{\color{cyan}87.80}&
{\color{green}88.81}&
{\color{cyan}88.82}&
{\color{green}88.83}&
{\color{cyan}88.84}&
{\color{cyan}88.85}&
{\color{cyan}88.86}&
{\color{green}89.87}&
{\color{green}89.88}&
{\color{cyan}89.89}&
{\color{green}89.90}
\\\hline
{\color{cyan}89.91}&
{\color{red}89.92}&
{\color{red}89.93}&
{\color{red}89.94}&
{\color{green}90.95}&
{\color{red}90.96}&
{\color{cyan}90.97}&
{\color{red}90.98}&
{\color{blue}90.99}&
{\color{red}90.100}&
{\color{green}90.101}&
{\color{red}90.102}&
{\color{green}91.103}&
{\color{red}91.104}&
{\color{green}91.105}
\\\hline
{\color{green}91.106}&
{\color{green}91.107}&
{\color{green}91.108}&
{\color{red}91.109}&
{\color{red}91.110}&
{\color{green}92.111}&
{\color{red}92.112}&
{\color{green}92.113}&
{\color{red}92.114}&
{\color{red}92.115}&
{\color{red}92.116}&
{\color{red}92.117}&
{\color{green}92.118}&
{\color{green}93.119}&
{\color{green}93.120}
\\\hline
{\color{cyan}93.121}&
{\color{green}93.122}&
{\color{cyan}93.123}&
{\color{red}93.124}&
{\color{red}93.125}&
{\color{red}93.126}&
{\color{green}94.127}&
{\color{red}94.128}&
{\color{cyan}94.129}&
{\color{red}94.130}&
{\color{blue}94.131}&
{\color{red}94.132}&
{\color{green}94.133}&
{\color{red}94.134}&
{\color{green}95.135}
\\\hline
{\color{red}95.136}&
{\color{green}95.137}&
{\color{green}95.138}&
{\color{green}95.139}&
{\color{green}95.140}&
{\color{red}95.141}&
{\color{red}95.142}&
{\color{green}96.143}&
{\color{red}96.144}&
{\color{green}96.145}&
{\color{red}96.146}&
{\color{red}96.147}&
{\color{red}96.148}&
{\color{red}96.149}&
{\color{green}96.150}
\\\hline
{\color{green}97.151}&
{\color{green}97.152}&
{\color{cyan}97.153}&
{\color{green}97.154}&
{\color{cyan}97.155}&
{\color{green}97.156}&
{\color{green}98.157}&
{\color{green}98.158}&
{\color{green}98.159}&
{\color{green}98.160}&
{\color{green}98.161}&
{\color{green}98.162}&
{\color{cyan}99.163}&
{\color{cyan}99.164}&
{\color{cyan}99.165}
\\\hline
{\color{cyan}99.166}&
{\color{green}99.167}&
{\color{blue}99.168}&
{\color{green}99.169}&
{\color{red}99.170}&
{\color{green}100.171}&
{\color{green}100.172}&
{\color{cyan}100.173}&
{\color{green}100.174}&
{\color{green}100.175}&
{\color{red}100.176}&
{\color{green}100.177}&
{\color{red}100.178}&
{\color{cyan}101.179}&
{\color{cyan}101.180}
\\\hline
\end{tabular}\newpage
\begin{tabular}{|c|c|c|c|c|c|c|c|c|c|c|c|c|c|c|}\hline
{\color{green}101.181}&
{\color{cyan}101.182}&
{\color{green}101.183}&
{\color{blue}101.184}&
{\color{green}101.185}&
{\color{red}101.186}&
{\color{green}102.187}&
{\color{green}102.188}&
{\color{cyan}102.189}&
{\color{green}102.190}&
{\color{green}102.191}&
{\color{red}102.192}&
{\color{green}102.193}&
{\color{red}102.194}&
{\color{cyan}103.195}
\\\hline
{\color{cyan}103.196}&
{\color{green}103.197}&
{\color{cyan}103.198}&
{\color{green}103.199}&
{\color{blue}103.200}&
{\color{green}103.201}&
{\color{red}103.202}&
{\color{green}104.203}&
{\color{green}104.204}&
{\color{cyan}104.205}&
{\color{green}104.206}&
{\color{green}104.207}&
{\color{red}104.208}&
{\color{green}104.209}&
{\color{red}104.210}
\\\hline
{\color{cyan}105.211}&
{\color{cyan}105.212}&
{\color{cyan}105.213}&
{\color{cyan}105.214}&
{\color{green}105.215}&
{\color{blue}105.216}&
{\color{green}105.217}&
{\color{red}105.218}&
{\color{green}106.219}&
{\color{green}106.220}&
{\color{cyan}106.221}&
{\color{green}106.222}&
{\color{green}106.223}&
{\color{red}106.224}&
{\color{green}106.225}
\\\hline
{\color{red}106.226}&
{\color{cyan}107.227}&
{\color{cyan}107.228}&
{\color{cyan}107.229}&
{\color{cyan}107.230}&
{\color{green}107.231}&
{\color{green}107.232}&
{\color{cyan}108.233}&
{\color{cyan}108.234}&
{\color{cyan}108.235}&
{\color{cyan}108.236}&
{\color{green}108.237}&
{\color{cyan}108.238}&
{\color{green}109.239}&
{\color{green}109.240}
\\\hline
{\color{green}109.241}&
{\color{green}109.242}&
{\color{green}109.243}&
{\color{green}109.244}&
{\color{green}110.245}&
{\color{green}110.246}&
{\color{green}110.247}&
{\color{green}110.248}&
{\color{green}110.249}&
{\color{green}110.250}&
{\color{cyan}111.251}&
{\color{cyan}111.252}&
{\color{cyan}111.253}&
111.254&
111.255
\\\hline
{\color{blue}111.256}&
{\color{blue}111.257}&
{\color{blue}111.258}&
{\color{cyan}112.259}&
{\color{cyan}112.260}&
{\color{cyan}112.261}&
{\color{cyan}112.262}&
{\color{green}112.263}&
{\color{blue}112.264}&
{\color{blue}112.265}&
{\color{blue}112.266}&
{\color{cyan}113.267}&
{\color{blue}113.268}&
{\color{blue}113.269}&
113.270
\\\hline
{\color{magenta}113.271}&
{\color{blue}113.272}&
{\color{cyan}113.273}&
{\color{blue}113.274}&
{\color{cyan}114.275}&
{\color{blue}114.276}&
{\color{blue}114.277}&
{\color{cyan}114.278}&
{\color{red}114.279}&
{\color{blue}114.280}&
{\color{cyan}114.281}&
{\color{blue}114.282}&
{\color{cyan}115.283}&
{\color{cyan}115.284}&
115.285
\\\hline
{\color{cyan}115.286}&
115.287&
{\color{blue}115.288}&
{\color{green}115.289}&
{\color{red}115.290}&
{\color{cyan}116.291}&
{\color{cyan}116.292}&
{\color{green}116.293}&
{\color{cyan}116.294}&
{\color{green}116.295}&
{\color{blue}116.296}&
{\color{green}116.297}&
{\color{red}116.298}&
{\color{green}117.299}&
{\color{green}117.300}
\\\hline
{\color{cyan}117.301}&
{\color{green}117.302}&
{\color{green}117.303}&
{\color{red}117.304}&
{\color{green}117.305}&
{\color{red}117.306}&
{\color{green}118.307}&
{\color{green}118.308}&
{\color{cyan}118.309}&
{\color{green}118.310}&
{\color{green}118.311}&
{\color{red}118.312}&
{\color{green}118.313}&
{\color{red}118.314}&
{\color{cyan}119.315}
\\\hline
{\color{cyan}119.316}&
119.317&
{\color{cyan}119.318}&
119.319&
{\color{green}119.320}&
{\color{cyan}120.321}&
{\color{green}120.322}&
{\color{green}120.323}&
{\color{cyan}120.324}&
{\color{green}120.325}&
{\color{cyan}120.326}&
{\color{cyan}121.327}&
{\color{cyan}121.328}&
{\color{cyan}121.329}&
121.330
\\\hline
121.331&
{\color{cyan}121.332}&
{\color{green}122.333}&
{\color{green}122.334}&
{\color{green}122.335}&
{\color{cyan}122.336}&
{\color{green}122.337}&
{\color{green}122.338}&
{\color{green}123.339}&
{\color{cyan}123.340}&
{\color{cyan}123.341}&
{\color{cyan}123.342}&
{\color{cyan}123.343}&
{\color{cyan}123.344}&
{\color{green}123.345}
\\\hline
{\color{cyan}123.346}&
{\color{cyan}123.347}&
{\color{gray}123.348}&
{\color{gray}123.349}&
{\color{gray}123.350}&
{\color{green}124.351}&
{\color{cyan}124.352}&
{\color{cyan}124.353}&
{\color{green}124.354}&
{\color{green}124.355}&
{\color{cyan}124.356}&
{\color{green}124.357}&
{\color{cyan}124.358}&
{\color{cyan}124.359}&
{\color{brown}124.360}
\\\hline
{\color{gray}124.361}&
{\color{gray}124.362}&
{\color{green}125.363}&
{\color{cyan}125.364}&
{\color{cyan}125.365}&
{\color{cyan}125.366}&
{\color{green}125.367}&
{\color{cyan}125.368}&
{\color{green}125.369}&
{\color{cyan}125.370}&
{\color{cyan}125.371}&
{\color{gray}125.372}&
{\color{brown}125.373}&
{\color{gray}125.374}&
{\color{green}126.375}
\\\hline
{\color{cyan}126.376}&
{\color{cyan}126.377}&
{\color{green}126.378}&
{\color{green}126.379}&
{\color{cyan}126.380}&
{\color{green}126.381}&
{\color{cyan}126.382}&
{\color{cyan}126.383}&
{\color{gray}126.384}&
{\color{brown}126.385}&
{\color{brown}126.386}&
{\color{green}127.387}&
{\color{brown}127.388}&
{\color{brown}127.389}&
{\color{blue}127.390}
\\\hline
{\color{green}127.391}&
{\color{cyan}127.392}&
{\color{red}127.393}&
{\color{brown}127.394}&
{\color{cyan}127.395}&
{\color{gray}127.396}&
{\color{cyan}127.397}&
{\color{gray}127.398}&
{\color{green}128.399}&
{\color{brown}128.400}&
{\color{brown}128.401}&
{\color{red}128.402}&
{\color{green}128.403}&
{\color{cyan}128.404}&
{\color{red}128.405}
\\\hline
{\color{brown}128.406}&
{\color{cyan}128.407}&
{\color{gray}128.408}&
{\color{cyan}128.409}&
{\color{brown}128.410}&
{\color{green}129.411}&
{\color{brown}129.412}&
{\color{brown}129.413}&
{\color{blue}129.414}&
{\color{green}129.415}&
{\color{cyan}129.416}&
{\color{red}129.417}&
{\color{brown}129.418}&
{\color{cyan}129.419}&
{\color{gray}129.420}
\\\hline
{\color{cyan}129.421}&
{\color{gray}129.422}&
{\color{green}130.423}&
{\color{gray}130.424}&
{\color{brown}130.425}&
{\color{red}130.426}&
{\color{green}130.427}&
{\color{cyan}130.428}&
{\color{red}130.429}&
{\color{brown}130.430}&
{\color{cyan}130.431}&
{\color{gray}130.432}&
{\color{cyan}130.433}&
{\color{gray}130.434}&
{\color{green}131.435}
\\\hline
{\color{cyan}131.436}&
{\color{cyan}131.437}&
{\color{green}131.438}&
{\color{cyan}131.439}&
{\color{cyan}131.440}&
{\color{green}131.441}&
{\color{cyan}131.442}&
{\color{cyan}131.443}&
{\color{gray}131.444}&
{\color{gray}131.445}&
{\color{gray}131.446}&
{\color{green}132.447}&
{\color{cyan}132.448}&
{\color{cyan}132.449}&
{\color{green}132.450}
\\\hline
{\color{green}132.451}&
{\color{cyan}132.452}&
{\color{green}132.453}&
{\color{cyan}132.454}&
{\color{cyan}132.455}&
{\color{gray}132.456}&
{\color{gray}132.457}&
{\color{gray}132.458}&
{\color{green}133.459}&
{\color{cyan}133.460}&
{\color{cyan}133.461}&
{\color{green}133.462}&
{\color{green}133.463}&
{\color{cyan}133.464}&
{\color{green}133.465}
\\\hline
{\color{cyan}133.466}&
{\color{cyan}133.467}&
{\color{gray}133.468}&
{\color{brown}133.469}&
{\color{gray}133.470}&
{\color{green}134.471}&
{\color{cyan}134.472}&
{\color{cyan}134.473}&
{\color{cyan}134.474}&
{\color{green}134.475}&
{\color{cyan}134.476}&
{\color{green}134.477}&
{\color{cyan}134.478}&
{\color{cyan}134.479}&
{\color{gray}134.480}
\\\hline
{\color{brown}134.481}&
{\color{gray}134.482}&
{\color{green}135.483}&
{\color{brown}135.484}&
{\color{brown}135.485}&
{\color{red}135.486}&
{\color{green}135.487}&
{\color{cyan}135.488}&
{\color{red}135.489}&
{\color{brown}135.490}&
{\color{cyan}135.491}&
{\color{gray}135.492}&
{\color{cyan}135.493}&
{\color{gray}135.494}&
{\color{green}136.495}
\\\hline
{\color{brown}136.496}&
{\color{brown}136.497}&
{\color{blue}136.498}&
{\color{green}136.499}&
{\color{cyan}136.500}&
{\color{red}136.501}&
{\color{brown}136.502}&
{\color{cyan}136.503}&
{\color{gray}136.504}&
{\color{cyan}136.505}&
{\color{gray}136.506}&
{\color{green}137.507}&
{\color{brown}137.508}&
{\color{brown}137.509}&
{\color{red}137.510}
\\\hline
{\color{green}137.511}&
{\color{cyan}137.512}&
{\color{red}137.513}&
{\color{brown}137.514}&
{\color{cyan}137.515}&
{\color{gray}137.516}&
{\color{cyan}137.517}&
{\color{gray}137.518}&
{\color{green}138.519}&
{\color{gray}138.520}&
{\color{brown}138.521}&
{\color{blue}138.522}&
{\color{green}138.523}&
{\color{cyan}138.524}&
{\color{red}138.525}
\\\hline
{\color{brown}138.526}&
{\color{cyan}138.527}&
{\color{gray}138.528}&
{\color{cyan}138.529}&
{\color{gray}138.530}&
{\color{green}139.531}&
{\color{cyan}139.532}&
{\color{cyan}139.533}&
{\color{cyan}139.534}&
{\color{cyan}139.535}&
{\color{cyan}139.536}&
{\color{green}139.537}&
{\color{cyan}139.538}&
{\color{cyan}139.539}&
{\color{cyan}139.540}
\\\hline
{\color{green}140.541}&
{\color{cyan}140.542}&
{\color{cyan}140.543}&
{\color{cyan}140.544}&
{\color{cyan}140.545}&
{\color{cyan}140.546}&
{\color{green}140.547}&
{\color{cyan}140.548}&
{\color{cyan}140.549}&
{\color{cyan}140.550}&
{\color{green}141.551}&
{\color{cyan}141.552}&
{\color{cyan}141.553}&
{\color{green}141.554}&
{\color{green}141.555}
\\\hline
{\color{cyan}141.556}&
{\color{green}141.557}&
{\color{cyan}141.558}&
{\color{cyan}141.559}&
{\color{green}141.560}&
{\color{green}142.561}&
{\color{green}142.562}&
{\color{cyan}142.563}&
{\color{green}142.564}&
{\color{green}142.565}&
{\color{green}142.566}&
{\color{green}142.567}&
{\color{cyan}142.568}&
{\color{cyan}142.569}&
{\color{cyan}142.570}
\\\hline
143.1&
{\color{green}143.2}&
{\color{green}143.3}&
144.4&
{\color{green}144.5}&
{\color{green}144.6}&
145.7&
{\color{green}145.8}&
{\color{green}145.9}&
146.10&
{\color{green}146.11}&
{\color{green}146.12}&
147.13&
147.14&
147.15
\\\hline
{\color{cyan}147.16}&
148.17&
148.18&
148.19&
{\color{cyan}148.20}&
149.21&
{\color{green}149.22}&
149.23&
{\color{green}149.24}&
150.25&
{\color{green}150.26}&
150.27&
{\color{green}150.28}&
151.29&
{\color{green}151.30}
\\\hline
151.31&
{\color{green}151.32}&
152.33&
{\color{green}152.34}&
152.35&
{\color{green}152.36}&
153.37&
{\color{green}153.38}&
153.39&
{\color{green}153.40}&
154.41&
{\color{green}154.42}&
154.43&
{\color{green}154.44}&
155.45
\\\hline
{\color{green}155.46}&
155.47&
{\color{green}155.48}&
{\color{green}156.49}&
{\color{green}156.50}&
156.51&
{\color{green}156.52}&
{\color{green}157.53}&
{\color{green}157.54}&
157.55&
{\color{green}157.56}&
{\color{green}158.57}&
{\color{green}158.58}&
{\color{green}158.59}&
{\color{green}158.60}
\\\hline
{\color{green}159.61}&
{\color{green}159.62}&
{\color{green}159.63}&
{\color{green}159.64}&
{\color{green}160.65}&
{\color{green}160.66}&
160.67&
{\color{green}160.68}&
{\color{green}161.69}&
{\color{green}161.70}&
{\color{green}161.71}&
{\color{green}161.72}&
{\color{green}162.73}&
{\color{cyan}162.74}&
{\color{cyan}162.75}
\\\hline
162.76&
162.77&
{\color{green}162.78}&
{\color{green}163.79}&
{\color{green}163.80}&
{\color{cyan}163.81}&
{\color{cyan}163.82}&
{\color{green}163.83}&
{\color{cyan}163.84}&
{\color{green}164.85}&
{\color{cyan}164.86}&
{\color{cyan}164.87}&
164.88&
164.89&
{\color{green}164.90}
\\\hline
{\color{green}165.91}&
{\color{green}165.92}&
{\color{cyan}165.93}&
{\color{cyan}165.94}&
{\color{green}165.95}&
{\color{cyan}165.96}&
{\color{green}166.97}&
{\color{cyan}166.98}&
{\color{cyan}166.99}&
166.100&
166.101&
{\color{green}166.102}&
{\color{green}167.103}&
{\color{green}167.104}&
{\color{cyan}167.105}
\\\hline
{\color{cyan}167.106}&
{\color{green}167.107}&
{\color{cyan}167.108}&
{\color{green}168.109}&
{\color{green}168.110}&
168.111&
{\color{red}168.112}&
{\color{green}169.113}&
{\color{red}169.114}&
{\color{magenta}169.115}&
{\color{green}169.116}&
{\color{green}170.117}&
{\color{red}170.118}&
{\color{magenta}170.119}&
{\color{green}170.120}
\\\hline
{\color{green}171.121}&
{\color{green}171.122}&
171.123&
{\color{red}171.124}&
{\color{green}172.125}&
{\color{green}172.126}&
172.127&
{\color{red}172.128}&
{\color{green}173.129}&
{\color{red}173.130}&
{\color{magenta}173.131}&
{\color{green}173.132}&
174.133&
{\color{yellow}174.134}&
174.135
\\\hline
{\color{green}174.136}&
{\color{green}175.137}&
{\color{green}175.138}&
175.139&
{\color{green}175.140}&
175.141&
{\color{gray}175.142}&
{\color{green}176.143}&
{\color{gray}176.144}&
{\color{orange}176.145}&
{\color{green}176.146}&
{\color{magenta}176.147}&
{\color{green}176.148}&
{\color{green}177.149}&
{\color{green}177.150}
\\\hline
177.151&
177.152&
{\color{green}177.153}&
{\color{red}177.154}&
{\color{green}178.155}&
{\color{red}178.156}&
{\color{magenta}178.157}&
{\color{magenta}178.158}&
{\color{green}178.159}&
{\color{green}178.160}&
{\color{green}179.161}&
{\color{red}179.162}&
{\color{magenta}179.163}&
{\color{magenta}179.164}&
{\color{green}179.165}
\\\hline
{\color{green}179.166}&
{\color{green}180.167}&
{\color{green}180.168}&
180.169&
180.170&
{\color{green}180.171}&
{\color{red}180.172}&
{\color{green}181.173}&
{\color{green}181.174}&
181.175&
181.176&
{\color{green}181.177}&
{\color{red}181.178}&
{\color{green}182.179}&
{\color{red}182.180}
\\\hline
{\color{magenta}182.181}&
{\color{magenta}182.182}&
{\color{green}182.183}&
{\color{green}182.184}&
{\color{green}183.185}&
{\color{green}183.186}&
{\color{green}183.187}&
{\color{green}183.188}&
{\color{green}183.189}&
{\color{red}183.190}&
{\color{green}184.191}&
{\color{green}184.192}&
{\color{green}184.193}&
{\color{green}184.194}&
{\color{green}184.195}
\\\hline
{\color{red}184.196}&
{\color{green}185.197}&
{\color{red}185.198}&
{\color{red}185.199}&
{\color{red}185.200}&
{\color{green}185.201}&
{\color{green}185.202}&
{\color{green}186.203}&
{\color{red}186.204}&
{\color{red}186.205}&
{\color{red}186.206}&
{\color{green}186.207}&
{\color{green}186.208}&
{\color{green}187.209}&
{\color{green}187.210}
\\\hline
187.211&
{\color{green}187.212}&
187.213&
{\color{green}187.214}&
{\color{green}188.215}&
{\color{green}188.216}&
{\color{green}188.217}&
{\color{green}188.218}&
{\color{green}188.219}&
{\color{green}188.220}&
{\color{green}189.221}&
{\color{green}189.222}&
{\color{green}189.223}&
189.224&
189.225
\\\hline
{\color{green}189.226}&
{\color{green}190.227}&
{\color{green}190.228}&
{\color{green}190.229}&
{\color{green}190.230}&
{\color{green}190.231}&
{\color{green}190.232}&
{\color{green}191.233}&
{\color{green}191.234}&
{\color{green}191.235}&
{\color{cyan}191.236}&
{\color{cyan}191.237}&
{\color{green}191.238}&
{\color{green}191.239}&
{\color{green}191.240}
\\\hline
{\color{green}191.241}&
{\color{gray}191.242}&
{\color{green}192.243}&
{\color{green}192.244}&
{\color{green}192.245}&
{\color{cyan}192.246}&
{\color{cyan}192.247}&
{\color{green}192.248}&
{\color{green}192.249}&
{\color{green}192.250}&
{\color{green}192.251}&
{\color{gray}192.252}&
{\color{green}193.253}&
{\color{gray}193.254}&
{\color{green}193.255}
\\\hline
{\color{brown}193.256}&
{\color{gray}193.257}&
{\color{red}193.258}&
{\color{red}193.259}&
{\color{green}193.260}&
{\color{green}193.261}&
{\color{green}193.262}&
{\color{green}194.263}&
{\color{gray}194.264}&
{\color{green}194.265}&
{\color{gray}194.266}&
{\color{brown}194.267}&
{\color{red}194.268}&
{\color{red}194.269}&
{\color{green}194.270}
\\\hline
{\color{green}194.271}&
{\color{green}194.272}&
195.1&
195.2&
{\color{magenta}195.3}&
196.4&
{\color{green}196.5}&
{\color{green}196.6}&
197.7&
197.8&
198.9&
{\color{magenta}198.10}&
198.11&
199.12&
199.13
\\\hline
{\color{cyan}200.14}&
{\color{cyan}200.15}&
200.16&
{\color{gray}200.17}&
{\color{green}201.18}&
{\color{cyan}201.19}&
{\color{cyan}201.20}&
{\color{brown}201.21}&
{\color{cyan}202.22}&
{\color{cyan}202.23}&
202.24&
{\color{cyan}202.25}&
{\color{green}203.26}&
{\color{cyan}203.27}&
{\color{cyan}203.28}
\\\hline
{\color{cyan}203.29}&
{\color{cyan}204.30}&
{\color{cyan}204.31}&
204.32&
{\color{green}205.33}&
{\color{gray}205.34}&
{\color{cyan}205.35}&
{\color{cyan}205.36}&
{\color{cyan}206.37}&
{\color{cyan}206.38}&
206.39&
{\color{green}207.40}&
{\color{green}207.41}&
{\color{cyan}207.42}&
{\color{red}207.43}
\\\hline
{\color{green}208.44}&
{\color{green}208.45}&
{\color{cyan}208.46}&
{\color{red}208.47}&
{\color{green}209.48}&
{\color{green}209.49}&
{\color{cyan}209.50}&
{\color{green}209.51}&
{\color{green}210.52}&
{\color{green}210.53}&
{\color{green}210.54}&
{\color{green}210.55}&
{\color{green}211.56}&
{\color{green}211.57}&
211.58
\\\hline
{\color{green}212.59}&
{\color{red}212.60}&
{\color{green}212.61}&
{\color{green}212.62}&
{\color{green}213.63}&
{\color{red}213.64}&
{\color{green}213.65}&
{\color{green}213.66}&
{\color{green}214.67}&
{\color{green}214.68}&
214.69&
{\color{green}215.70}&
{\color{green}215.71}&
215.72&
{\color{red}215.73}
\\\hline
{\color{green}216.74}&
{\color{green}216.75}&
216.76&
{\color{green}216.77}&
{\color{green}217.78}&
{\color{green}217.79}&
217.80&
{\color{green}218.81}&
{\color{green}218.82}&
{\color{cyan}218.83}&
{\color{red}218.84}&
{\color{green}219.85}&
{\color{green}219.86}&
{\color{green}219.87}&
{\color{green}219.88}
\\\hline
{\color{green}220.89}&
{\color{green}220.90}&
{\color{cyan}220.91}&
{\color{green}221.92}&
{\color{cyan}221.93}&
{\color{cyan}221.94}&
{\color{cyan}221.95}&
{\color{cyan}221.96}&
{\color{gray}221.97}&
{\color{green}222.98}&
{\color{cyan}222.99}&
{\color{cyan}222.100}&
{\color{green}222.101}&
{\color{cyan}222.102}&
{\color{brown}222.103}
\\\hline
{\color{green}223.104}&
{\color{cyan}223.105}&
{\color{cyan}223.106}&
{\color{cyan}223.107}&
{\color{cyan}223.108}&
{\color{gray}223.109}&
{\color{green}224.110}&
{\color{cyan}224.111}&
{\color{cyan}224.112}&
{\color{green}224.113}&
{\color{cyan}224.114}&
{\color{gray}224.115}&
{\color{green}225.116}&
{\color{cyan}225.117}&
{\color{cyan}225.118}
\\\hline
{\color{cyan}225.119}&
{\color{cyan}225.120}&
{\color{green}225.121}&
{\color{green}226.122}&
{\color{green}226.123}&
{\color{cyan}226.124}&
{\color{green}226.125}&
{\color{cyan}226.126}&
{\color{cyan}226.127}&
{\color{green}227.128}&
{\color{cyan}227.129}&
{\color{cyan}227.130}&
{\color{green}227.131}&
{\color{cyan}227.132}&
{\color{green}227.133}
\\\hline
{\color{green}228.134}&
{\color{green}228.135}&
{\color{cyan}228.136}&
{\color{green}228.137}&
{\color{cyan}228.138}&
{\color{cyan}228.139}&
{\color{green}229.140}&
{\color{cyan}229.141}&
{\color{cyan}229.142}&
{\color{cyan}229.143}&
{\color{cyan}229.144}&
{\color{green}230.145}&
{\color{cyan}230.146}&
{\color{cyan}230.147}&
{\color{cyan}230.148}
\\\hline
{\color{cyan}230.149}&
\\\hline\end{tabular}

}
\section{Page numbers for any MSG in \textbf{Supplementary Information III}}\label{page}
The following table tabulates all page numbers in \textbf{Supplementary Information III} for any given MSG, corresponding to results for all possible BNs in all special $k$ points in both spinful and spinless settings. For example, for the MSG 2.7, from the following table, we can know that it's of type IV (see the parentheses ()). The results are given from page 8191 to page 8192 and from page 39935 to page 39936. These results are all in the spinless setting, thus there are no BNs for this MSG in the spinful settings. Note that the results from page 39935 to page 39936 are for the MLG, related with the MSG 2.7 with the layer stacking along $\mathbf{a}$ or $\mathbf{c}$ direction.\clearpage
\small
\begin{tabular}{|llllllllll|}\hline
1.2(II):&P2-3&
1.2(II):&P39412-39414&
1.3(IV):&P3-4&
1.3(IV):&P8191-8191&
1.3(IV):&P39414-39415\\\hline
1.3(IV):&P39935-39935&
2.7(IV):&P8191-8192&
2.7(IV):&P39935-39936&
3.1(I):&P19087-19089&
3.1(I):&P26008-26010\\\hline
3.1(I):&P40530-40531&
3.1(I):&P40858-40859&
3.2(II):&P4-6&
3.2(II):&P19089-19090&
3.2(II):&P26010-26011\\\hline
3.2(II):&P39415-39417&
3.2(II):&P40532-40533&
3.2(II):&P40859-40861&
3.4(IV):&P6-7&
3.4(IV):&P8192-8193\\\hline
3.4(IV):&P19091-19092&
3.4(IV):&P26011-26013&
3.4(IV):&P39418-39418&
3.4(IV):&P39936-39937&
3.4(IV):&P40533-40534\\\hline
3.4(IV):&P40861-40861&
3.5(IV):&P8-10&
3.5(IV):&P8194-8196&
3.5(IV):&P19092-19094&
3.5(IV):&P26013-26014\\\hline
3.5(IV):&P39419-39421&
3.5(IV):&P39937-39940&
3.5(IV):&P40534-40535&
3.5(IV):&P40862-40863&
3.6(IV):&P10-12\\\hline
3.6(IV):&P8196-8199&
3.6(IV):&P19094-19095&
3.6(IV):&P26015-26016&
3.6(IV):&P39421-39422&
3.6(IV):&P39940-39941\\\hline
3.6(IV):&P40536-40536&
3.6(IV):&P40863-40864&
4.7(I):&P19095-19097&
4.7(I):&P26016-26018&
4.7(I):&P40536-40538\\\hline
4.7(I):&P40864-40865&
4.8(II):&P13-16&
4.8(II):&P8199-8200&
4.8(II):&P19097-19098&
4.8(II):&P26018-26019\\\hline
4.8(II):&P39422-39426&
4.8(II):&P39941-39943&
4.8(II):&P40538-40539&
4.8(II):&P40866-40867&
4.9(III):&P16-18\\\hline
4.9(III):&P8200-8202&
4.9(III):&P39426-39428&
4.9(III):&P39943-39945&
4.10(IV):&P18-20&
4.10(IV):&P8202-8204\\\hline
4.10(IV):&P19099-19100&
4.10(IV):&P26019-26021&
4.10(IV):&P39428-39429&
4.10(IV):&P39945-39946&
4.10(IV):&P40540-40540\\\hline
4.10(IV):&P40867-40868&
4.11(IV):&P20-21&
4.11(IV):&P8204-8205&
4.11(IV):&P19100-19102&
4.11(IV):&P26021-26022\\\hline
4.11(IV):&P39430-39430&
4.11(IV):&P39946-39947&
4.11(IV):&P40540-40542&
4.11(IV):&P40868-40869&
4.12(IV):&P21-22\\\hline
4.12(IV):&P8206-8207&
4.12(IV):&P19102-19103&
4.12(IV):&P26023-26024&
4.12(IV):&P39431-39431&
4.12(IV):&P39947-39948\\\hline
4.12(IV):&P40542-40543&
4.12(IV):&P40870-40870&
5.13(I):&P19103-19104&
5.13(I):&P26024-26025&
5.13(I):&P40543-40543\\\hline
5.13(I):&P40870-40871&
5.14(II):&P22-24&
5.14(II):&P19104-19105&
5.14(II):&P26025-26026&
5.14(II):&P39431-39432\\\hline
5.14(II):&P40543-40543&
5.14(II):&P40871-40871&
5.16(IV):&P24-24&
5.16(IV):&P8207-8207&
5.16(IV):&P19105-19106\\\hline
5.16(IV):&P26026-26026&
5.17(IV):&P25-25&
5.17(IV):&P8208-8208&
5.17(IV):&P19106-19106&
5.17(IV):&P26027-26027\\\hline
5.17(IV):&P39432-39432&
5.17(IV):&P39948-39948&
5.17(IV):&P40544-40544&
5.17(IV):&P40871-40871&
6.18(I):&P19107-19107\\\hline
6.18(I):&P26027-26028&
6.18(I):&P40544-40546&
6.18(I):&P40872-40873&
6.19(II):&P26-28&
6.19(II):&P19107-19108\\\hline
6.19(II):&P26028-26029&
6.19(II):&P39432-39435&
6.19(II):&P40546-40548&
6.19(II):&P40874-40875&
6.21(IV):&P28-30\\\hline
6.21(IV):&P8208-8211&
6.21(IV):&P19108-19109&
6.21(IV):&P26029-26030&
6.21(IV):&P39435-39437&
6.21(IV):&P39948-39950\\\hline
6.21(IV):&P40548-40549&
6.21(IV):&P40876-40877&
6.22(IV):&P31-31&
6.22(IV):&P8211-8212&
6.22(IV):&P19109-19110\\\hline
6.22(IV):&P26030-26030&
6.22(IV):&P39438-39438&
6.22(IV):&P39950-39951&
6.22(IV):&P40549-40551&
6.22(IV):&P40877-40878\\\hline
6.23(IV):&P32-34&
6.23(IV):&P8212-8215&
6.23(IV):&P19110-19110&
6.23(IV):&P26031-26031&
6.23(IV):&P39439-39440\\\hline
6.23(IV):&P39952-39953&
6.23(IV):&P40551-40551&
6.23(IV):&P40878-40879&
7.24(I):&P19111-19111&
7.24(I):&P26031-26032\\\hline
7.24(I):&P40552-40553&
7.24(I):&P40879-40880&
7.25(II):&P34-38&
7.25(II):&P8215-8216&
7.25(II):&P19111-19112\\\hline
7.25(II):&P26032-26033&
7.25(II):&P39440-39443&
7.25(II):&P39953-39954&
7.25(II):&P40553-40554&
7.25(II):&P40880-40881\\\hline
7.26(III):&P39-41&
7.26(III):&P8216-8218&
7.26(III):&P39443-39445&
7.26(III):&P39954-39956&
7.27(IV):&P41-43\\\hline
7.27(IV):&P8218-8221&
7.27(IV):&P19112-19113&
7.27(IV):&P26033-26034&
7.27(IV):&P39445-39445&
7.27(IV):&P39956-39956\\\hline
7.27(IV):&P40554-40554&
7.27(IV):&P40882-40882&
7.28(IV):&P44-44&
7.28(IV):&P8221-8222&
7.28(IV):&P19113-19114\\\hline
7.28(IV):&P26034-26034&
7.28(IV):&P39446-39446&
7.28(IV):&P39957-39957&
7.28(IV):&P40554-40555&
7.28(IV):&P40882-40883\\\hline
7.29(IV):&P45-47&
7.29(IV):&P8223-8225&
7.29(IV):&P19114-19114&
7.29(IV):&P26035-26035&
7.29(IV):&P39447-39448\\\hline
7.29(IV):&P39958-39959&
7.29(IV):&P40556-40556&
7.29(IV):&P40883-40884&
7.30(IV):&P47-50&
7.30(IV):&P8226-8228\\\hline
7.30(IV):&P19115-19115&
7.30(IV):&P26035-26036&
7.31(IV):&P50-51&
7.31(IV):&P8229-8230&
7.31(IV):&P19115-19116\\\hline
7.31(IV):&P26036-26037&
7.31(IV):&P39448-39448&
7.31(IV):&P39959-39959&
7.31(IV):&P40556-40557&
7.31(IV):&P40884-40885\\\hline
8.32(I):&P19116-19116&
8.32(I):&P26037-26037&
8.32(I):&P40557-40557&
8.32(I):&P40885-40885&
8.33(II):&P51-53\\\hline
8.33(II):&P19117-19117&
8.33(II):&P26037-26038&
8.33(II):&P39448-39449&
8.33(II):&P40558-40558&
8.33(II):&P40885-40885\\\hline
8.35(IV):&P53-54&
8.35(IV):&P8230-8231&
8.35(IV):&P19117-19117&
8.35(IV):&P26038-26038&
8.36(IV):&P54-55\\\hline
8.36(IV):&P8232-8232&
8.36(IV):&P19117-19118&
8.36(IV):&P26038-26038&
8.36(IV):&P39449-39449&
8.36(IV):&P39960-39960\\\hline
8.36(IV):&P40558-40558&
8.36(IV):&P40886-40886&
9.37(I):&P19118-19118&
9.37(I):&P26039-26039&
9.38(II):&P55-57\\\hline
9.38(II):&P8232-8233&
9.38(II):&P19118-19118&
9.38(II):&P26039-26039&
9.39(III):&P58-58&
9.39(III):&P8233-8234\\\hline
9.40(IV):&P59-59&
9.40(IV):&P8234-8234&
9.40(IV):&P19119-19119&
9.40(IV):&P26039-26040&
9.41(IV):&P59-61\\\hline
9.41(IV):&P8235-8236&
9.41(IV):&P19119-19119&
9.41(IV):&P26040-26040&
10.42(I):&P19119-19120&
10.42(I):&P26040-26041\\\hline
10.42(I):&P40558-40560&
10.42(I):&P40886-40888&
10.43(II):&P26041-26042&
10.43(II):&P40888-40889&
10.44(III):&P26043-26043\\\hline
10.44(III):&P40890-40891&
10.45(III):&P26043-26045&
10.45(III):&P40892-40893&
10.47(IV):&P8236-8238&
10.47(IV):&P19121-19121\\\hline
10.47(IV):&P26045-26046&
10.47(IV):&P39960-39962&
10.47(IV):&P40560-40560&
10.47(IV):&P40893-40894&
10.48(IV):&P8239-8241\\\hline
10.48(IV):&P19121-19121&
10.48(IV):&P26046-26047&
10.48(IV):&P39962-39964&
10.48(IV):&P40560-40561&
10.48(IV):&P40894-40895\\\hline
10.49(IV):&P8241-8246&
10.49(IV):&P19121-19122&
10.49(IV):&P26047-26047&
10.49(IV):&P39965-39967&
10.49(IV):&P40561-40561\\\hline
10.49(IV):&P40895-40895&
11.50(I):&P61-62&
11.50(I):&P8246-8247&
11.50(I):&P19122-19123&
11.50(I):&P26047-26048\\\hline
11.50(I):&P39450-39451&
11.50(I):&P39967-39968&
11.50(I):&P40561-40563&
11.50(I):&P40896-40897&
11.51(II):&P62-63\\\hline
11.51(II):&P8247-8248&
11.51(II):&P19123-19123&
11.51(II):&P26049-26049&
11.51(II):&P39451-39452&
11.51(II):&P39968-39969\\\hline
11.51(II):&P40563-40563&
11.51(II):&P40897-40898&
11.52(III):&P8248-8250&
11.52(III):&P19123-19123&
11.52(III):&P26050-26050\\\hline
\end{tabular}\newpage\begin{tabular}{|llllllllll|}\hline
11.52(III):&P39969-39971&
11.52(III):&P40563-40564&
11.52(III):&P40898-40899&
11.53(III):&P26050-26051&
11.53(III):&P40899-40901\\\hline
11.54(III):&P64-65&
11.54(III):&P8250-8251&
11.54(III):&P39452-39454&
11.54(III):&P39971-39973&
11.55(IV):&P65-66\\\hline
11.55(IV):&P8251-8254&
11.55(IV):&P19124-19124&
11.55(IV):&P26052-26052&
11.55(IV):&P39454-39454&
11.55(IV):&P39973-39974\\\hline
11.55(IV):&P40564-40564&
11.55(IV):&P40901-40901&
11.56(IV):&P8254-8255&
11.56(IV):&P26052-26053&
11.56(IV):&P39974-39975\\\hline
11.56(IV):&P40901-40903&
11.57(IV):&P66-67&
11.57(IV):&P8255-8258&
11.57(IV):&P19124-19124&
11.57(IV):&P26053-26054\\\hline
11.57(IV):&P39454-39454&
11.57(IV):&P39976-39977&
11.57(IV):&P40564-40564&
11.57(IV):&P40903-40903&
12.58(I):&P19125-19125\\\hline
12.58(I):&P26054-26055&
12.58(I):&P40565-40565&
12.58(I):&P40903-40904&
12.59(II):&P26055-26055&
12.59(II):&P40904-40904\\\hline
12.60(III):&P26055-26056&
12.60(III):&P40904-40904&
12.61(III):&P26056-26056&
12.61(III):&P40905-40905&
12.63(IV):&P8258-8260\\
\hline
\end{tabular}\newpage\begin{tabular}{|llllllllll|}\hline
17.14(IV):&P19164-19165&
17.14(IV):&P26104-26105&
17.15(IV):&P157-164&
17.15(IV):&P8371-8378&
17.15(IV):&P19165-19167\\\hline
17.15(IV):&P26105-26107&
18.16(I):&P165-166&
18.16(I):&P8379-8380&
18.16(I):&P19167-19170&
18.16(I):&P26107-26110\\\hline
18.16(I):&P39484-39484&
18.16(I):&P40010-40011&
18.16(I):&P40585-40586&
18.16(I):&P40930-40931&
18.17(II):&P167-175\\\hline
18.17(II):&P8381-8387&
18.17(II):&P19170-19171&
18.17(II):&P26110-26111&
18.17(II):&P39484-39486&
18.17(II):&P40011-40012\\\hline
18.17(II):&P40586-40586&
18.17(II):&P40931-40931&
18.18(III):&P176-184&
18.18(III):&P8388-8396&
18.18(III):&P19172-19172\\\hline
18.18(III):&P26112-26112&
18.18(III):&P39486-39488&
18.18(III):&P40013-40014&
18.19(III):&P184-188&
18.19(III):&P8396-8400\\\hline
18.19(III):&P19173-19173&
18.19(III):&P26113-26113&
18.19(III):&P39488-39489&
18.19(III):&P40014-40015&
18.19(III):&P40586-40586\\\hline
18.19(III):&P40932-40932&
18.20(IV):&P188-193&
18.20(IV):&P8401-8406&
18.20(IV):&P19174-19175&
18.20(IV):&P26114-26115\\
\hline
\end{tabular}\newpage\begin{tabular}{|llllllllll|}\hline
26.70(III):&P26377-26379&
26.70(III):&P39528-39529&
26.70(III):&P40039-40040&
26.70(III):&P40614-40615&
26.70(III):&P41017-41018\\\hline
26.71(IV):&P388-396&
26.71(IV):&P8555-8563&
26.71(IV):&P19307-19310&
26.71(IV):&P26379-26397&
26.71(IV):&P39529-39531\\\hline
26.71(IV):&P40040-40042&
26.71(IV):&P40615-40616&
26.71(IV):&P41018-41024&
26.72(IV):&P396-405&
26.72(IV):&P8563-8573\\\hline
26.72(IV):&P19310-19313&
26.72(IV):&P26397-26415&
26.72(IV):&P39531-39533&
26.72(IV):&P40042-40045&
26.72(IV):&P40616-40617\\\hline
26.72(IV):&P41024-41029&
26.73(IV):&P405-412&
26.73(IV):&P8574-8578&
26.73(IV):&P19313-19316&
26.73(IV):&P26415-26433\\\hline
26.73(IV):&P39534-39537&
26.73(IV):&P40045-40048&
26.73(IV):&P40617-40618&
26.73(IV):&P41029-41040&
26.74(IV):&P412-418\\\hline
26.74(IV):&P8578-8582&
26.74(IV):&P19316-19320&
26.74(IV):&P26433-26451&
26.74(IV):&P39537-39539&
26.74(IV):&P40048-40049\\\hline
26.74(IV):&P40618-40619&
26.74(IV):&P41040-41046&
26.75(IV):&P419-426&
26.75(IV):&P8583-8590&
26.75(IV):&P19320-19322\\
\hline
\end{tabular}\newpage\begin{tabular}{|llllllllll|}\hline
29.110(IV):&P654-661&
29.110(IV):&P8804-8811&
29.110(IV):&P19569-19579&
29.110(IV):&P26826-26836&
30.111(I):&P661-664\\\hline
30.111(I):&P8811-8814&
30.111(I):&P19579-19591&
30.111(I):&P26836-26846&
30.111(I):&P39594-39595&
30.111(I):&P40093-40094\\\hline
30.111(I):&P40674-40677&
30.111(I):&P41144-41147&
30.112(II):&P664-672&
30.112(II):&P8814-8820&
30.112(II):&P19591-19601\\\hline
30.112(II):&P26846-26855&
30.112(II):&P39595-39598&
30.112(II):&P40094-40096&
30.112(II):&P40677-40680&
30.112(II):&P41147-41150\\\hline
30.113(III):&P672-675&
30.113(III):&P8821-8824&
30.113(III):&P19601-19606&
30.113(III):&P26855-26860&
30.113(III):&P39598-39598\\\hline
30.113(III):&P40096-40096&
30.113(III):&P40680-40680&
30.113(III):&P41150-41151&
30.114(III):&P675-678&
30.114(III):&P8824-8827\\\hline
30.114(III):&P19606-19611&
30.114(III):&P26860-26865&
30.114(III):&P39598-39599&
30.114(III):&P40097-40097&
30.114(III):&P40681-40682\\\hline
30.114(III):&P41151-41153&
30.115(III):&P678-683&
30.115(III):&P8827-8832&
30.115(III):&P19611-19613&
30.115(III):&P26865-26867\\
\hline
\end{tabular}\newpage\begin{tabular}{|llllllllll|}\hline
34.157(II):&P9106-9112&
34.157(II):&P20006-20016&
34.157(II):&P27221-27230&
34.158(III):&P969-972&
34.158(III):&P9112-9115\\\hline
34.158(III):&P20016-20021&
34.158(III):&P27230-27235&
34.159(III):&P972-977&
34.159(III):&P9115-9120&
34.159(III):&P20021-20024\\\hline
34.159(III):&P27235-27238&
34.160(IV):&P977-984&
34.160(IV):&P9120-9127&
34.160(IV):&P20024-20034&
34.160(IV):&P27238-27247\\\hline
34.161(IV):&P984-993&
34.161(IV):&P9128-9137&
34.161(IV):&P20034-20045&
34.161(IV):&P27247-27257&
34.162(IV):&P994-1001\\\hline
34.162(IV):&P9138-9146&
34.162(IV):&P20045-20055&
34.162(IV):&P27257-27266&
34.163(IV):&P1002-1008&
34.163(IV):&P9146-9152\\\hline
34.163(IV):&P20055-20066&
34.163(IV):&P27266-27276&
34.164(IV):&P1008-1014&
34.164(IV):&P9153-9159&
34.164(IV):&P20066-20076\\\hline
34.164(IV):&P27276-27285&
35.165(I):&P1014-1017&
35.165(I):&P20076-20079&
35.165(I):&P27286-27295&
35.165(I):&P39626-39626\\\hline
35.165(I):&P40714-40715&
35.165(I):&P41184-41185&
35.166(II):&P1017-1020&
35.166(II):&P20080-20083&
35.166(II):&P27296-27305\\
\hline
\end{tabular}\newpage\begin{tabular}{|llllllllll|}\hline
42.219(I):&P1189-1194&
42.219(I):&P20320-20323&
42.219(I):&P27767-27782&
42.220(II):&P1195-1199&
42.220(II):&P20324-20327\\\hline
42.220(II):&P27782-27798&
42.221(III):&P20327-20333&
42.221(III):&P27798-27803&
42.222(III):&P20333-20335&
42.222(III):&P27804-27805\\\hline
42.223(IV):&P1200-1204&
42.223(IV):&P9271-9271&
42.223(IV):&P20335-20338&
42.223(IV):&P27806-27821&
43.224(I):&P1205-1206\\\hline
43.224(I):&P9272-9274&
43.224(I):&P20339-20350&
43.224(I):&P27821-27829&
43.225(II):&P1207-1213&
43.225(II):&P9275-9281\\\hline
43.225(II):&P20350-20360&
43.225(II):&P27829-27835&
43.226(III):&P1214-1216&
43.226(III):&P9281-9283&
43.226(III):&P20361-20365\\\hline
43.226(III):&P27835-27839&
43.227(III):&P1217-1221&
43.227(III):&P9284-9288&
43.227(III):&P20365-20367&
43.227(III):&P27840-27841\\\hline
43.228(IV):&P1222-1228&
43.228(IV):&P9289-9295&
43.228(IV):&P20367-20377&
43.228(IV):&P27842-27848&
44.229(I):&P1229-1231\\\hline
44.229(I):&P20377-20379&
44.229(I):&P27848-27858&
44.230(II):&P1231-1234&
44.230(II):&P20380-20382&
44.230(II):&P27858-27868\\
\hline
\end{tabular}\newpage\begin{tabular}{|llllllllll|}\hline
50.281(III):&P9532-9537&
50.281(III):&P20508-20511&
50.281(III):&P28311-28314&
50.281(III):&P39667-39669&
50.281(III):&P40162-40164\\\hline
50.281(III):&P40742-40742&
50.281(III):&P41341-41342&
50.282(III):&P1395-1400&
50.282(III):&P9537-9542&
50.282(III):&P20511-20514\\\hline
50.282(III):&P28314-28317&
50.282(III):&P39669-39670&
50.282(III):&P40164-40165&
50.282(III):&P40743-40743&
50.282(III):&P41342-41343\\\hline
50.283(III):&P1400-1402&
50.283(III):&P9542-9549&
50.283(III):&P20514-20516&
50.283(III):&P28317-28319&
50.283(III):&P39670-39671\\\hline
50.283(III):&P40165-40167&
50.283(III):&P40743-40743&
50.283(III):&P41343-41343&
50.284(IV):&P1403-1405&
50.284(IV):&P9550-9557\\\hline
50.284(IV):&P20516-20518&
50.284(IV):&P28319-28328&
50.284(IV):&P39671-39671&
50.284(IV):&P40167-40169&
50.284(IV):&P40744-40744\\\hline
50.284(IV):&P41343-41346&
50.285(IV):&P1405-1408&
50.285(IV):&P9558-9568&
50.285(IV):&P20518-20520&
50.285(IV):&P28328-28336\\\hline
50.286(IV):&P1408-1413&
50.286(IV):&P9568-9578&
50.286(IV):&P20520-20524&
50.286(IV):&P28336-28342&
50.287(IV):&P1413-1414\\
\hline
\end{tabular}\newpage\begin{tabular}{|llllllllll|}\hline
53.328(III):&P28672-28674&
53.328(III):&P39696-39698&
53.328(III):&P40213-40215&
53.328(III):&P40758-40758&
53.328(III):&P41426-41426\\\hline
53.329(III):&P1562-1562&
53.329(III):&P9896-9901&
53.329(III):&P20637-20638&
53.329(III):&P28674-28676&
53.329(III):&P39699-39699\\\hline
53.329(III):&P40216-40217&
53.329(III):&P40759-40759&
53.329(III):&P41426-41426&
53.330(IV):&P1563-1566&
53.330(IV):&P9901-9912\\\hline
53.330(IV):&P20638-20642&
53.330(IV):&P28676-28682&
53.331(IV):&P1567-1568&
53.331(IV):&P9912-9919&
53.331(IV):&P20642-20646\\\hline
53.331(IV):&P28682-28691&
53.331(IV):&P39699-39699&
53.331(IV):&P40217-40219&
53.331(IV):&P40759-40759&
53.331(IV):&P41427-41429\\\hline
53.332(IV):&P1568-1569&
53.332(IV):&P9920-9926&
53.332(IV):&P20646-20648&
53.332(IV):&P28691-28701&
53.332(IV):&P39699-39699\\\hline
53.332(IV):&P40219-40220&
53.332(IV):&P40759-40759&
53.332(IV):&P41429-41432&
53.333(IV):&P9927-9934&
53.333(IV):&P20648-20650\\\hline
53.333(IV):&P28701-28711&
53.333(IV):&P40221-40223&
53.333(IV):&P40760-40760&
53.333(IV):&P41432-41434&
53.334(IV):&P1569-1571\\
\hline
\end{tabular}\newpage\begin{tabular}{|llllllllll|}\hline
56.374(IV):&P10292-10303&
56.374(IV):&P20748-20749&
56.374(IV):&P29030-29038&
56.375(IV):&P1756-1757&
56.375(IV):&P10304-10311\\\hline
56.375(IV):&P20750-20750&
56.375(IV):&P29038-29053&
56.376(IV):&P1757-1759&
56.376(IV):&P10311-10321&
56.376(IV):&P20751-20752\\\hline
56.376(IV):&P29053-29064&
57.377(I):&P1760-1769&
57.377(I):&P10322-10329&
57.377(I):&P20752-20761&
57.377(I):&P29064-29078\\\hline
57.377(I):&P39723-39726&
57.377(I):&P40263-40264&
57.377(I):&P40768-40769&
57.377(I):&P41484-41488&
57.378(II):&P1769-1774\\\hline
57.378(II):&P10329-10337&
57.378(II):&P20761-20764&
57.378(II):&P29079-29087&
57.378(II):&P39726-39727&
57.378(II):&P40265-40267\\\hline
57.378(II):&P40769-40769&
57.378(II):&P41488-41491&
57.379(III):&P1774-1777&
57.379(III):&P10338-10345&
57.379(III):&P20764-20765\\\hline
57.379(III):&P29088-29098&
57.379(III):&P39727-39728&
57.379(III):&P40267-40269&
57.379(III):&P41491-41494&
57.380(III):&P10346-10350\\\hline
57.380(III):&P20765-20767&
57.380(III):&P29098-29107&
57.380(III):&P40269-40270&
57.380(III):&P40769-40770&
57.380(III):&P41494-41497\\
\hline
\end{tabular}\newpage\begin{tabular}{|llllllllll|}\hline
60.424(III):&P10717-10724&
60.424(III):&P20910-20912&
60.424(III):&P29410-29412&
60.425(III):&P1960-1961&
60.425(III):&P10725-10731\\\hline
60.425(III):&P20912-20914&
60.425(III):&P29412-29413&
60.426(IV):&P1961-1964&
60.426(IV):&P10732-10740&
60.426(IV):&P20914-20916\\\hline
60.426(IV):&P29414-29421&
60.427(IV):&P1964-1966&
60.427(IV):&P10741-10749&
60.427(IV):&P20916-20918&
60.427(IV):&P29421-29430\\\hline
60.428(IV):&P1966-1970&
60.428(IV):&P10750-10759&
60.428(IV):&P20918-20924&
60.428(IV):&P29431-29436&
60.429(IV):&P1970-1973\\\hline
60.429(IV):&P10760-10768&
60.429(IV):&P20924-20927&
60.429(IV):&P29437-29444&
60.430(IV):&P1973-1975&
60.430(IV):&P10769-10778\\\hline
60.430(IV):&P20928-20930&
60.430(IV):&P29444-29451&
60.431(IV):&P1975-1976&
60.431(IV):&P10779-10787&
60.431(IV):&P20930-20931\\\hline
60.431(IV):&P29451-29462&
60.432(IV):&P1977-1978&
60.432(IV):&P10787-10795&
60.432(IV):&P20932-20933&
60.432(IV):&P29462-29471\\\hline
61.433(I):&P1978-1986&
61.433(I):&P10796-10804&
61.433(I):&P20934-20946&
61.433(I):&P29471-29481&
61.434(II):&P1987-1992\\
\hline
\end{tabular}\newpage\begin{tabular}{|llllllllll|}\hline
66.497(III):&P29967-29969&
66.498(IV):&P11126-11129&
66.498(IV):&P21072-21072&
66.498(IV):&P29969-29978&
66.499(IV):&P2153-2153\\\hline
66.499(IV):&P11130-11134&
66.499(IV):&P21073-21073&
66.499(IV):&P29978-29988&
66.500(IV):&P11134-11138&
66.500(IV):&P21074-21075\\\hline
66.500(IV):&P29988-29996&
67.501(I):&P2154-2160&
67.501(I):&P11138-11139&
67.501(I):&P21075-21076&
67.501(I):&P29997-30012\\\hline
67.501(I):&P39745-39747&
67.501(I):&P40299-40299&
67.501(I):&P40781-40781&
67.501(I):&P41545-41548&
67.502(II):&P2161-2161\\\hline
67.502(II):&P11139-11139&
67.502(II):&P21076-21076&
67.502(II):&P30012-30027&
67.502(II):&P39747-39747&
67.502(II):&P40299-40299\\\hline
67.502(II):&P41548-41552&
67.503(III):&P30027-30036&
67.503(III):&P41552-41555&
67.504(III):&P11140-11140&
67.504(III):&P21076-21076\\\hline
67.504(III):&P30036-30045&
67.504(III):&P40299-40299&
67.504(III):&P41555-41556&
67.505(III):&P2161-2162&
67.505(III):&P11141-11142\\\hline
67.505(III):&P21077-21080&
67.505(III):&P30046-30049&
67.505(III):&P39747-39747&
67.505(III):&P40300-40300&
67.505(III):&P40781-40782\\
\hline
\end{tabular}\newpage\begin{tabular}{|llllllllll|}\hline
76.11(IV):&P21232-21241&
76.11(IV):&P30555-30563&
76.12(IV):&P2292-2295&
76.12(IV):&P11304-11307&
76.12(IV):&P21241-21249\\\hline
76.12(IV):&P30564-30572&
77.13(I):&P21250-21258&
77.13(I):&P30572-30581&
77.14(II):&P2295-2300&
77.14(II):&P11308-11309\\\hline
77.14(II):&P21258-21267&
77.14(II):&P30581-30589&
77.15(III):&P2300-2302&
77.15(III):&P11310-11311&
77.15(III):&P21267-21269\\\hline
77.15(III):&P30590-30592&
77.16(IV):&P2302-2308&
77.16(IV):&P11312-11318&
77.16(IV):&P21270-21278&
77.16(IV):&P30592-30601\\\hline
77.17(IV):&P2309-2313&
77.17(IV):&P11319-11321&
77.17(IV):&P21278-21287&
77.17(IV):&P30601-30609&
77.18(IV):&P2313-2319\\\hline
77.18(IV):&P11322-11328&
77.18(IV):&P21287-21295&
77.18(IV):&P30610-30618&
78.19(I):&P21296-21304&
78.19(I):&P30618-30627\\\hline
78.20(II):&P2320-2328&
78.20(II):&P11329-11334&
78.20(II):&P21304-21313&
78.20(II):&P30627-30635&
78.21(III):&P2328-2330\\\hline
78.21(III):&P11334-11336&
78.21(III):&P21313-21315&
78.21(III):&P30636-30638&
78.22(IV):&P2330-2333&
78.22(IV):&P11336-11339\\
\hline
\end{tabular}\newpage\begin{tabular}{|llllllllll|}\hline
87.80(IV):&P21520-21521&
87.80(IV):&P30918-30922&
88.81(I):&P2464-2466&
88.81(I):&P11603-11605&
88.81(I):&P21521-21528\\\hline
88.81(I):&P30922-30928&
88.82(II):&P2467-2467&
88.82(II):&P11606-11611&
88.82(II):&P21528-21529&
88.82(II):&P30929-30932\\\hline
88.83(III):&P2468-2474&
88.83(III):&P11612-11618&
88.83(III):&P21529-21532&
88.83(III):&P30932-30936&
88.84(III):&P11618-11622\\\hline
88.84(III):&P21533-21535&
88.84(III):&P30936-30941&
88.85(III):&P11623-11627&
88.85(III):&P21535-21535&
88.85(III):&P30942-30942\\\hline
88.86(IV):&P2474-2475&
88.86(IV):&P11628-11633&
88.86(IV):&P21536-21536&
88.86(IV):&P30943-30946&
89.87(I):&P2476-2481\\\hline
89.87(I):&P11634-11635&
89.87(I):&P21537-21542&
89.87(I):&P30946-30952&
89.87(I):&P39760-39762&
89.87(I):&P40320-40320\\\hline
89.87(I):&P40789-40789&
89.87(I):&P41572-41572&
89.88(II):&P2481-2486&
89.88(II):&P11636-11637&
89.88(II):&P21542-21548\\\hline
89.88(II):&P30952-30958&
89.88(II):&P39762-39763&
89.88(II):&P40320-40321&
89.88(II):&P40789-40790&
89.88(II):&P41573-41573\\
\hline
\end{tabular}\newpage\begin{tabular}{|llllllllll|}\hline
95.141(IV):&P11894-11901&
95.141(IV):&P21781-21784&
95.141(IV):&P31191-31194&
95.142(IV):&P2748-2755&
95.142(IV):&P11902-11908\\\hline
95.142(IV):&P21784-21787&
95.142(IV):&P31194-31197&
96.143(I):&P2756-2759&
96.143(I):&P11909-11912&
96.143(I):&P21787-21793\\\hline
96.143(I):&P31197-31203&
96.144(II):&P2759-2768&
96.144(II):&P11913-11920&
96.144(II):&P21793-21796&
96.144(II):&P31203-31206\\\hline
96.145(III):&P2769-2771&
96.145(III):&P11921-11923&
96.145(III):&P21796-21799&
96.145(III):&P31206-31209&
96.146(III):&P2772-2778\\\hline
96.146(III):&P11924-11930&
96.146(III):&P21800-21804&
96.146(III):&P31210-31214&
96.147(III):&P2778-2785&
96.147(III):&P11930-11936\\\hline
96.147(III):&P21804-21806&
96.147(III):&P31214-31216&
96.148(IV):&P2786-2793&
96.148(IV):&P11937-11943&
96.148(IV):&P21806-21809\\\hline
96.148(IV):&P31216-31219&
96.149(IV):&P2794-2800&
96.149(IV):&P11944-11949&
96.149(IV):&P21809-21814&
96.149(IV):&P31219-31224\\\hline
96.150(IV):&P2801-2804&
96.150(IV):&P11950-11953&
96.150(IV):&P21814-21820&
96.150(IV):&P31224-31230&
97.151(I):&P2804-2808\\
\hline
\end{tabular}\newpage\begin{tabular}{|llllllllll|}\hline
103.201(IV):&P3107-3115&
103.201(IV):&P12229-12238&
103.201(IV):&P22109-22112&
103.201(IV):&P31892-31910&
103.202(IV):&P3115-3124\\\hline
103.202(IV):&P12238-12249&
103.202(IV):&P22112-22115&
103.202(IV):&P31911-31929&
104.203(I):&P3124-3131&
104.203(I):&P12249-12254\\\hline
104.203(I):&P22116-22123&
104.203(I):&P31930-31945&
104.204(II):&P3131-3139&
104.204(II):&P12254-12264&
104.204(II):&P22123-22130\\\hline
104.204(II):&P31946-31961&
104.205(III):&P3140-3144&
104.205(III):&P12264-12269&
104.205(III):&P22130-22132&
104.205(III):&P31961-31971\\\hline
104.206(III):&P3145-3148&
104.206(III):&P12270-12277&
104.206(III):&P22133-22144&
104.206(III):&P31971-31983&
104.207(III):&P3149-3155\\\hline
104.207(III):&P12278-12284&
104.207(III):&P22144-22152&
104.207(III):&P31983-31991&
104.208(IV):&P3155-3164&
104.208(IV):&P12284-12294\\\hline
104.208(IV):&P22153-22159&
104.208(IV):&P31992-32007&
104.209(IV):&P3165-3173&
104.209(IV):&P12295-12303&
104.209(IV):&P22160-22166\\\hline
104.209(IV):&P32007-32022&
104.210(IV):&P3174-3183&
104.210(IV):&P12303-12312&
104.210(IV):&P22167-22173&
104.210(IV):&P32022-32037\\
\hline
\end{tabular}\newpage\begin{tabular}{|llllllllll|}\hline
112.265(IV):&P32770-32772&
112.266(IV):&P3526-3533&
112.266(IV):&P12617-12627&
112.266(IV):&P22445-22446&
112.266(IV):&P32773-32776\\\hline
113.267(I):&P3533-3538&
113.267(I):&P12628-12630&
113.267(I):&P22447-22449&
113.267(I):&P32776-32782&
113.267(I):&P39797-39798\\\hline
113.267(I):&P40351-40352&
113.267(I):&P40803-40803&
113.267(I):&P41586-41586&
113.268(II):&P3538-3544&
113.268(II):&P12631-12638\\\hline
113.268(II):&P22449-22450&
113.268(II):&P32782-32786&
113.268(II):&P39798-39800&
113.268(II):&P40352-40354&
113.268(II):&P40803-40803\\\hline
113.268(II):&P41587-41587&
113.269(III):&P3545-3550&
113.269(III):&P12639-12648&
113.269(III):&P22450-22452&
113.269(III):&P32786-32791\\\hline
113.269(III):&P39800-39801&
113.269(III):&P40354-40356&
113.269(III):&P40804-40804&
113.269(III):&P41587-41587&
113.270(III):&P3551-3553\\\hline
113.270(III):&P12649-12653&
113.270(III):&P22452-22455&
113.270(III):&P32791-32795&
113.270(III):&P39802-39802&
113.270(III):&P40356-40357\\\hline
113.270(III):&P40804-40805&
113.270(III):&P41587-41588&
113.271(III):&P3554-3559&
113.271(III):&P12653-12658&
113.271(III):&P22455-22456\\
\hline
\end{tabular}\newpage\begin{tabular}{|llllllllll|}\hline
119.319(III):&P22604-22605&
119.319(III):&P33048-33049&
119.320(IV):&P3824-3829&
119.320(IV):&P12949-12952&
119.320(IV):&P22605-22607\\\hline
119.320(IV):&P33050-33053&
120.321(I):&P3830-3834&
120.321(I):&P12953-12954&
120.321(I):&P22607-22608&
120.321(I):&P33054-33060\\\hline
120.322(II):&P3835-3841&
120.322(II):&P12955-12958&
120.322(II):&P22608-22610&
120.322(II):&P33061-33064&
120.323(III):&P3842-3847\\\hline
120.323(III):&P12959-12963&
120.323(III):&P22610-22612&
120.323(III):&P33065-33067&
120.324(III):&P3847-3850&
120.324(III):&P12963-12966\\\hline
120.324(III):&P22612-22614&
120.324(III):&P33067-33074&
120.325(III):&P3850-3853&
120.325(III):&P12966-12969&
120.325(III):&P22614-22615\\\hline
120.325(III):&P33074-33075&
120.326(IV):&P3853-3857&
120.326(IV):&P12969-12971&
120.326(IV):&P22615-22617&
120.326(IV):&P33076-33079\\\hline
121.327(I):&P3858-3865&
121.327(I):&P12972-12974&
121.327(I):&P22617-22619&
121.327(I):&P33080-33091&
121.328(II):&P3865-3872\\\hline
121.328(II):&P12975-12978&
121.328(II):&P22619-22621&
121.328(II):&P33091-33100&
121.329(III):&P3872-3879&
121.329(III):&P12979-12982\\
\hline
\end{tabular}\newpage\begin{tabular}{|llllllllll|}\hline
126.376(II):&P13327-13339&
126.376(II):&P22773-22775&
126.376(II):&P33571-33583&
126.377(III):&P13339-13344&
126.377(III):&P22775-22777\\\hline
126.377(III):&P33583-33599&
126.378(III):&P4077-4082&
126.378(III):&P13344-13352&
126.378(III):&P22777-22783&
126.378(III):&P33600-33606\\\hline
126.379(III):&P4083-4091&
126.379(III):&P13352-13365&
126.379(III):&P22783-22796&
126.379(III):&P33607-33617&
126.380(III):&P4092-4093\\\hline
126.380(III):&P13365-13372&
126.380(III):&P22797-22797&
126.380(III):&P33618-33623&
126.381(III):&P4094-4100&
126.381(III):&P13372-13378\\\hline
126.381(III):&P22797-22803&
126.381(III):&P33623-33629&
126.382(III):&P4100-4101&
126.382(III):&P13378-13385&
126.382(III):&P22803-22804\\\hline
126.382(III):&P33629-33636&
126.383(III):&P4101-4104&
126.383(III):&P13385-13395&
126.383(III):&P22804-22806&
126.383(III):&P33636-33638\\\hline
126.384(IV):&P4104-4106&
126.384(IV):&P13395-13406&
126.384(IV):&P22806-22808&
126.384(IV):&P33639-33649&
126.385(IV):&P4107-4109\\\hline
126.385(IV):&P13406-13416&
126.385(IV):&P22808-22810&
126.385(IV):&P33649-33657&
126.386(IV):&P4110-4111&
126.386(IV):&P13417-13429\\
\hline
\end{tabular}\newpage\begin{tabular}{|llllllllll|}\hline
130.428(III):&P4317-4318&
130.428(III):&P13850-13857&
130.428(III):&P22945-22945&
130.428(III):&P34067-34072&
130.429(III):&P4318-4325\\\hline
130.429(III):&P13857-13864&
130.429(III):&P22946-22950&
130.429(III):&P34072-34076&
130.430(III):&P4326-4330&
130.430(III):&P13865-13875\\\hline
130.430(III):&P22950-22951&
130.430(III):&P34076-34080&
130.431(III):&P4330-4332&
130.431(III):&P13876-13883&
130.431(III):&P22951-22953\\\hline
130.431(III):&P34081-34083&
130.432(IV):&P4332-4335&
130.432(IV):&P13883-13895&
130.432(IV):&P22953-22954&
130.432(IV):&P34083-34092\\\hline
130.433(IV):&P4335-4337&
130.433(IV):&P13896-13906&
130.433(IV):&P22955-22956&
130.433(IV):&P34092-34106&
130.434(IV):&P4337-4339\\\hline
130.434(IV):&P13907-13918&
130.434(IV):&P22956-22958&
130.434(IV):&P34107-34119&
131.435(I):&P4339-4349&
131.435(I):&P13919-13927\\\hline
131.435(I):&P22958-22962&
131.435(I):&P34119-34137&
131.436(II):&P4349-4350&
131.436(II):&P13927-13935&
131.436(II):&P22962-22963\\\hline
131.436(II):&P34137-34155&
131.437(III):&P13935-13938&
131.437(III):&P22963-22964&
131.437(III):&P34155-34175&
131.438(III):&P4350-4355\\
\hline
\end{tabular}\newpage\begin{tabular}{|llllllllll|}\hline
136.499(III):&P4606-4614&
136.499(III):&P14500-14510&
136.499(III):&P23154-23167&
136.499(III):&P34789-34799&
136.500(III):&P4614-4615\\\hline
136.500(III):&P14511-14516&
136.500(III):&P23168-23168&
136.500(III):&P34800-34805&
136.501(III):&P4615-4623&
136.501(III):&P14516-14524\\\hline
136.501(III):&P23168-23172&
136.501(III):&P34806-34810&
136.502(III):&P4624-4625&
136.502(III):&P14525-14532&
136.502(III):&P23172-23174\\\hline
136.502(III):&P34810-34814&
136.503(III):&P4625-4626&
136.503(III):&P14533-14539&
136.503(III):&P23174-23175&
136.503(III):&P34814-34817\\\hline
136.504(IV):&P4626-4629&
136.504(IV):&P14539-14552&
136.504(IV):&P23175-23178&
136.504(IV):&P34817-34825&
136.505(IV):&P4630-4630\\\hline
136.505(IV):&P14553-14563&
136.505(IV):&P23179-23180&
136.505(IV):&P34826-34839&
136.506(IV):&P4631-4632&
136.506(IV):&P14563-14573\\\hline
136.506(IV):&P23180-23182&
136.506(IV):&P34839-34852&
137.507(I):&P4632-4642&
137.507(I):&P14574-14585&
137.507(I):&P23182-23190\\\hline
137.507(I):&P34852-34866&
137.508(II):&P4643-4646&
137.508(II):&P14586-14598&
137.508(II):&P23190-23191&
137.508(II):&P34866-34876\\
\hline
\end{tabular}\newpage\begin{tabular}{|llllllllll|}\hline
143.2(II):&P23361-23370&
143.2(II):&P35538-35546&
143.2(II):&P39868-39869&
143.2(II):&P40454-40454&
143.3(IV):&P4880-4882\\\hline
143.3(IV):&P15049-15052&
143.3(IV):&P23370-23379&
143.3(IV):&P35547-35555&
144.4(I):&P23379-23388&
144.4(I):&P35556-35564\\\hline
144.5(II):&P4883-4888&
144.5(II):&P15053-15054&
144.5(II):&P23388-23397&
144.5(II):&P35565-35573&
144.6(IV):&P4888-4891\\\hline
144.6(IV):&P15055-15058&
144.6(IV):&P23397-23406&
144.6(IV):&P35574-35582&
145.7(I):&P23406-23415&
145.7(I):&P35583-35591\\\hline
145.8(II):&P4892-4896&
145.8(II):&P15058-15059&
145.8(II):&P23415-23424&
145.8(II):&P35592-35600&
145.9(IV):&P4897-4899\\\hline
145.9(IV):&P15060-15063&
145.9(IV):&P23424-23433&
145.9(IV):&P35601-35609&
146.10(I):&P23433-23439&
146.10(I):&P35610-35615\\\hline
146.11(II):&P4900-4904&
146.11(II):&P15064-15065&
146.11(II):&P23439-23445&
146.11(II):&P35616-35621&
146.12(IV):&P4905-4908\\\hline
146.12(IV):&P15066-15069&
146.12(IV):&P23445-23451&
146.12(IV):&P35622-35627&
147.13(I):&P23451-23455&
147.13(I):&P35628-35632\\
\hline
\end{tabular}\newpage\begin{tabular}{|llllllllll|}\hline
161.71(III):&P15326-15329&
161.71(III):&P23860-23865&
161.71(III):&P36037-36042&
161.72(IV):&P5166-5171&
161.72(IV):&P15330-15335\\\hline
161.72(IV):&P23866-23873&
161.72(IV):&P36043-36050&
162.73(I):&P5171-5178&
162.73(I):&P15335-15342&
162.73(I):&P23873-23880\\\hline
162.73(I):&P36050-36057&
162.73(I):&P39882-39883&
162.73(I):&P40466-40467&
162.73(I):&P40835-40835&
162.73(I):&P41676-41677\\\hline
162.74(II):&P15343-15350&
162.74(II):&P23880-23881&
162.74(II):&P36057-36064&
162.74(II):&P40467-40468&
162.74(II):&P41677-41677\\\hline
162.75(III):&P15350-15358&
162.75(III):&P23882-23884&
162.75(III):&P36064-36077&
162.75(III):&P40469-40470&
162.75(III):&P41678-41678\\\hline
162.76(III):&P15359-15364&
162.76(III):&P23885-23886&
162.76(III):&P36078-36080&
162.76(III):&P40470-40471&
162.76(III):&P41678-41679\\\hline
162.77(III):&P23886-23890&
162.77(III):&P36080-36084&
162.78(IV):&P5179-5179&
162.78(IV):&P15364-15373&
162.78(IV):&P23891-23892\\\hline
162.78(IV):&P36085-36091&
163.79(I):&P5180-5188&
163.79(I):&P15374-15382&
163.79(I):&P23893-23900&
163.79(I):&P36092-36099\\
\hline
\end{tabular}\newpage\begin{tabular}{|llllllllll|}\hline
175.140(III):&P24408-24412&
175.140(III):&P36684-36691&
175.140(III):&P40487-40488&
175.141(III):&P5439-5445&
175.141(III):&P15770-15776\\\hline
175.141(III):&P24412-24415&
175.141(III):&P36691-36694&
175.141(III):&P39891-39892&
175.141(III):&P40489-40490&
175.142(IV):&P15777-15795\\\hline
175.142(IV):&P24416-24419&
175.142(IV):&P36695-36699&
176.143(I):&P5445-5448&
176.143(I):&P15796-15799&
176.143(I):&P24419-24430\\\hline
176.143(I):&P36700-36711&
176.144(II):&P5449-5451&
176.144(II):&P15799-15813&
176.144(II):&P24431-24434&
176.144(II):&P36712-36716\\\hline
176.145(III):&P15814-15833&
176.145(III):&P24434-24438&
176.145(III):&P36717-36720&
176.146(III):&P15833-15842&
176.146(III):&P24438-24442\\\hline
176.146(III):&P36721-36728&
176.147(III):&P5451-5461&
176.147(III):&P15842-15852&
176.147(III):&P24442-24445&
176.147(III):&P36728-36731\\\hline
176.148(IV):&P5462-5462&
176.148(IV):&P15853-15864&
176.148(IV):&P24446-24447&
176.148(IV):&P36732-36738&
177.149(I):&P5463-5471\\\hline
177.149(I):&P15864-15869&
177.149(I):&P24448-24459&
177.149(I):&P36738-36749&
177.149(I):&P39893-39895&
177.149(I):&P40490-40491\\
\hline
\end{tabular}\newpage\begin{tabular}{|llllllllll|}\hline
186.204(II):&P37474-37497&
186.205(III):&P5964-5975&
186.205(III):&P16308-16319&
186.205(III):&P25034-25042&
186.205(III):&P37498-37506\\\hline
186.206(III):&P5976-5989&
186.206(III):&P16320-16333&
186.206(III):&P25042-25054&
186.206(III):&P37506-37518&
186.207(III):&P5990-5993\\\hline
186.207(III):&P16334-16337&
186.207(III):&P25054-25072&
186.207(III):&P37518-37536&
186.208(IV):&P5994-6007&
186.208(IV):&P16338-16350\\\hline
186.208(IV):&P25073-25080&
186.208(IV):&P37537-37560&
187.209(I):&P6008-6016&
187.209(I):&P16351-16355&
187.209(I):&P25081-25086\\\hline
187.209(I):&P37561-37573&
187.209(I):&P39908-39910&
187.209(I):&P40501-40502&
187.209(I):&P40846-40847&
187.209(I):&P41694-41697\\\hline
187.210(II):&P6016-6024&
187.210(II):&P16355-16359&
187.210(II):&P25087-25092&
187.210(II):&P37573-37585&
187.210(II):&P39910-39912\\\hline
187.210(II):&P40502-40503&
187.210(II):&P40847-40847&
187.210(II):&P41697-41700&
187.211(III):&P6025-6027&
187.211(III):&P16360-16362\\\hline
187.211(III):&P25093-25097&
187.211(III):&P37585-37589&
187.211(III):&P39912-39912&
187.211(III):&P40503-40503&
187.211(III):&P40848-40848\\
\hline
\end{tabular}\newpage\begin{tabular}{|llllllllll|}\hline
193.262(IV):&P16890-16903&
193.262(IV):&P25425-25427&
193.262(IV):&P38224-38245&
194.263(I):&P6424-6443&
194.263(I):&P16904-16918\\\hline
194.263(I):&P25427-25435&
194.263(I):&P38246-38264&
194.264(II):&P6444-6446&
194.264(II):&P16919-16934&
194.264(II):&P25435-25438\\\hline
194.264(II):&P38264-38281&
194.265(III):&P16935-16943&
194.265(III):&P25438-25442&
194.265(III):&P38282-38306&
194.266(III):&P6447-6449\\\hline
194.266(III):&P16944-16958&
194.266(III):&P25442-25444&
194.266(III):&P38307-38314&
194.267(III):&P16958-16977&
194.267(III):&P25444-25447\\\hline
194.267(III):&P38314-38327&
194.268(III):&P6450-6460&
194.268(III):&P16977-16987&
194.268(III):&P25448-25452&
194.268(III):&P38328-38332\\\hline
194.269(III):&P6461-6472&
194.269(III):&P16988-16999&
194.269(III):&P25453-25459&
194.269(III):&P38333-38339&
194.270(III):&P6472-6476\\\hline
194.270(III):&P16999-17003&
194.270(III):&P25459-25470&
194.270(III):&P38339-38350&
194.271(III):&P6476-6476&
194.271(III):&P17003-17011\\\hline
194.271(III):&P25470-25472&
194.271(III):&P38350-38355&
194.272(IV):&P6477-6478&
194.272(IV):&P17012-17025&
194.272(IV):&P25473-25475\\
\hline
\end{tabular}\newpage\begin{tabular}{|llllllllll|}\hline
203.26(I):&P6737-6737&
203.26(I):&P6739-6739&
203.26(I):&P6741-6743&
203.26(I):&P17265-17265&
203.26(I):&P17267-17267\\\hline
203.26(I):&P17269-17272&
203.26(I):&P25552-25558&
203.26(I):&P38508-38513&
203.27(II):&P6743-6743&
203.27(II):&P6745-6745\\\hline
203.27(II):&P6747-6747&
203.27(II):&P17273-17273&
203.27(II):&P17275-17275&
203.27(II):&P17277-17277&
203.27(II):&P17279-17279\\\hline
203.27(II):&P17281-17285&
203.27(II):&P25559-25559&
203.27(II):&P38513-38518&
203.28(III):&P6748-6748&
203.28(III):&P6750-6750\\\hline
203.28(III):&P17286-17286&
203.28(III):&P17288-17288&
203.28(III):&P17290-17294&
203.28(III):&P25560-25561&
203.28(III):&P38518-38519\\\hline
203.29(IV):&P6751-6751&
203.29(IV):&P6753-6753&
203.29(IV):&P6755-6755&
203.29(IV):&P17294-17294&
203.29(IV):&P17296-17296\\\hline
203.29(IV):&P17298-17298&
203.29(IV):&P17300-17300&
203.29(IV):&P17302-17307&
203.29(IV):&P25561-25562&
203.29(IV):&P38519-38524\\\hline
204.30(I):&P6756-6756&
204.30(I):&P6758-6758&
204.30(I):&P6760-6760&
204.30(I):&P6762-6762&
204.30(I):&P6764-6764\\
\hline
\end{tabular}\newpage\begin{tabular}{|llllllllll|}\hline
209.51(IV):&P17570-17570&
209.51(IV):&P17572-17572&
209.51(IV):&P17574-17576&
209.51(IV):&P25647-25652&
209.51(IV):&P38628-38633\\\hline
210.52(I):&P7010-7010&
210.52(I):&P7012-7012&
210.52(I):&P7014-7014&
210.52(I):&P7016-7018&
210.52(I):&P17576-17576\\\hline
210.52(I):&P17578-17578&
210.52(I):&P17580-17580&
210.52(I):&P17582-17584&
210.52(I):&P25653-25658&
210.52(I):&P38633-38638\\\hline
210.53(II):&P7019-7019&
210.53(II):&P7021-7021&
210.53(II):&P7023-7023&
210.53(II):&P7025-7028&
210.53(II):&P17584-17584\\\hline
210.53(II):&P17586-17586&
210.53(II):&P17588-17588&
210.53(II):&P17590-17592&
210.53(II):&P25658-25663&
210.53(II):&P38639-38644\\\hline
210.54(III):&P7028-7028&
210.54(III):&P7030-7030&
210.54(III):&P7032-7032&
210.54(III):&P7034-7035&
210.54(III):&P17592-17592\\\hline
210.54(III):&P17594-17595&
210.54(III):&P25664-25666&
210.54(III):&P38644-38647&
210.55(IV):&P7036-7036&
210.55(IV):&P7038-7038\\\hline
210.55(IV):&P7040-7040&
210.55(IV):&P7042-7044&
210.55(IV):&P17596-17596&
210.55(IV):&P17598-17598&
210.55(IV):&P17600-17600\\
\hline
\end{tabular}\newpage\begin{tabular}{|llllllllll|}\hline
216.77(IV):&P7316-7316&
216.77(IV):&P7318-7318&
216.77(IV):&P7320-7325&
216.77(IV):&P17844-17844&
216.77(IV):&P17846-17846\\\hline
216.77(IV):&P17848-17848&
216.77(IV):&P17850-17855&
216.77(IV):&P25781-25785&
216.77(IV):&P38776-38781&
217.78(I):&P7326-7326\\\hline
217.78(I):&P7328-7328&
217.78(I):&P7330-7330&
217.78(I):&P7332-7332&
217.78(I):&P7334-7334&
217.78(I):&P7336-7336\\\hline
217.78(I):&P7338-7339&
217.78(I):&P7341-7342&
217.78(I):&P7344-7345&
217.78(I):&P7347-7350&
217.78(I):&P17855-17855\\\hline
217.78(I):&P17857-17857&
217.78(I):&P17859-17859&
217.78(I):&P17861-17861&
217.78(I):&P17863-17863&
217.78(I):&P17865-17865\\\hline
217.78(I):&P17867-17867&
217.78(I):&P17869-17870&
217.78(I):&P17872-17873&
217.78(I):&P17875-17878&
217.78(I):&P25785-25793\\\hline
217.78(I):&P38781-38798&
217.79(II):&P7351-7351&
217.79(II):&P7353-7353&
217.79(II):&P7355-7355&
217.79(II):&P7357-7357\\\hline
217.79(II):&P7359-7359&
217.79(II):&P7361-7361&
217.79(II):&P7363-7364&
217.79(II):&P7366-7367&
217.79(II):&P7369-7370\\
\hline
\end{tabular}\newpage\begin{tabular}{|llllllllll|}\hline
222.98(I):&P18186-18196&
222.98(I):&P18198-18199&
222.98(I):&P18201-18202&
222.98(I):&P25878-25884&
222.98(I):&P38952-38963\\\hline
222.99(II):&P7649-7649&
222.99(II):&P7651-7656&
222.99(II):&P18203-18204&
222.99(II):&P18206-18207&
222.99(II):&P18209-18210\\\hline
222.99(II):&P18212-18220&
222.99(II):&P18222-18224&
222.99(II):&P25885-25886&
222.99(II):&P38963-38975&
222.100(III):&P7658-7661\\\hline
222.100(III):&P18225-18226&
222.100(III):&P18228-18236&
222.100(III):&P18238-18240&
222.100(III):&P25886-25887&
222.100(III):&P38975-38983\\\hline
222.101(III):&P7662-7662&
222.101(III):&P7664-7665&
222.101(III):&P7667-7668&
222.101(III):&P7670-7675&
222.101(III):&P7677-7679\\\hline
222.101(III):&P18240-18240&
222.101(III):&P18242-18248&
222.101(III):&P18250-18251&
222.101(III):&P18253-18254&
222.101(III):&P25887-25893\\\hline
222.101(III):&P38983-38988&
222.102(III):&P7679-7679&
222.102(III):&P7681-7685&
222.102(III):&P18255-18256&
222.102(III):&P18258-18266\\\hline
222.102(III):&P18268-18270&
222.102(III):&P25893-25895&
222.102(III):&P38988-38991&
222.103(IV):&P7687-7688&
222.103(IV):&P7690-7692\\
\hline
\end{tabular}\newpage\begin{tabular}{|llllllllll|}\hline
226.125(III):&P7921-7926&
226.125(III):&P18650-18650&
226.125(III):&P18652-18652&
226.125(III):&P18654-18659&
226.125(III):&P25936-25938\\\hline
226.125(III):&P39195-39202&
226.126(III):&P7927-7927&
226.126(III):&P7929-7929&
226.126(III):&P18660-18660&
226.126(III):&P18662-18662\\\hline
226.126(III):&P18664-18664&
226.126(III):&P18666-18670&
226.126(III):&P25939-25940&
226.126(III):&P39203-39205&
226.127(IV):&P7930-7930\\\hline
226.127(IV):&P7932-7932&
226.127(IV):&P18670-18670&
226.127(IV):&P18672-18672&
226.127(IV):&P18674-18674&
226.127(IV):&P18676-18676\\\hline
226.127(IV):&P18678-18678&
226.127(IV):&P18680-18680&
226.127(IV):&P18682-18687&
226.127(IV):&P25940-25940&
226.127(IV):&P39205-39218\\\hline
227.128(I):&P7934-7934&
227.128(I):&P7936-7936&
227.128(I):&P7938-7938&
227.128(I):&P7940-7940&
227.128(I):&P7942-7942\\\hline
227.128(I):&P7944-7944&
227.128(I):&P7946-7950&
227.128(I):&P18687-18687&
227.128(I):&P18689-18689&
227.128(I):&P18691-18691\\\hline
227.128(I):&P18693-18693&
227.128(I):&P18695-18695&
227.128(I):&P18697-18697&
227.128(I):&P18699-18706&
227.128(I):&P25941-25945\\
\hline
\end{tabular}\newpage\begin{tabular}{|llllllllll|}\hline
230.145(I):&P18979-18979&
230.145(I):&P18981-18981&
230.145(I):&P18983-18983&
230.145(I):&P18985-18985&
230.145(I):&P18987-18987\\\hline
230.145(I):&P18989-18989&
230.145(I):&P18991-18991&
230.145(I):&P18993-18993&
230.145(I):&P18995-18995&
230.145(I):&P18997-18997\\\hline
230.145(I):&P18999-19000&
230.145(I):&P19002-19005&
230.145(I):&P25987-25996&
230.145(I):&P39370-39381&
230.146(II):&P8145-8145\\\hline
230.146(II):&P8147-8147&
230.146(II):&P8149-8149&
230.146(II):&P8151-8151&
230.146(II):&P8153-8154&
230.146(II):&P19005-19005\\\hline
230.146(II):&P19007-19007&
230.146(II):&P19009-19009&
230.146(II):&P19011-19011&
230.146(II):&P19013-19013&
230.146(II):&P19015-19015\\\hline
230.146(II):&P19017-19017&
230.146(II):&P19019-19019&
230.146(II):&P19021-19021&
230.146(II):&P19023-19023&
230.146(II):&P19025-19029\\\hline
230.146(II):&P25996-25998&
230.146(II):&P39381-39392&
230.147(III):&P8155-8155&
230.147(III):&P8157-8157&
230.147(III):&P8159-8160\\\hline
230.147(III):&P19030-19030&
230.147(III):&P19032-19032&
230.147(III):&P19034-19034&
230.147(III):&P19036-19036&
230.147(III):&P19038-19038\\\hline
230.147(III):&P19040-19040&
230.147(III):&P19042-19043&
230.147(III):&P19045-19046&
230.147(III):&P19048-19051&
230.147(III):&P25998-26001\\\hline
230.147(III):&P39392-39403&
230.148(III):&P8162-8162&
230.148(III):&P8164-8164&
230.148(III):&P8166-8166&
230.148(III):&P8168-8168\\\hline
230.148(III):&P8170-8170&
230.148(III):&P8172-8172&
230.148(III):&P8174-8174&
230.148(III):&P8176-8176&
230.148(III):&P8178-8178\\\hline
230.148(III):&P8180-8182&
230.148(III):&P19051-19051&
230.148(III):&P19053-19053&
230.148(III):&P19055-19055&
230.148(III):&P19057-19057\\\hline
230.148(III):&P19059-19059&
230.148(III):&P19061-19061&
230.148(III):&P19063-19063&
230.148(III):&P19065-19066&
230.148(III):&P19068-19069\\\hline
230.148(III):&P26001-26005&
230.148(III):&P39403-39408&
230.149(III):&P8183-8183&
230.149(III):&P8185-8185&
230.149(III):&P8187-8187\\\hline
230.149(III):&P8189-8190&
230.149(III):&P19069-19069&
230.149(III):&P19071-19071&
230.149(III):&P19073-19073&
230.149(III):&P19075-19075\\\hline
230.149(III):&P19077-19077&
230.149(III):&P19079-19079&
230.149(III):&P19081-19081&
230.149(III):&P19083-19087&
230.149(III):&P26006-26008\\\hline
230.149(III):&P39409-39411&
\end{tabular}
\end{widetext}

\bibliography{allnodes}

\begin{thebibliography}{78}%
\makeatletter
\providecommand \@ifxundefined [1]{%
 \@ifx{#1\undefined}
}%
\providecommand \@ifnum [1]{%
 \ifnum #1\expandafter \@firstoftwo
 \else \expandafter \@secondoftwo
 \fi
}%
\providecommand \@ifx [1]{%
 \ifx #1\expandafter \@firstoftwo
 \else \expandafter \@secondoftwo
 \fi
}%
\providecommand \natexlab [1]{#1}%
\providecommand \enquote  [1]{``#1''}%
\providecommand \bibnamefont  [1]{#1}%
\providecommand \bibfnamefont [1]{#1}%
\providecommand \citenamefont [1]{#1}%
\providecommand \href@noop [0]{\@secondoftwo}%
\providecommand \href [0]{\begingroup \@sanitize@url \@href}%
\providecommand \@href[1]{\@@startlink{#1}\@@href}%
\providecommand \@@href[1]{\endgroup#1\@@endlink}%
\providecommand \@sanitize@url [0]{\catcode `\\12\catcode `\$12\catcode
  `\&12\catcode `\#12\catcode `\^12\catcode `\_12\catcode `\%12\relax}%
\providecommand \@@startlink[1]{}%
\providecommand \@@endlink[0]{}%
\providecommand \url  [0]{\begingroup\@sanitize@url \@url }%
\providecommand \@url [1]{\endgroup\@href {#1}{\urlprefix }}%
\providecommand \urlprefix  [0]{URL }%
\providecommand \Eprint [0]{\href }%
\providecommand \doibase [0]{http://dx.doi.org/}%
\providecommand \selectlanguage [0]{\@gobble}%
\providecommand \bibinfo  [0]{\@secondoftwo}%
\providecommand \bibfield  [0]{\@secondoftwo}%
\providecommand \translation [1]{[#1]}%
\providecommand \BibitemOpen [0]{}%
\providecommand \bibitemStop [0]{}%
\providecommand \bibitemNoStop [0]{.\EOS\space}%
\providecommand \EOS [0]{\spacefactor3000\relax}%
\providecommand \BibitemShut  [1]{\csname bibitem#1\endcsname}%
\let\auto@bib@innerbib\@empty
%</preamble>
\bibitem [{\citenamefont {Wan}\ \emph {et~al.}(2011)\citenamefont {Wan},
  \citenamefont {Turner}, \citenamefont {Vishwanath},\ and\ \citenamefont
  {Savrasov}}]{PRB-Wan}%
  \BibitemOpen
  \bibfield  {author} {\bibinfo {author} {\bibfnamefont {Xiangang}\
  \bibnamefont {Wan}}, \bibinfo {author} {\bibfnamefont {Ari~M.}\ \bibnamefont
  {Turner}}, \bibinfo {author} {\bibfnamefont {Ashvin}\ \bibnamefont
  {Vishwanath}}, \ and\ \bibinfo {author} {\bibfnamefont {Sergey~Y.}\
  \bibnamefont {Savrasov}},\ }\bibfield  {title} {\enquote {\bibinfo {title}
  {Topological semimetal and {F}ermi-arc surface states in the electronic
  structure of pyrochlore iridates},}\ }\href@noop {} {\bibfield  {journal}
  {\bibinfo  {journal} {Phys. Rev. B}\ }\textbf {\bibinfo {volume} {83}},\
  \bibinfo {pages} {205101} (\bibinfo {year} {2011})}\BibitemShut {NoStop}%
\bibitem [{\citenamefont {Weng}\ \emph {et~al.}(2015)\citenamefont {Weng},
  \citenamefont {Fang}, \citenamefont {Fang}, \citenamefont {Bernevig},\ and\
  \citenamefont {Dai}}]{TaAs-PRX}%
  \BibitemOpen
  \bibfield  {author} {\bibinfo {author} {\bibfnamefont {Hongming}\
  \bibnamefont {Weng}}, \bibinfo {author} {\bibfnamefont {Chen}\ \bibnamefont
  {Fang}}, \bibinfo {author} {\bibfnamefont {Zhong}\ \bibnamefont {Fang}},
  \bibinfo {author} {\bibfnamefont {B.~Andrei}\ \bibnamefont {Bernevig}}, \
  and\ \bibinfo {author} {\bibfnamefont {Xi}~\bibnamefont {Dai}},\ }\bibfield
  {title} {\enquote {\bibinfo {title} {Weyl {S}emimetal {P}hase in
  {N}oncentrosymmetric {T}ransition-{M}etal {M}onophosphides},}\ }\href@noop {}
  {\bibfield  {journal} {\bibinfo  {journal} {Phys. Rev. X}\ }\textbf {\bibinfo
  {volume} {5}},\ \bibinfo {pages} {011029} (\bibinfo {year}
  {2015})}\BibitemShut {NoStop}%
\bibitem [{\citenamefont {et~al.}(2015)}]{TaAs-NC}%
  \BibitemOpen
  \bibfield  {author} {\bibinfo {author} {\bibfnamefont {Shin-Ming~Huang}\
  \bibnamefont {et~al.}},\ }\bibfield  {title} {\enquote {\bibinfo {title} {A
  {W}eyl {F}ermion semimetal with surface {F}ermi arcs in the transition metal
  monopnictide {T}a{A}s class},}\ }\href@noop {} {\bibfield  {journal}
  {\bibinfo  {journal} {Nature Communications}\ }\textbf {\bibinfo {volume}
  {6}},\ \bibinfo {pages} {7373} (\bibinfo {year} {2015})}\BibitemShut
  {NoStop}%
\bibitem [{\citenamefont {Fang}\ \emph {et~al.}(2012)\citenamefont {Fang},
  \citenamefont {Gilbert}, \citenamefont {Dai},\ and\ \citenamefont
  {Bernevig}}]{ChenFang-largeChern}%
  \BibitemOpen
  \bibfield  {author} {\bibinfo {author} {\bibfnamefont {Chen}\ \bibnamefont
  {Fang}}, \bibinfo {author} {\bibfnamefont {Matthew~J.}\ \bibnamefont
  {Gilbert}}, \bibinfo {author} {\bibfnamefont {Xi}~\bibnamefont {Dai}}, \ and\
  \bibinfo {author} {\bibfnamefont {B.~Andrei}\ \bibnamefont {Bernevig}},\
  }\bibfield  {title} {\enquote {\bibinfo {title} {Multi-{W}eyl {T}opological
  {S}emimetals {S}tabilized by {P}oint {G}roup {S}ymmetry},}\ }\href@noop {}
  {\bibfield  {journal} {\bibinfo  {journal} {Phys. Rev. Lett.}\ }\textbf
  {\bibinfo {volume} {108}},\ \bibinfo {pages} {266802} (\bibinfo {year}
  {2012})}\BibitemShut {NoStop}%
\bibitem [{\citenamefont {Bradlyn}\ \emph {et~al.}(2016)\citenamefont
  {Bradlyn}, \citenamefont {Cano}, \citenamefont {Wang}, \citenamefont
  {Vergniory}, \citenamefont {Felser}, \citenamefont {Cava},\ and\
  \citenamefont {Bernevig}}]{new-fermions}%
  \BibitemOpen
  \bibfield  {author} {\bibinfo {author} {\bibfnamefont {Barry}\ \bibnamefont
  {Bradlyn}}, \bibinfo {author} {\bibfnamefont {Jennifer}\ \bibnamefont
  {Cano}}, \bibinfo {author} {\bibfnamefont {Zhijun}\ \bibnamefont {Wang}},
  \bibinfo {author} {\bibfnamefont {M.~G.}\ \bibnamefont {Vergniory}}, \bibinfo
  {author} {\bibfnamefont {C.}~\bibnamefont {Felser}}, \bibinfo {author}
  {\bibfnamefont {R.~J.}\ \bibnamefont {Cava}}, \ and\ \bibinfo {author}
  {\bibfnamefont {B.~Andrei}\ \bibnamefont {Bernevig}},\ }\bibfield  {title}
  {\enquote {\bibinfo {title} {Beyond {D}irac and {W}eyl fermions:
  {U}nconventional quasiparticles in conventional crystals},}\ }\href@noop {}
  {\bibfield  {journal} {\bibinfo  {journal} {Science}\ }\textbf {\bibinfo
  {volume} {353}},\ \bibinfo {pages} {aaf5037} (\bibinfo {year}
  {2016})}\BibitemShut {NoStop}%
\bibitem [{\citenamefont {Tang}\ \emph {et~al.}(2017)\citenamefont {Tang},
  \citenamefont {Zhou},\ and\ \citenamefont {Zhang}}]{CoSi-Tang}%
  \BibitemOpen
  \bibfield  {author} {\bibinfo {author} {\bibfnamefont {Peizhe}\ \bibnamefont
  {Tang}}, \bibinfo {author} {\bibfnamefont {Quan}\ \bibnamefont {Zhou}}, \
  and\ \bibinfo {author} {\bibfnamefont {Shou-Cheng}\ \bibnamefont {Zhang}},\
  }\bibfield  {title} {\enquote {\bibinfo {title} {Multiple {T}ypes of
  {T}opological {F}ermions in {T}ransition {M}etal {S}ilicides},}\ }\href@noop
  {} {\bibfield  {journal} {\bibinfo  {journal} {Phys. Rev. Lett.}\ }\textbf
  {\bibinfo {volume} {119}},\ \bibinfo {pages} {206402} (\bibinfo {year}
  {2017})}\BibitemShut {NoStop}%
\bibitem [{\citenamefont {Zhang}\ \emph {et~al.}(2018)\citenamefont {Zhang},
  \citenamefont {Song}, \citenamefont {Alexandradinata}, \citenamefont {Weng},
  \citenamefont {Fang}, \citenamefont {Lu},\ and\ \citenamefont
  {Fang}}]{CoSi-Zhang}%
  \BibitemOpen
  \bibfield  {author} {\bibinfo {author} {\bibfnamefont {Tiantian}\
  \bibnamefont {Zhang}}, \bibinfo {author} {\bibfnamefont {Zhida}\ \bibnamefont
  {Song}}, \bibinfo {author} {\bibfnamefont {A.}~\bibnamefont
  {Alexandradinata}}, \bibinfo {author} {\bibfnamefont {Hongming}\ \bibnamefont
  {Weng}}, \bibinfo {author} {\bibfnamefont {Chen}\ \bibnamefont {Fang}},
  \bibinfo {author} {\bibfnamefont {Ling}\ \bibnamefont {Lu}}, \ and\ \bibinfo
  {author} {\bibfnamefont {Zhong}\ \bibnamefont {Fang}},\ }\bibfield  {title}
  {\enquote {\bibinfo {title} {Double-{W}eyl {P}honons in {T}ransition-{M}etal
  {M}onosilicides},}\ }\href@noop {} {\bibfield  {journal} {\bibinfo  {journal}
  {Phys. Rev. Lett.}\ }\textbf {\bibinfo {volume} {120}},\ \bibinfo {pages}
  {016401} (\bibinfo {year} {2018})}\BibitemShut {NoStop}%
\bibitem [{\citenamefont {Wu}\ \emph {et~al.}(2019)\citenamefont {Wu},
  \citenamefont {Soluyanov},\ and\ \citenamefont
  {Bzdu$\check{\mathrm{s}}$ek}}]{QuanShengWu-S}%
  \BibitemOpen
  \bibfield  {author} {\bibinfo {author} {\bibfnamefont {QuanSheng}\
  \bibnamefont {Wu}}, \bibinfo {author} {\bibfnamefont {Alexey~A.}\
  \bibnamefont {Soluyanov}}, \ and\ \bibinfo {author} {\bibfnamefont
  {Tom$\acute{\mathrm{a}}\check{\mathrm{s}}$}\ \bibnamefont
  {Bzdu$\check{\mathrm{s}}$ek}},\ }\bibfield  {title} {\enquote {\bibinfo
  {title} {Non-{A}belian band topologyin noninteracting metals},}\ }\href@noop
  {} {\bibfield  {journal} {\bibinfo  {journal} {Science}\ }\textbf {\bibinfo
  {volume} {365}},\ \bibinfo {pages} {1273--1277} (\bibinfo {year}
  {2019})}\BibitemShut {NoStop}%
\bibitem [{\citenamefont {Young}\ \emph {et~al.}(2012)\citenamefont {Young},
  \citenamefont {Zaheer}, \citenamefont {Teo}, \citenamefont {Kane},
  \citenamefont {Mele},\ and\ \citenamefont {Rappe}}]{SMYoung}%
  \BibitemOpen
  \bibfield  {author} {\bibinfo {author} {\bibfnamefont {S.~M.}\ \bibnamefont
  {Young}}, \bibinfo {author} {\bibfnamefont {S.}~\bibnamefont {Zaheer}},
  \bibinfo {author} {\bibfnamefont {J.~C.~Y.}\ \bibnamefont {Teo}}, \bibinfo
  {author} {\bibfnamefont {C.~L.}\ \bibnamefont {Kane}}, \bibinfo {author}
  {\bibfnamefont {E.~J.}\ \bibnamefont {Mele}}, \ and\ \bibinfo {author}
  {\bibfnamefont {A.~M.}\ \bibnamefont {Rappe}},\ }\bibfield  {title} {\enquote
  {\bibinfo {title} {Dirac {S}emimetal in {T}hree {D}imensions},}\ }\href@noop
  {} {\bibfield  {journal} {\bibinfo  {journal} {Phys. Rev. Lett.}\ }\textbf
  {\bibinfo {volume} {108}},\ \bibinfo {pages} {140405} (\bibinfo {year}
  {2012})}\BibitemShut {NoStop}%
\bibitem [{\citenamefont {Wang}\ \emph {et~al.}(2012)\citenamefont {Wang},
  \citenamefont {Sun}, \citenamefont {Chen}, \citenamefont {Franchini},
  \citenamefont {Xu}, \citenamefont {Weng}, \citenamefont {Dai},\ and\
  \citenamefont {Fang}}]{Na3Bi}%
  \BibitemOpen
  \bibfield  {author} {\bibinfo {author} {\bibfnamefont {Zhijun}\ \bibnamefont
  {Wang}}, \bibinfo {author} {\bibfnamefont {Yan}\ \bibnamefont {Sun}},
  \bibinfo {author} {\bibfnamefont {Xing-Qiu}\ \bibnamefont {Chen}}, \bibinfo
  {author} {\bibfnamefont {Cesare}\ \bibnamefont {Franchini}}, \bibinfo
  {author} {\bibfnamefont {Gang}\ \bibnamefont {Xu}}, \bibinfo {author}
  {\bibfnamefont {Hongming}\ \bibnamefont {Weng}}, \bibinfo {author}
  {\bibfnamefont {Xi}~\bibnamefont {Dai}}, \ and\ \bibinfo {author}
  {\bibfnamefont {Zhong}\ \bibnamefont {Fang}},\ }\bibfield  {title} {\enquote
  {\bibinfo {title} {Dirac semimetal and topological phase transitions in
  ${A}_{3}${B}i (${A}=\text{{N}a}$, {K}, {R}b)},}\ }\href@noop {} {\bibfield
  {journal} {\bibinfo  {journal} {Phys. Rev. B}\ }\textbf {\bibinfo {volume}
  {85}},\ \bibinfo {pages} {195320} (\bibinfo {year} {2012})}\BibitemShut
  {NoStop}%
\bibitem [{\citenamefont {Wang}\ \emph {et~al.}(2013)\citenamefont {Wang},
  \citenamefont {Weng}, \citenamefont {Wu}, \citenamefont {Dai},\ and\
  \citenamefont {Fang}}]{Cd3As2}%
  \BibitemOpen
  \bibfield  {author} {\bibinfo {author} {\bibfnamefont {Zhijun}\ \bibnamefont
  {Wang}}, \bibinfo {author} {\bibfnamefont {Hongming}\ \bibnamefont {Weng}},
  \bibinfo {author} {\bibfnamefont {Quansheng}\ \bibnamefont {Wu}}, \bibinfo
  {author} {\bibfnamefont {Xi}~\bibnamefont {Dai}}, \ and\ \bibinfo {author}
  {\bibfnamefont {Zhong}\ \bibnamefont {Fang}},\ }\bibfield  {title} {\enquote
  {\bibinfo {title} {Three-dimensional {D}irac semimetal and quantum transport
  in {C}d${}_{3}${A}s${}_{2}$},}\ }\href@noop {} {\bibfield  {journal}
  {\bibinfo  {journal} {Phys. Rev. B}\ }\textbf {\bibinfo {volume} {88}},\
  \bibinfo {pages} {125427} (\bibinfo {year} {2013})}\BibitemShut {NoStop}%
\bibitem [{\citenamefont {Yang}\ and\ \citenamefont
  {Nagaosa}(2014)}]{Nagaosa-NC}%
  \BibitemOpen
  \bibfield  {author} {\bibinfo {author} {\bibfnamefont {Bohm-Jung}\
  \bibnamefont {Yang}}\ and\ \bibinfo {author} {\bibfnamefont {Naoto}\
  \bibnamefont {Nagaosa}},\ }\bibfield  {title} {\enquote {\bibinfo {title}
  {Classification of stable three-dimensional {D}irac semimetals with
  nontrivial topology},}\ }\href@noop {} {\bibfield  {journal} {\bibinfo
  {journal} {Nature Communications}\ }\textbf {\bibinfo {volume} {5}},\
  \bibinfo {pages} {4898} (\bibinfo {year} {2014})}\BibitemShut {NoStop}%
\bibitem [{\citenamefont {Du}\ \emph {et~al.}(2015)\citenamefont {Du},
  \citenamefont {Wan}, \citenamefont {Wang}, \citenamefont {Sheng},
  \citenamefont {Duan},\ and\ \citenamefont {Wan}}]{Du-1}%
  \BibitemOpen
  \bibfield  {author} {\bibinfo {author} {\bibfnamefont {Yongping}\
  \bibnamefont {Du}}, \bibinfo {author} {\bibfnamefont {Bo}~\bibnamefont
  {Wan}}, \bibinfo {author} {\bibfnamefont {Di}~\bibnamefont {Wang}}, \bibinfo
  {author} {\bibfnamefont {Li}~\bibnamefont {Sheng}}, \bibinfo {author}
  {\bibfnamefont {Chun-Gang}\ \bibnamefont {Duan}}, \ and\ \bibinfo {author}
  {\bibfnamefont {Xiangang}\ \bibnamefont {Wan}},\ }\bibfield  {title}
  {\enquote {\bibinfo {title} {Dirac and {W}eyl {S}emimetal in {X}{Y}{B}i
  ({X}={B}a, {E}u; {Y}={C}u, {A}g and {A}u)},}\ }\href@noop {} {\bibfield
  {journal} {\bibinfo  {journal} {Scientific Reports}\ }\textbf {\bibinfo
  {volume} {5}},\ \bibinfo {pages} {14423} (\bibinfo {year}
  {2015})}\BibitemShut {NoStop}%
\bibitem [{\citenamefont {Du}\ \emph {et~al.}(2017)\citenamefont {Du},
  \citenamefont {Tang}, \citenamefont {Wang}, \citenamefont {Sheng},
  \citenamefont {Kan}, \citenamefont {Duan}, \citenamefont {Savrasov},\ and\
  \citenamefont {Wan}}]{Du-2}%
  \BibitemOpen
  \bibfield  {author} {\bibinfo {author} {\bibfnamefont {Yongping}\
  \bibnamefont {Du}}, \bibinfo {author} {\bibfnamefont {Feng}\ \bibnamefont
  {Tang}}, \bibinfo {author} {\bibfnamefont {Di}~\bibnamefont {Wang}}, \bibinfo
  {author} {\bibfnamefont {Li}~\bibnamefont {Sheng}}, \bibinfo {author}
  {\bibfnamefont {Er-Jun}\ \bibnamefont {Kan}}, \bibinfo {author}
  {\bibfnamefont {Chun-Gang}\ \bibnamefont {Duan}}, \bibinfo {author}
  {\bibfnamefont {Sergey~Y.}\ \bibnamefont {Savrasov}}, \ and\ \bibinfo
  {author} {\bibfnamefont {Xiangang}\ \bibnamefont {Wan}},\ }\bibfield  {title}
  {\enquote {\bibinfo {title} {{C}a{T}e: a new topological node-line and
  {D}irac semimetal},}\ }\href@noop {} {\bibfield  {journal} {\bibinfo
  {journal} {npj Quantum Materials}\ }\textbf {\bibinfo {volume} {2}},\
  \bibinfo {pages} {3} (\bibinfo {year} {2017})}\BibitemShut {NoStop}%
\bibitem [{\citenamefont {Cano}\ \emph {et~al.}(2019)\citenamefont {Cano},
  \citenamefont {Bradlyn},\ and\ \citenamefont {Vergniory}}]{APL}%
  \BibitemOpen
  \bibfield  {author} {\bibinfo {author} {\bibfnamefont {J.}~\bibnamefont
  {Cano}}, \bibinfo {author} {\bibfnamefont {B.}~\bibnamefont {Bradlyn}}, \
  and\ \bibinfo {author} {\bibfnamefont {M.~G.}\ \bibnamefont {Vergniory}},\
  }\bibfield  {title} {\enquote {\bibinfo {title} {Multifold nodal points in
  magnetic materials},}\ }\href@noop {} {\bibfield  {journal} {\bibinfo
  {journal} {APL Materials}\ }\textbf {\bibinfo {volume} {7}},\ \bibinfo
  {pages} {101125} (\bibinfo {year} {2019})}\BibitemShut {NoStop}%
\bibitem [{\citenamefont {Yang}\ \emph {et~al.}(2021)\citenamefont {Yang},
  \citenamefont {Fang},\ and\ \citenamefont {Liu}}]{zxl}%
  \BibitemOpen
  \bibfield  {author} {\bibinfo {author} {\bibfnamefont {Jian}\ \bibnamefont
  {Yang}}, \bibinfo {author} {\bibfnamefont {Chen}\ \bibnamefont {Fang}}, \
  and\ \bibinfo {author} {\bibfnamefont {Zheng-Xin}\ \bibnamefont {Liu}},\
  }\bibfield  {title} {\enquote {\bibinfo {title} {Symmetry-protected nodal
  points and nodal lines in magnetic materials},}\ }\href@noop {} {\bibfield
  {journal} {\bibinfo  {journal} {Phys. Rev. B}\ }\textbf {\bibinfo {volume}
  {103}},\ \bibinfo {pages} {245141} (\bibinfo {year} {2021})}\BibitemShut
  {NoStop}%
\bibitem [{\citenamefont {Wieder}\ \emph {et~al.}(2016)\citenamefont {Wieder},
  \citenamefont {Kim}, \citenamefont {Rappe},\ and\ \citenamefont
  {Kane}}]{DoubleDSM}%
  \BibitemOpen
  \bibfield  {author} {\bibinfo {author} {\bibfnamefont {Benjamin~J.}\
  \bibnamefont {Wieder}}, \bibinfo {author} {\bibfnamefont {Youngkuk}\
  \bibnamefont {Kim}}, \bibinfo {author} {\bibfnamefont {A.~M.}\ \bibnamefont
  {Rappe}}, \ and\ \bibinfo {author} {\bibfnamefont {C.~L.}\ \bibnamefont
  {Kane}},\ }\bibfield  {title} {\enquote {\bibinfo {title} {Double {D}irac
  {S}emimetals in {T}hree {D}imensions},}\ }\href@noop {} {\bibfield  {journal}
  {\bibinfo  {journal} {Phys. Rev. Lett.}\ }\textbf {\bibinfo {volume} {116}},\
  \bibinfo {pages} {186402} (\bibinfo {year} {2016})}\BibitemShut {NoStop}%
\bibitem [{\citenamefont {Weng}\ \emph {et~al.}(2016)\citenamefont {Weng},
  \citenamefont {Fang}, \citenamefont {Fang},\ and\ \citenamefont
  {Dai}}]{triple-Weng}%
  \BibitemOpen
  \bibfield  {author} {\bibinfo {author} {\bibfnamefont {Hongming}\
  \bibnamefont {Weng}}, \bibinfo {author} {\bibfnamefont {Chen}\ \bibnamefont
  {Fang}}, \bibinfo {author} {\bibfnamefont {Zhong}\ \bibnamefont {Fang}}, \
  and\ \bibinfo {author} {\bibfnamefont {Xi}~\bibnamefont {Dai}},\ }\bibfield
  {title} {\enquote {\bibinfo {title} {Topological semimetals with triply
  degenerate nodal points in $\theta$-phase tantalum nitride},}\ }\href@noop {}
  {\bibfield  {journal} {\bibinfo  {journal} {Phys. Rev. B}\ }\textbf {\bibinfo
  {volume} {93}},\ \bibinfo {pages} {241202(R)} (\bibinfo {year}
  {2016})}\BibitemShut {NoStop}%
\bibitem [{\citenamefont {Burkov}\ \emph {et~al.}(2011)\citenamefont {Burkov},
  \citenamefont {Hook},\ and\ \citenamefont {Balents}}]{nodal-line-balents}%
  \BibitemOpen
  \bibfield  {author} {\bibinfo {author} {\bibfnamefont {A.~A.}\ \bibnamefont
  {Burkov}}, \bibinfo {author} {\bibfnamefont {M.~D.}\ \bibnamefont {Hook}}, \
  and\ \bibinfo {author} {\bibfnamefont {Leon}\ \bibnamefont {Balents}},\
  }\bibfield  {title} {\enquote {\bibinfo {title} {Topological nodal
  semimetals},}\ }\href@noop {} {\bibfield  {journal} {\bibinfo  {journal}
  {Phys. Rev. B}\ }\textbf {\bibinfo {volume} {84}},\ \bibinfo {pages} {235126}
  (\bibinfo {year} {2011})}\BibitemShut {NoStop}%
\bibitem [{\citenamefont {Fang}\ \emph {et~al.}(2015)\citenamefont {Fang},
  \citenamefont {Chen}, \citenamefont {Kee},\ and\ \citenamefont
  {Fu}}]{nodal-line-fangchen}%
  \BibitemOpen
  \bibfield  {author} {\bibinfo {author} {\bibfnamefont {Chen}\ \bibnamefont
  {Fang}}, \bibinfo {author} {\bibfnamefont {Yige}\ \bibnamefont {Chen}},
  \bibinfo {author} {\bibfnamefont {Hae-Young}\ \bibnamefont {Kee}}, \ and\
  \bibinfo {author} {\bibfnamefont {Liang}\ \bibnamefont {Fu}},\ }\bibfield
  {title} {\enquote {\bibinfo {title} {Topological nodal line semimetals with
  and without spin-orbital coupling},}\ }\href@noop {} {\bibfield  {journal}
  {\bibinfo  {journal} {Phys. Rev. B}\ }\textbf {\bibinfo {volume} {92}},\
  \bibinfo {pages} {081201} (\bibinfo {year} {2015})}\BibitemShut {NoStop}%
\bibitem [{\citenamefont {Yu}\ \emph {et~al.}(2015)\citenamefont {Yu},
  \citenamefont {Weng}, \citenamefont {Fang}, \citenamefont {Dai},\ and\
  \citenamefont {Hu}}]{cupdn-1}%
  \BibitemOpen
  \bibfield  {author} {\bibinfo {author} {\bibfnamefont {Rui}\ \bibnamefont
  {Yu}}, \bibinfo {author} {\bibfnamefont {Hongming}\ \bibnamefont {Weng}},
  \bibinfo {author} {\bibfnamefont {Zhong}\ \bibnamefont {Fang}}, \bibinfo
  {author} {\bibfnamefont {Xi}~\bibnamefont {Dai}}, \ and\ \bibinfo {author}
  {\bibfnamefont {Xiao}\ \bibnamefont {Hu}},\ }\bibfield  {title} {\enquote
  {\bibinfo {title} {Topological {N}ode-{L}ine {S}emimetal and {D}irac
  {S}emimetal {S}tate in {A}ntiperovskite {C}u$_{3}\mathrm{{P}d{N}}$},}\
  }\href@noop {} {\bibfield  {journal} {\bibinfo  {journal} {Phys. Rev. Lett.}\
  }\textbf {\bibinfo {volume} {115}},\ \bibinfo {pages} {036807} (\bibinfo
  {year} {2015})}\BibitemShut {NoStop}%
\bibitem [{\citenamefont {Kim}\ \emph {et~al.}(2015)\citenamefont {Kim},
  \citenamefont {Wieder}, \citenamefont {Kane},\ and\ \citenamefont
  {Rappe}}]{cupdn-2}%
  \BibitemOpen
  \bibfield  {author} {\bibinfo {author} {\bibfnamefont {Youngkuk}\
  \bibnamefont {Kim}}, \bibinfo {author} {\bibfnamefont {Benjamin~J.}\
  \bibnamefont {Wieder}}, \bibinfo {author} {\bibfnamefont {C.~L.}\
  \bibnamefont {Kane}}, \ and\ \bibinfo {author} {\bibfnamefont {Andrew~M.}\
  \bibnamefont {Rappe}},\ }\bibfield  {title} {\enquote {\bibinfo {title}
  {Dirac {L}ine {N}odes in {I}nversion-{S}ymmetric {C}rystals},}\ }\href@noop
  {} {\bibfield  {journal} {\bibinfo  {journal} {Phys. Rev. Lett.}\ }\textbf
  {\bibinfo {volume} {115}},\ \bibinfo {pages} {036806} (\bibinfo {year}
  {2015})}\BibitemShut {NoStop}%
\bibitem [{\citenamefont {Wang}\ \emph {et~al.}(2016)\citenamefont {Wang},
  \citenamefont {Alexandradinata}, \citenamefont {Cava},\ and\ \citenamefont
  {Bernevig}}]{hourglass-n}%
  \BibitemOpen
  \bibfield  {author} {\bibinfo {author} {\bibfnamefont {Zhijun}\ \bibnamefont
  {Wang}}, \bibinfo {author} {\bibfnamefont {A.}~\bibnamefont
  {Alexandradinata}}, \bibinfo {author} {\bibfnamefont {R.~J.}\ \bibnamefont
  {Cava}}, \ and\ \bibinfo {author} {\bibfnamefont {B.~Andrei}\ \bibnamefont
  {Bernevig}},\ }\bibfield  {title} {\enquote {\bibinfo {title} {Hourglass
  fermions},}\ }\href@noop {} {\bibfield  {journal} {\bibinfo  {journal}
  {Nature}\ }\textbf {\bibinfo {volume} {532}},\ \bibinfo {pages} {189--194}
  (\bibinfo {year} {2016})}\BibitemShut {NoStop}%
\bibitem [{\citenamefont {Bzdu$\check{\mathrm{s}}$ek}\ \emph
  {et~al.}(2016)\citenamefont {Bzdu$\check{\mathrm{s}}$ek}, \citenamefont {Wu},
  \citenamefont {R$\ddot{\mathrm{u}}$egg}, \citenamefont {Sigrist},\ and\
  \citenamefont {Soluyanov}}]{nodal-chain-n}%
  \BibitemOpen
  \bibfield  {author} {\bibinfo {author} {\bibfnamefont
  {Tom$\acute{\mathrm{a}}\check{\mathrm{s}}$}\ \bibnamefont
  {Bzdu$\check{\mathrm{s}}$ek}}, \bibinfo {author} {\bibfnamefont {Quansheng}\
  \bibnamefont {Wu}}, \bibinfo {author} {\bibfnamefont {Andreas}\ \bibnamefont
  {R$\ddot{\mathrm{u}}$egg}}, \bibinfo {author} {\bibfnamefont {Manfred}\
  \bibnamefont {Sigrist}}, \ and\ \bibinfo {author} {\bibfnamefont {Alexey~A.}\
  \bibnamefont {Soluyanov}},\ }\bibfield  {title} {\enquote {\bibinfo {title}
  {Nodal-chain metals},}\ }\href@noop {} {\bibfield  {journal} {\bibinfo
  {journal} {Nature}\ }\textbf {\bibinfo {volume} {538}},\ \bibinfo {pages}
  {75--78} (\bibinfo {year} {2016})}\BibitemShut {NoStop}%
\bibitem [{\citenamefont {Chen}\ \emph {et~al.}(2017)\citenamefont {Chen},
  \citenamefont {Lu},\ and\ \citenamefont {Hou}}]{hopf-link-1}%
  \BibitemOpen
  \bibfield  {author} {\bibinfo {author} {\bibfnamefont {Wei}\ \bibnamefont
  {Chen}}, \bibinfo {author} {\bibfnamefont {Hai-Zhou}\ \bibnamefont {Lu}}, \
  and\ \bibinfo {author} {\bibfnamefont {Jing-Min}\ \bibnamefont {Hou}},\
  }\bibfield  {title} {\enquote {\bibinfo {title} {Topological semimetals with
  a double-helix nodal link},}\ }\href@noop {} {\bibfield  {journal} {\bibinfo
  {journal} {Phys. Rev. B}\ }\textbf {\bibinfo {volume} {96}},\ \bibinfo
  {pages} {041102} (\bibinfo {year} {2017})}\BibitemShut {NoStop}%
\bibitem [{\citenamefont {Yan}\ \emph {et~al.}(2017)\citenamefont {Yan},
  \citenamefont {Bi}, \citenamefont {Shen}, \citenamefont {Lu}, \citenamefont
  {Zhang},\ and\ \citenamefont {Wang}}]{hopf-link-2}%
  \BibitemOpen
  \bibfield  {author} {\bibinfo {author} {\bibfnamefont {Zhongbo}\ \bibnamefont
  {Yan}}, \bibinfo {author} {\bibfnamefont {Ren}\ \bibnamefont {Bi}}, \bibinfo
  {author} {\bibfnamefont {Huitao}\ \bibnamefont {Shen}}, \bibinfo {author}
  {\bibfnamefont {Ling}\ \bibnamefont {Lu}}, \bibinfo {author} {\bibfnamefont
  {Shou-Cheng}\ \bibnamefont {Zhang}}, \ and\ \bibinfo {author} {\bibfnamefont
  {Zhong}\ \bibnamefont {Wang}},\ }\bibfield  {title} {\enquote {\bibinfo
  {title} {Nodal-link semimetals},}\ }\href@noop {} {\bibfield  {journal}
  {\bibinfo  {journal} {Phys. Rev. B}\ }\textbf {\bibinfo {volume} {96}},\
  \bibinfo {pages} {041103} (\bibinfo {year} {2017})}\BibitemShut {NoStop}%
\bibitem [{\citenamefont {Ezawa}(2017)}]{hopf-link-3}%
  \BibitemOpen
  \bibfield  {author} {\bibinfo {author} {\bibfnamefont {Motohiko}\
  \bibnamefont {Ezawa}},\ }\bibfield  {title} {\enquote {\bibinfo {title}
  {Topological semimetals carrying arbitrary {H}opf numbers: {F}ermi surface
  topologies of a {H}opf link, {S}olomon's knot, trefoil knot, and other linked
  nodal varieties},}\ }\href@noop {} {\bibfield  {journal} {\bibinfo  {journal}
  {Phys. Rev. B}\ }\textbf {\bibinfo {volume} {96}},\ \bibinfo {pages} {041202}
  (\bibinfo {year} {2017})}\BibitemShut {NoStop}%
\bibitem [{\citenamefont {Chang}\ and\ \citenamefont
  {Yee}(2017)}]{hopf-link-4}%
  \BibitemOpen
  \bibfield  {author} {\bibinfo {author} {\bibfnamefont {Po-Yao}\ \bibnamefont
  {Chang}}\ and\ \bibinfo {author} {\bibfnamefont {Chuck-Hou}\ \bibnamefont
  {Yee}},\ }\bibfield  {title} {\enquote {\bibinfo {title} {Weyl-link
  semimetals},}\ }\href@noop {} {\bibfield  {journal} {\bibinfo  {journal}
  {Phys. Rev. B}\ }\textbf {\bibinfo {volume} {96}},\ \bibinfo {pages} {081114}
  (\bibinfo {year} {2017})}\BibitemShut {NoStop}%
\bibitem [{\citenamefont {Bi}\ \emph {et~al.}(2017)\citenamefont {Bi},
  \citenamefont {Yan}, \citenamefont {Lu},\ and\ \citenamefont
  {Wang}}]{hopf-link-5}%
  \BibitemOpen
  \bibfield  {author} {\bibinfo {author} {\bibfnamefont {Ren}\ \bibnamefont
  {Bi}}, \bibinfo {author} {\bibfnamefont {Zhongbo}\ \bibnamefont {Yan}},
  \bibinfo {author} {\bibfnamefont {Ling}\ \bibnamefont {Lu}}, \ and\ \bibinfo
  {author} {\bibfnamefont {Zhong}\ \bibnamefont {Wang}},\ }\bibfield  {title}
  {\enquote {\bibinfo {title} {Nodal-knot semimetals},}\ }\href@noop {}
  {\bibfield  {journal} {\bibinfo  {journal} {Phys. Rev. B}\ }\textbf {\bibinfo
  {volume} {96}},\ \bibinfo {pages} {201305} (\bibinfo {year}
  {2017})}\BibitemShut {NoStop}%
\bibitem [{\citenamefont {Chang}\ and\ \citenamefont
  {et~al.}(2017)}]{hopf-link-6}%
  \BibitemOpen
  \bibfield  {author} {\bibinfo {author} {\bibfnamefont {Guoqing}\ \bibnamefont
  {Chang}}\ and\ \bibinfo {author} {\bibnamefont {et~al.}},\ }\bibfield
  {title} {\enquote {\bibinfo {title} {Topological {H}opf and {C}hain {L}ink
  {S}emimetal {S}tates and {T}heir {A}pplication to
  ${\mathrm{{c}o}}_{2}\mathrm{{M}n}\text{{G}}\text{a}$},}\ }\href@noop {}
  {\bibfield  {journal} {\bibinfo  {journal} {Phys. Rev. Lett.}\ }\textbf
  {\bibinfo {volume} {119}},\ \bibinfo {pages} {156401} (\bibinfo {year}
  {2017})}\BibitemShut {NoStop}%
\bibitem [{\citenamefont {Zhou}\ \emph {et~al.}(2018)\citenamefont {Zhou},
  \citenamefont {Xiong}, \citenamefont {Wan},\ and\ \citenamefont
  {An}}]{hopf-link-7}%
  \BibitemOpen
  \bibfield  {author} {\bibinfo {author} {\bibfnamefont {Yao}\ \bibnamefont
  {Zhou}}, \bibinfo {author} {\bibfnamefont {Feng}\ \bibnamefont {Xiong}},
  \bibinfo {author} {\bibfnamefont {Xiangang}\ \bibnamefont {Wan}}, \ and\
  \bibinfo {author} {\bibfnamefont {Jin}\ \bibnamefont {An}},\ }\bibfield
  {title} {\enquote {\bibinfo {title} {Hopf-link topological nodal-loop
  semimetals},}\ }\href@noop {} {\bibfield  {journal} {\bibinfo  {journal}
  {Phys. Rev. B}\ }\textbf {\bibinfo {volume} {97}},\ \bibinfo {pages} {155140}
  (\bibinfo {year} {2018})}\BibitemShut {NoStop}%
\bibitem [{\citenamefont {et~al.}(2018{\natexlab{a}})}]{wwk}%
  \BibitemOpen
  \bibfield  {author} {\bibinfo {author} {\bibfnamefont {Weikang~Wu}\
  \bibnamefont {et~al.}},\ }\bibfield  {title} {\enquote {\bibinfo {title}
  {Nodal surface semimetals: {T}heory and material realization},}\ }\href@noop
  {} {\bibfield  {journal} {\bibinfo  {journal} {Phys. Rev. B}\ }\textbf
  {\bibinfo {volume} {97}},\ \bibinfo {pages} {115125} (\bibinfo {year}
  {2018}{\natexlab{a}})}\BibitemShut {NoStop}%
\bibitem [{\citenamefont {et~al.}(2021)}]{n-sym-enf}%
  \BibitemOpen
  \bibfield  {author} {\bibinfo {author} {\bibfnamefont {Marc A.~Wilde}\
  \bibnamefont {et~al.}},\ }\bibfield  {title} {\enquote {\bibinfo {title}
  {Symmetry-enforced topological nodal planes at the {F}ermi surface of a
  chiral magnet},}\ }\href@noop {} {\bibfield  {journal} {\bibinfo  {journal}
  {Nature}\ }\textbf {\bibinfo {volume} {594}},\ \bibinfo {pages} {374--379}
  (\bibinfo {year} {2021})}\BibitemShut {NoStop}%
\bibitem [{\citenamefont {Son}\ and\ \citenamefont {Spivak}(2013)}]{DTson}%
  \BibitemOpen
  \bibfield  {author} {\bibinfo {author} {\bibfnamefont {D.~T.}\ \bibnamefont
  {Son}}\ and\ \bibinfo {author} {\bibfnamefont {B.~Z.}\ \bibnamefont
  {Spivak}},\ }\bibfield  {title} {\enquote {\bibinfo {title} {Chiral anomaly
  and classical negative magnetoresistance of {W}eyl metals},}\ }\href@noop {}
  {\bibfield  {journal} {\bibinfo  {journal} {Phys. Rev. B}\ }\textbf {\bibinfo
  {volume} {88}},\ \bibinfo {pages} {104412} (\bibinfo {year}
  {2013})}\BibitemShut {NoStop}%
\bibitem [{\citenamefont {Katsnelson}\ \emph {et~al.}(2006)\citenamefont
  {Katsnelson}, \citenamefont {Novoselov},\ and\ \citenamefont {Geim}}]{klein}%
  \BibitemOpen
  \bibfield  {author} {\bibinfo {author} {\bibfnamefont {M.~I.}\ \bibnamefont
  {Katsnelson}}, \bibinfo {author} {\bibfnamefont {K.~S.}\ \bibnamefont
  {Novoselov}}, \ and\ \bibinfo {author} {\bibfnamefont {A.~K.}\ \bibnamefont
  {Geim}},\ }\bibfield  {title} {\enquote {\bibinfo {title} {Chiral tunneling
  and the {K}lein {P}aradox in {G}raphene},}\ }\href@noop {} {\bibfield
  {journal} {\bibinfo  {journal} {Nature Physics}\ }\textbf {\bibinfo {volume}
  {2}},\ \bibinfo {pages} {620--625} (\bibinfo {year} {2006})}\BibitemShut
  {NoStop}%
\bibitem [{\citenamefont {Bradley}\ and\ \citenamefont
  {Cracknell}()}]{bradley}%
  \BibitemOpen
  \bibfield  {author} {\bibinfo {author} {\bibfnamefont {C.~J.}\ \bibnamefont
  {Bradley}}\ and\ \bibinfo {author} {\bibfnamefont {A.~P.}\ \bibnamefont
  {Cracknell}},\ }\href@noop {} {\emph {\bibinfo {title} {The {M}athematical
  {T}heory of {S}ymmetry in {S}olids ({O}xford {U}niv. {P}ress,
  1972)}}}\BibitemShut {NoStop}%
\bibitem [{\citenamefont {Aroyo}\ \emph {et~al.}(2006)\citenamefont {Aroyo},
  \citenamefont {Kirov}, \citenamefont {Capillas}, \citenamefont {Perez-Mato},\
  and\ \citenamefont {Wondratschek}}]{bilbao}%
  \BibitemOpen
  \bibfield  {author} {\bibinfo {author} {\bibfnamefont {M.~I.}\ \bibnamefont
  {Aroyo}}, \bibinfo {author} {\bibfnamefont {A.}~\bibnamefont {Kirov}},
  \bibinfo {author} {\bibfnamefont {C.}~\bibnamefont {Capillas}}, \bibinfo
  {author} {\bibfnamefont {J.~M.}\ \bibnamefont {Perez-Mato}}, \ and\ \bibinfo
  {author} {\bibfnamefont {H.}~\bibnamefont {Wondratschek}},\ }\bibfield
  {title} {\enquote {\bibinfo {title} {Bilbaocrystallographic server. {II}.
  {R}epresentations of crystallographic point groups and spacegroups},}\
  }\href@noop {} {\bibfield  {journal} {\bibinfo  {journal} {Acta Crystallogr.
  A}\ }\textbf {\bibinfo {volume} {62}},\ \bibinfo {pages} {115--128} (\bibinfo
  {year} {2006})}\BibitemShut {NoStop}%
\bibitem [{\citenamefont {Armitage}\ \emph {et~al.}(2018)\citenamefont
  {Armitage}, \citenamefont {Mele},\ and\ \citenamefont {Vishwanath}}]{AV-RMP}%
  \BibitemOpen
  \bibfield  {author} {\bibinfo {author} {\bibfnamefont {N.~P.}\ \bibnamefont
  {Armitage}}, \bibinfo {author} {\bibfnamefont {E.~J.}\ \bibnamefont {Mele}},
  \ and\ \bibinfo {author} {\bibfnamefont {Ashvin}\ \bibnamefont
  {Vishwanath}},\ }\bibfield  {title} {\enquote {\bibinfo {title} {Weyl and
  {D}irac semimetals in three-dimensional solids},}\ }\href@noop {} {\bibfield
  {journal} {\bibinfo  {journal} {Rev. Mod. Phys.}\ }\textbf {\bibinfo {volume}
  {90}},\ \bibinfo {pages} {015001} (\bibinfo {year} {2018})}\BibitemShut
  {NoStop}%
\bibitem [{\citenamefont {Hasan}\ and\ \citenamefont {Kane}(2010)}]{TI-RMP-1}%
  \BibitemOpen
  \bibfield  {author} {\bibinfo {author} {\bibfnamefont {M.~Z.}\ \bibnamefont
  {Hasan}}\ and\ \bibinfo {author} {\bibfnamefont {C.~L.}\ \bibnamefont
  {Kane}},\ }\bibfield  {title} {\enquote {\bibinfo {title} {Colloquium:
  {T}opological insulators},}\ }\href@noop {} {\bibfield  {journal} {\bibinfo
  {journal} {Rev. Mod. Phys.}\ }\textbf {\bibinfo {volume} {82}},\ \bibinfo
  {pages} {3045--3067} (\bibinfo {year} {2010})}\BibitemShut {NoStop}%
\bibitem [{\citenamefont {Qi}\ and\ \citenamefont {Zhang}(2011)}]{TI-RMP-2}%
  \BibitemOpen
  \bibfield  {author} {\bibinfo {author} {\bibfnamefont {Xiao-Liang}\
  \bibnamefont {Qi}}\ and\ \bibinfo {author} {\bibfnamefont {Shou-Cheng}\
  \bibnamefont {Zhang}},\ }\bibfield  {title} {\enquote {\bibinfo {title}
  {Topological insulators and superconductors},}\ }\href@noop {} {\bibfield
  {journal} {\bibinfo  {journal} {Rev. Mod. Phys.}\ }\textbf {\bibinfo {volume}
  {83}},\ \bibinfo {pages} {1057--1110} (\bibinfo {year} {2011})}\BibitemShut
  {NoStop}%
\bibitem [{\citenamefont {Xiao}\ \emph {et~al.}(2010)\citenamefont {Xiao},
  \citenamefont {Chang},\ and\ \citenamefont {Niu}}]{berryphase}%
  \BibitemOpen
  \bibfield  {author} {\bibinfo {author} {\bibfnamefont {Di}~\bibnamefont
  {Xiao}}, \bibinfo {author} {\bibfnamefont {Ming-Che}\ \bibnamefont {Chang}},
  \ and\ \bibinfo {author} {\bibfnamefont {Qian}\ \bibnamefont {Niu}},\
  }\bibfield  {title} {\enquote {\bibinfo {title} {Berry phase effects on
  electronic properties},}\ }\href@noop {} {\bibfield  {journal} {\bibinfo
  {journal} {Rev. Mod. Phys.}\ }\textbf {\bibinfo {volume} {82}},\ \bibinfo
  {pages} {1959} (\bibinfo {year} {2010})}\BibitemShut {NoStop}%
\bibitem [{\citenamefont {Schaibley}\ \emph {et~al.}(2016)\citenamefont
  {Schaibley}, \citenamefont {Yu}, \citenamefont {Clark}, \citenamefont
  {Rivera}, \citenamefont {Ross}, \citenamefont {Seyler}, \citenamefont {Yao},\
  and\ \citenamefont {Xu}}]{valley}%
  \BibitemOpen
  \bibfield  {author} {\bibinfo {author} {\bibfnamefont {John~R.}\ \bibnamefont
  {Schaibley}}, \bibinfo {author} {\bibfnamefont {Hongyi}\ \bibnamefont {Yu}},
  \bibinfo {author} {\bibfnamefont {Genevieve}\ \bibnamefont {Clark}}, \bibinfo
  {author} {\bibfnamefont {Pasqual}\ \bibnamefont {Rivera}}, \bibinfo {author}
  {\bibfnamefont {Jason~S.}\ \bibnamefont {Ross}}, \bibinfo {author}
  {\bibfnamefont {Kyle~L.}\ \bibnamefont {Seyler}}, \bibinfo {author}
  {\bibfnamefont {Wang}\ \bibnamefont {Yao}}, \ and\ \bibinfo {author}
  {\bibfnamefont {Xiaodong}\ \bibnamefont {Xu}},\ }\bibfield  {title} {\enquote
  {\bibinfo {title} {Valleytronics in 2{D} materials},}\ }\href@noop {}
  {\bibfield  {journal} {\bibinfo  {journal} {Nature Reviews Materials}\
  }\textbf {\bibinfo {volume} {1}},\ \bibinfo {pages} {16055} (\bibinfo {year}
  {2016})}\BibitemShut {NoStop}%
\bibitem [{\citenamefont {Chiu}\ \emph {et~al.}(2016)\citenamefont {Chiu},
  \citenamefont {Teo}, \citenamefont {Schnyder},\ and\ \citenamefont
  {Ryu}}]{Chiu-RMP}%
  \BibitemOpen
  \bibfield  {author} {\bibinfo {author} {\bibfnamefont {Ching-Kai}\
  \bibnamefont {Chiu}}, \bibinfo {author} {\bibfnamefont {Jeffrey C.~Y.}\
  \bibnamefont {Teo}}, \bibinfo {author} {\bibfnamefont {Andreas~P.}\
  \bibnamefont {Schnyder}}, \ and\ \bibinfo {author} {\bibfnamefont {Shinsei}\
  \bibnamefont {Ryu}},\ }\bibfield  {title} {\enquote {\bibinfo {title}
  {Classification of topological quantum matter with symmetries},}\ }\href@noop
  {} {\bibfield  {journal} {\bibinfo  {journal} {Rev. Mod. Phys.}\ }\textbf
  {\bibinfo {volume} {88}},\ \bibinfo {pages} {035005} (\bibinfo {year}
  {2016})}\BibitemShut {NoStop}%
\bibitem [{\citenamefont {Lv}\ \emph {et~al.}(2021)\citenamefont {Lv},
  \citenamefont {Qian},\ and\ \citenamefont {Ding}}]{LBQ-RMP}%
  \BibitemOpen
  \bibfield  {author} {\bibinfo {author} {\bibfnamefont {B.~Q.}\ \bibnamefont
  {Lv}}, \bibinfo {author} {\bibfnamefont {T.}~\bibnamefont {Qian}}, \ and\
  \bibinfo {author} {\bibfnamefont {H.}~\bibnamefont {Ding}},\ }\bibfield
  {title} {\enquote {\bibinfo {title} {Experimental perspective on
  three-dimensional topological semimetals},}\ }\href@noop {} {\bibfield
  {journal} {\bibinfo  {journal} {Rev. Mod. Phys.}\ }\textbf {\bibinfo {volume}
  {93}},\ \bibinfo {pages} {025002} (\bibinfo {year} {2021})}\BibitemShut
  {NoStop}%
\bibitem [{\citenamefont {Ando}(2013)}]{Ando}%
  \BibitemOpen
  \bibfield  {author} {\bibinfo {author} {\bibfnamefont {Yoichi}\ \bibnamefont
  {Ando}},\ }\bibfield  {title} {\enquote {\bibinfo {title} {Topological
  insulator materials},}\ }\href@noop {} {\bibfield  {journal} {\bibinfo
  {journal} {J. Phys. Soc. Jpn.}\ }\textbf {\bibinfo {volume} {82}},\ \bibinfo
  {pages} {102001} (\bibinfo {year} {2013})}\BibitemShut {NoStop}%
\bibitem [{\citenamefont {Wehling}\ \emph {et~al.}(2014)\citenamefont
  {Wehling}, \citenamefont {Black-Schaffer},\ and\ \citenamefont
  {Balatsky}}]{Bala}%
  \BibitemOpen
  \bibfield  {author} {\bibinfo {author} {\bibfnamefont {T.~O.}\ \bibnamefont
  {Wehling}}, \bibinfo {author} {\bibfnamefont {A.~M.}\ \bibnamefont
  {Black-Schaffer}}, \ and\ \bibinfo {author} {\bibfnamefont {A.~V.}\
  \bibnamefont {Balatsky}},\ }\bibfield  {title} {\enquote {\bibinfo {title}
  {Dirac materials},}\ }\href@noop {} {\bibfield  {journal} {\bibinfo
  {journal} {Advances in Physics}\ }\textbf {\bibinfo {volume} {63}},\ \bibinfo
  {pages} {1--76} (\bibinfo {year} {2014})}\BibitemShut {NoStop}%
\bibitem [{\citenamefont {Bradlyn}\ \emph {et~al.}(2017)\citenamefont
  {Bradlyn}, \citenamefont {Elcoro}, \citenamefont {Cano}, \citenamefont
  {Vergniory}, \citenamefont {Wang}, \citenamefont {Felser}, \citenamefont
  {Aroyo},\ and\ \citenamefont {Bernevig}}]{tqc}%
  \BibitemOpen
  \bibfield  {author} {\bibinfo {author} {\bibfnamefont {Barry}\ \bibnamefont
  {Bradlyn}}, \bibinfo {author} {\bibfnamefont {L.}~\bibnamefont {Elcoro}},
  \bibinfo {author} {\bibfnamefont {Jennifer}\ \bibnamefont {Cano}}, \bibinfo
  {author} {\bibfnamefont {M.~G.}\ \bibnamefont {Vergniory}}, \bibinfo {author}
  {\bibfnamefont {Zhijun}\ \bibnamefont {Wang}}, \bibinfo {author}
  {\bibfnamefont {C.}~\bibnamefont {Felser}}, \bibinfo {author} {\bibfnamefont
  {M.~I.}\ \bibnamefont {Aroyo}}, \ and\ \bibinfo {author} {\bibfnamefont
  {B.~Andrei}\ \bibnamefont {Bernevig}},\ }\bibfield  {title} {\enquote
  {\bibinfo {title} {Topological quantum chemistry},}\ }\href@noop {}
  {\bibfield  {journal} {\bibinfo  {journal} {Nature}\ }\textbf {\bibinfo
  {volume} {547}},\ \bibinfo {pages} {298--305} (\bibinfo {year}
  {2017})}\BibitemShut {NoStop}%
\bibitem [{\citenamefont {Po}\ \emph {et~al.}(2017)\citenamefont {Po},
  \citenamefont {Vishwanath},\ and\ \citenamefont {Watanabe}}]{si}%
  \BibitemOpen
  \bibfield  {author} {\bibinfo {author} {\bibfnamefont {Hoi~Chun}\
  \bibnamefont {Po}}, \bibinfo {author} {\bibfnamefont {Ashvin}\ \bibnamefont
  {Vishwanath}}, \ and\ \bibinfo {author} {\bibfnamefont {Haruki}\ \bibnamefont
  {Watanabe}},\ }\bibfield  {title} {\enquote {\bibinfo {title} {Symmetry-based
  indicators of band topology in the 230 space groups},}\ }\href@noop {}
  {\bibfield  {journal} {\bibinfo  {journal} {Nature Commnunications}\ }\textbf
  {\bibinfo {volume} {8}},\ \bibinfo {pages} {50} (\bibinfo {year}
  {2017})}\BibitemShut {NoStop}%
\bibitem [{\citenamefont {Kruthoff}\ \emph {et~al.}(2017)\citenamefont
  {Kruthoff}, \citenamefont {de~Boer}, \citenamefont {van Wezel}, \citenamefont
  {Kane},\ and\ \citenamefont {Slager}}]{slager-prx}%
  \BibitemOpen
  \bibfield  {author} {\bibinfo {author} {\bibfnamefont {Jorrit}\ \bibnamefont
  {Kruthoff}}, \bibinfo {author} {\bibfnamefont {Jan}\ \bibnamefont {de~Boer}},
  \bibinfo {author} {\bibfnamefont {Jasper}\ \bibnamefont {van Wezel}},
  \bibinfo {author} {\bibfnamefont {Charles~L.}\ \bibnamefont {Kane}}, \ and\
  \bibinfo {author} {\bibfnamefont {Robert-Jan}\ \bibnamefont {Slager}},\
  }\bibfield  {title} {\enquote {\bibinfo {title} {Topological {C}lassification
  of {C}rystalline {I}nsulators through {B}and {S}tructure {C}ombinatorics},}\
  }\href@noop {} {\bibfield  {journal} {\bibinfo  {journal} {Phys. Rev. X}\
  }\textbf {\bibinfo {volume} {7}},\ \bibinfo {pages} {041069} (\bibinfo {year}
  {2017})}\BibitemShut {NoStop}%
\bibitem [{\citenamefont {Zhang}\ \emph {et~al.}(2019)\citenamefont {Zhang},
  \citenamefont {Jiang}, \citenamefont {Song}, \citenamefont {Huang},
  \citenamefont {He}, \citenamefont {Fang}, \citenamefont {Weng},\ and\
  \citenamefont {Fang}}]{n-1}%
  \BibitemOpen
  \bibfield  {author} {\bibinfo {author} {\bibfnamefont {Tiantian}\
  \bibnamefont {Zhang}}, \bibinfo {author} {\bibfnamefont {Yi}~\bibnamefont
  {Jiang}}, \bibinfo {author} {\bibfnamefont {Zhida}\ \bibnamefont {Song}},
  \bibinfo {author} {\bibfnamefont {He}~\bibnamefont {Huang}}, \bibinfo
  {author} {\bibfnamefont {Yuqing}\ \bibnamefont {He}}, \bibinfo {author}
  {\bibfnamefont {Zhong}\ \bibnamefont {Fang}}, \bibinfo {author}
  {\bibfnamefont {Hongming}\ \bibnamefont {Weng}}, \ and\ \bibinfo {author}
  {\bibfnamefont {Chen}\ \bibnamefont {Fang}},\ }\bibfield  {title} {\enquote
  {\bibinfo {title} {Catalogue of topological elecronic materials},}\
  }\href@noop {} {\bibfield  {journal} {\bibinfo  {journal} {Nature}\ }\textbf
  {\bibinfo {volume} {566}},\ \bibinfo {pages} {475--479} (\bibinfo {year}
  {2019})}\BibitemShut {NoStop}%
\bibitem [{\citenamefont {Vergniory}\ \emph {et~al.}(2019)\citenamefont
  {Vergniory}, \citenamefont {Elcoro}, \citenamefont {Felser}, \citenamefont
  {Regnault}, \citenamefont {Bernevig},\ and\ \citenamefont {Wang}}]{n-2}%
  \BibitemOpen
  \bibfield  {author} {\bibinfo {author} {\bibfnamefont {M.~G.}\ \bibnamefont
  {Vergniory}}, \bibinfo {author} {\bibfnamefont {L.}~\bibnamefont {Elcoro}},
  \bibinfo {author} {\bibfnamefont {Claudia}\ \bibnamefont {Felser}}, \bibinfo
  {author} {\bibfnamefont {Nicolas}\ \bibnamefont {Regnault}}, \bibinfo
  {author} {\bibfnamefont {B.~Andrei}\ \bibnamefont {Bernevig}}, \ and\
  \bibinfo {author} {\bibfnamefont {Zhijun}\ \bibnamefont {Wang}},\ }\bibfield
  {title} {\enquote {\bibinfo {title} {A complete catalogue of high-quality
  topological materials},}\ }\href@noop {} {\bibfield  {journal} {\bibinfo
  {journal} {Nature}\ }\textbf {\bibinfo {volume} {566}},\ \bibinfo {pages}
  {480--485} (\bibinfo {year} {2019})}\BibitemShut {NoStop}%
\bibitem [{\citenamefont {Tang}\ \emph {et~al.}(2019)\citenamefont {Tang},
  \citenamefont {Po}, \citenamefont {Vishwanath},\ and\ \citenamefont
  {Wan}}]{n-3}%
  \BibitemOpen
  \bibfield  {author} {\bibinfo {author} {\bibfnamefont {Feng}\ \bibnamefont
  {Tang}}, \bibinfo {author} {\bibfnamefont {Hoi~Chun}\ \bibnamefont {Po}},
  \bibinfo {author} {\bibfnamefont {Ashvin}\ \bibnamefont {Vishwanath}}, \ and\
  \bibinfo {author} {\bibfnamefont {Xiangang}\ \bibnamefont {Wan}},\ }\bibfield
   {title} {\enquote {\bibinfo {title} {Comprehensive search for topological
  materials using symmetry indicators},}\ }\href@noop {} {\bibfield  {journal}
  {\bibinfo  {journal} {Nature}\ }\textbf {\bibinfo {volume} {566}},\ \bibinfo
  {pages} {486--489} (\bibinfo {year} {2019})}\BibitemShut {NoStop}%
\bibitem [{\citenamefont {Xu}\ \emph {et~al.}(2020)\citenamefont {Xu},
  \citenamefont {Elcoro}, \citenamefont {Song}, \citenamefont {Wieder},
  \citenamefont {Vergniory}, \citenamefont {Regnault}, \citenamefont {Chen},
  \citenamefont {Felser},\ and\ \citenamefont {Bernevig}}]{n-4}%
  \BibitemOpen
  \bibfield  {author} {\bibinfo {author} {\bibfnamefont {Yuanfeng}\
  \bibnamefont {Xu}}, \bibinfo {author} {\bibfnamefont {Luis}\ \bibnamefont
  {Elcoro}}, \bibinfo {author} {\bibfnamefont {Zhi-Da}\ \bibnamefont {Song}},
  \bibinfo {author} {\bibfnamefont {Benjamin~J.}\ \bibnamefont {Wieder}},
  \bibinfo {author} {\bibfnamefont {M.~G.}\ \bibnamefont {Vergniory}}, \bibinfo
  {author} {\bibfnamefont {Nicolas}\ \bibnamefont {Regnault}}, \bibinfo
  {author} {\bibfnamefont {Yulin}\ \bibnamefont {Chen}}, \bibinfo {author}
  {\bibfnamefont {Claudia}\ \bibnamefont {Felser}}, \ and\ \bibinfo {author}
  {\bibfnamefont {B.~Andrei}\ \bibnamefont {Bernevig}},\ }\bibfield  {title}
  {\enquote {\bibinfo {title} {High-throughput calculations of magnetic
  topological materials},}\ }\href@noop {} {\bibfield  {journal} {\bibinfo
  {journal} {Nature}\ }\textbf {\bibinfo {volume} {586}},\ \bibinfo {pages}
  {702--707} (\bibinfo {year} {2020})}\BibitemShut {NoStop}%
\bibitem [{\citenamefont {Ma\~nes}(2012)}]{Manes}%
  \BibitemOpen
  \bibfield  {author} {\bibinfo {author} {\bibfnamefont {J.~L.}\ \bibnamefont
  {Ma\~nes}},\ }\bibfield  {title} {\enquote {\bibinfo {title} {Existence of
  bulk chiral fermions and crystal symmetry},}\ }\href@noop {} {\bibfield
  {journal} {\bibinfo  {journal} {Phys. Rev. B}\ }\textbf {\bibinfo {volume}
  {85}},\ \bibinfo {pages} {155118} (\bibinfo {year} {2012})}\BibitemShut
  {NoStop}%
\bibitem [{\citenamefont {et~al.}(2018{\natexlab{b}})}]{KramersWeyl}%
  \BibitemOpen
  \bibfield  {author} {\bibinfo {author} {\bibfnamefont {G.~Chang}\
  \bibnamefont {et~al.}},\ }\bibfield  {title} {\enquote {\bibinfo {title}
  {Topological quantum properties of chiral crystals},}\ }\href@noop {}
  {\bibfield  {journal} {\bibinfo  {journal} {Nature Materials}\ }\textbf
  {\bibinfo {volume} {17}},\ \bibinfo {pages} {978} (\bibinfo {year}
  {2018}{\natexlab{b}})}\BibitemShut {NoStop}%
\bibitem [{\citenamefont {Yu}\ \emph {et~al.}(2021)\citenamefont {Yu},
  \citenamefont {Zhang}, \citenamefont {Liu}, \citenamefont {Wu}, \citenamefont
  {Li}, \citenamefont {Zhang}, \citenamefont {Yang},\ and\ \citenamefont
  {Yao}}]{yao}%
  \BibitemOpen
  \bibfield  {author} {\bibinfo {author} {\bibfnamefont {Zhi-Ming}\
  \bibnamefont {Yu}}, \bibinfo {author} {\bibfnamefont {Zeying}\ \bibnamefont
  {Zhang}}, \bibinfo {author} {\bibfnamefont {Gui-Bin}\ \bibnamefont {Liu}},
  \bibinfo {author} {\bibfnamefont {Weikang}\ \bibnamefont {Wu}}, \bibinfo
  {author} {\bibfnamefont {Xiao-Ping}\ \bibnamefont {Li}}, \bibinfo {author}
  {\bibfnamefont {Run-Wu}\ \bibnamefont {Zhang}}, \bibinfo {author}
  {\bibfnamefont {Shengyuan~A}\ \bibnamefont {Yang}}, \ and\ \bibinfo {author}
  {\bibfnamefont {Yugui}\ \bibnamefont {Yao}},\ }\bibfield  {title} {\enquote
  {\bibinfo {title} {Encyclopedia of emergent particles in three-dimensional
  crystals},}\ }\href@noop {} {\bibfield  {journal} {\bibinfo  {journal} {arXiv
  2102.01517}\ } (\bibinfo {year} {2021})}\BibitemShut {NoStop}%
\bibitem [{\citenamefont {Zhang}\ \emph {et~al.}(2020)\citenamefont {Zhang},
  \citenamefont {Takahashi}, \citenamefont {Fang},\ and\ \citenamefont
  {Murakami}}]{TT}%
  \BibitemOpen
  \bibfield  {author} {\bibinfo {author} {\bibfnamefont {Tiantian}\
  \bibnamefont {Zhang}}, \bibinfo {author} {\bibfnamefont {Ryo}\ \bibnamefont
  {Takahashi}}, \bibinfo {author} {\bibfnamefont {Chen}\ \bibnamefont {Fang}},
  \ and\ \bibinfo {author} {\bibfnamefont {Shuichi}\ \bibnamefont {Murakami}},\
  }\bibfield  {title} {\enquote {\bibinfo {title} {Twofold quadruple {W}eyl
  nodes in chiral cubic crystals},}\ }\href@noop {} {\bibfield  {journal}
  {\bibinfo  {journal} {Phys. Rev. B}\ }\textbf {\bibinfo {volume} {102}},\
  \bibinfo {pages} {125148} (\bibinfo {year} {2020})}\BibitemShut {NoStop}%
\bibitem [{\citenamefont {Novoselov}\ \emph {et~al.}(2004)\citenamefont
  {Novoselov}, \citenamefont {Geim}, \citenamefont {Morozov}, \citenamefont
  {Jiang}, \citenamefont {Zhang}, \citenamefont {Dubonos}, \citenamefont
  {Grigorieva},\ and\ \citenamefont {Firsov}}]{graphene-S}%
  \BibitemOpen
  \bibfield  {author} {\bibinfo {author} {\bibfnamefont {K.~S.}\ \bibnamefont
  {Novoselov}}, \bibinfo {author} {\bibfnamefont {A.~K.}\ \bibnamefont {Geim}},
  \bibinfo {author} {\bibfnamefont {S.~V.}\ \bibnamefont {Morozov}}, \bibinfo
  {author} {\bibfnamefont {D.}~\bibnamefont {Jiang}}, \bibinfo {author}
  {\bibfnamefont {Y.}~\bibnamefont {Zhang}}, \bibinfo {author} {\bibfnamefont
  {S.~V.}\ \bibnamefont {Dubonos}}, \bibinfo {author} {\bibfnamefont {I.~V.}\
  \bibnamefont {Grigorieva}}, \ and\ \bibinfo {author} {\bibfnamefont {A.~A.}\
  \bibnamefont {Firsov}},\ }\bibfield  {title} {\enquote {\bibinfo {title}
  {Electric field effect in atomically thin carbon film},}\ }\href@noop {}
  {\bibfield  {journal} {\bibinfo  {journal} {Science}\ }\textbf {\bibinfo
  {volume} {306}},\ \bibinfo {pages} {666--669} (\bibinfo {year}
  {2004})}\BibitemShut {NoStop}%
\bibitem [{\citenamefont {et~al.}(2018{\natexlab{c}})}]{wallpaper}%
  \BibitemOpen
  \bibfield  {author} {\bibinfo {author} {\bibfnamefont {Benjamin J.~Wieder}\
  \bibnamefont {et~al.}},\ }\bibfield  {title} {\enquote {\bibinfo {title}
  {Wallpaper fermions and the nonsymmorphic {D}irac insulator},}\ }\href@noop
  {} {\bibfield  {journal} {\bibinfo  {journal} {Science}\ }\textbf {\bibinfo
  {volume} {361}},\ \bibinfo {pages} {246--251} (\bibinfo {year}
  {2018}{\natexlab{c}})}\BibitemShut {NoStop}%
\bibitem [{\citenamefont {et~al.}(2016)}]{NL-exp-2}%
  \BibitemOpen
  \bibfield  {author} {\bibinfo {author} {\bibfnamefont {Leslie M.~Schoop}\
  \bibnamefont {et~al.}},\ }\bibfield  {title} {\enquote {\bibinfo {title}
  {Dirac cone protected by non-symmorphic symmetry and three-dimensional
  {D}irac line node in {Z}r{S}i{S}},}\ }\href@noop {} {\bibfield  {journal}
  {\bibinfo  {journal} {Nature Communications}\ }\textbf {\bibinfo {volume}
  {7}},\ \bibinfo {pages} {11696} (\bibinfo {year} {2016})}\BibitemShut
  {NoStop}%
\bibitem [{\citenamefont {et~al.}(2019{\natexlab{a}})}]{NL-exp-1}%
  \BibitemOpen
  \bibfield  {author} {\bibinfo {author} {\bibfnamefont {B.-B.~Fu}\
  \bibnamefont {et~al.}},\ }\bibfield  {title} {\enquote {\bibinfo {title}
  {Dirac nodal surfaces and nodal lines in {Z}r{S}i{S}},}\ }\href@noop {}
  {\bibfield  {journal} {\bibinfo  {journal} {Science Advances}\ }\textbf
  {\bibinfo {volume} {5}},\ \bibinfo {pages} {aau6459} (\bibinfo {year}
  {2019}{\natexlab{a}})}\BibitemShut {NoStop}%
\bibitem [{\citenamefont {et~al.}(2019{\natexlab{b}})}]{NL-exp-3}%
  \BibitemOpen
  \bibfield  {author} {\bibinfo {author} {\bibfnamefont {Ilya~Belopolski}\
  \bibnamefont {et~al.}},\ }\bibfield  {title} {\enquote {\bibinfo {title}
  {Discovery of topological {W}eyl fermion lines and drumhead surface states in
  a room temperature magnet},}\ }\href@noop {} {\bibfield  {journal} {\bibinfo
  {journal} {Science}\ }\textbf {\bibinfo {volume} {365}},\ \bibinfo {pages}
  {1278--1281} (\bibinfo {year} {2019}{\natexlab{b}})}\BibitemShut {NoStop}%
\bibitem [{\citenamefont {Young}\ and\ \citenamefont {Kim}(2009)}]{klein-exp}%
  \BibitemOpen
  \bibfield  {author} {\bibinfo {author} {\bibfnamefont {Andrea~F.}\
  \bibnamefont {Young}}\ and\ \bibinfo {author} {\bibfnamefont {Philip}\
  \bibnamefont {Kim}},\ }\bibfield  {title} {\enquote {\bibinfo {title}
  {Quantum interference and {K}lein tunnelling in graphene heterojunctions},}\
  }\href@noop {} {\bibfield  {journal} {\bibinfo  {journal} {Nature Physics}\
  }\textbf {\bibinfo {volume} {5}},\ \bibinfo {pages} {222--226} (\bibinfo
  {year} {2009})}\BibitemShut {NoStop}%
\bibitem [{pri()}]{private}%
  \BibitemOpen
  \bibfield  {title} {\enquote {\bibinfo {title} {Private communication},}\
  }\href@noop {} {\ }\BibitemShut {NoStop}%
\bibitem [{\citenamefont {Tang}\ and\ \citenamefont {Wan}(2021)}]{tang-kp}%
  \BibitemOpen
  \bibfield  {author} {\bibinfo {author} {\bibfnamefont {Feng}\ \bibnamefont
  {Tang}}\ and\ \bibinfo {author} {\bibfnamefont {Xiangang}\ \bibnamefont
  {Wan}},\ }\bibfield  {title} {\enquote {\bibinfo {title} {Exhaustive
  construction of effective models in 1651 magnetic space groups},}\
  }\href@noop {} {\bibfield  {journal} {\bibinfo  {journal} {Phys. Rev. B}\
  }\textbf {\bibinfo {volume} {104}},\ \bibinfo {pages} {085137} (\bibinfo
  {year} {2021})}\BibitemShut {NoStop}%
\bibitem [{rem()}]{remark-3}%
  \BibitemOpen
  \href@noop {} {\enquote {\bibinfo {title} {When the little group of a {B}{N}
  contains antiunitary element like time-reversal or its combination with a
  spatial operation, we should use co-representations.}}\ }\BibitemShut
  {NoStop}%
\bibitem [{\citenamefont {Belov}\ \emph {et~al.}(1957)\citenamefont {Belov},
  \citenamefont {Neronova},\ and\ \citenamefont {Smirnova}}]{bns}%
  \BibitemOpen
  \bibfield  {author} {\bibinfo {author} {\bibfnamefont {N.~V.}\ \bibnamefont
  {Belov}}, \bibinfo {author} {\bibfnamefont {N.~N.}\ \bibnamefont {Neronova}},
  \ and\ \bibinfo {author} {\bibfnamefont {T.~S.}\ \bibnamefont {Smirnova}},\
  }\bibfield  {title} {\enquote {\bibinfo {title} {Shubnikov groups},}\
  }\href@noop {} {\bibfield  {journal} {\bibinfo  {journal} {Sov. Phys.
  Crystallogr.}\ }\textbf {\bibinfo {volume} {2}},\ \bibinfo {pages} {311}
  (\bibinfo {year} {1957})}\BibitemShut {NoStop}%
\bibitem [{mag()}]{magndata}%
  \BibitemOpen
  \bibfield  {title} {\enquote {\bibinfo {title}
  {http://webbdcrista1.ehu.es/magndata/},}\ }\href@noop {} {\ }\BibitemShut
  {NoStop}%
\bibitem [{\citenamefont {Bansil}\ \emph {et~al.}(2016)\citenamefont {Bansil},
  \citenamefont {Lin},\ and\ \citenamefont {Das}}]{TI-RMP-3}%
  \BibitemOpen
  \bibfield  {author} {\bibinfo {author} {\bibfnamefont {A.}~\bibnamefont
  {Bansil}}, \bibinfo {author} {\bibfnamefont {Hsin}\ \bibnamefont {Lin}}, \
  and\ \bibinfo {author} {\bibfnamefont {Tanmoy}\ \bibnamefont {Das}},\
  }\bibfield  {title} {\enquote {\bibinfo {title} {Colloquium: {T}opological
  band theory},}\ }\href@noop {} {\bibfield  {journal} {\bibinfo  {journal}
  {Rev. Mod. Phys.}\ }\textbf {\bibinfo {volume} {88}},\ \bibinfo {pages}
  {021004} (\bibinfo {year} {2016})}\BibitemShut {NoStop}%
\bibitem [{\citenamefont {$\acute{\mathrm{E}}$tienne Lantagne-Hurtubise}\ and\
  \citenamefont {Franz}(2019)}]{franz}%
  \BibitemOpen
  \bibfield  {author} {\bibinfo {author} {\bibnamefont
  {$\acute{\mathrm{E}}$tienne Lantagne-Hurtubise}}\ and\ \bibinfo {author}
  {\bibfnamefont {Marcel}\ \bibnamefont {Franz}},\ }\bibfield  {title}
  {\enquote {\bibinfo {title} {Topology in abundance},}\ }\href@noop {}
  {\bibfield  {journal} {\bibinfo  {journal} {Nature Review Physics}\ }\textbf
  {\bibinfo {volume} {1}},\ \bibinfo {pages} {183--184} (\bibinfo {year}
  {2019})}\BibitemShut {NoStop}%
\bibitem [{\citenamefont {Tokura}\ \emph {et~al.}(2019)\citenamefont {Tokura},
  \citenamefont {Yasuda},\ and\ \citenamefont {Tsukazaki}}]{tokura}%
  \BibitemOpen
  \bibfield  {author} {\bibinfo {author} {\bibfnamefont {Yoshinori}\
  \bibnamefont {Tokura}}, \bibinfo {author} {\bibfnamefont {Kenji}\
  \bibnamefont {Yasuda}}, \ and\ \bibinfo {author} {\bibfnamefont {Atsushi}\
  \bibnamefont {Tsukazaki}},\ }\bibfield  {title} {\enquote {\bibinfo {title}
  {Magnetic topological insulators},}\ }\href@noop {} {\bibfield  {journal}
  {\bibinfo  {journal} {Nature Review Physics}\ }\textbf {\bibinfo {volume}
  {1}},\ \bibinfo {pages} {126--143} (\bibinfo {year} {2019})}\BibitemShut
  {NoStop}%
\bibitem [{\citenamefont {Watanabe}\ \emph {et~al.}(2016)\citenamefont
  {Watanabe}, \citenamefont {Po}, \citenamefont {Zaletel},\ and\ \citenamefont
  {Vishwanath}}]{filling-L}%
  \BibitemOpen
  \bibfield  {author} {\bibinfo {author} {\bibfnamefont {Haruki}\ \bibnamefont
  {Watanabe}}, \bibinfo {author} {\bibfnamefont {Hoi~Chun}\ \bibnamefont {Po}},
  \bibinfo {author} {\bibfnamefont {Michael~P.}\ \bibnamefont {Zaletel}}, \
  and\ \bibinfo {author} {\bibfnamefont {Ashvin}\ \bibnamefont {Vishwanath}},\
  }\bibfield  {title} {\enquote {\bibinfo {title} {Filling-enforced gaplessness
  in band structures of the 230 space groups},}\ }\href@noop {} {\bibfield
  {journal} {\bibinfo  {journal} {Phys. Rev. Lett.}\ }\textbf {\bibinfo
  {volume} {117}},\ \bibinfo {pages} {096404} (\bibinfo {year}
  {2016})}\BibitemShut {NoStop}%
\bibitem [{\citenamefont {Watanabe}\ \emph {et~al.}(2018)\citenamefont
  {Watanabe}, \citenamefont {Po},\ and\ \citenamefont {Vishwanath}}]{m-si}%
  \BibitemOpen
  \bibfield  {author} {\bibinfo {author} {\bibfnamefont {Haruki}\ \bibnamefont
  {Watanabe}}, \bibinfo {author} {\bibfnamefont {Hoi.~Chun.}\ \bibnamefont
  {Po}}, \ and\ \bibinfo {author} {\bibfnamefont {Ashvin}\ \bibnamefont
  {Vishwanath}},\ }\bibfield  {title} {\enquote {\bibinfo {title} {Structure
  and topology of band structures in the 1651 magnetic space groups},}\
  }\href@noop {} {\bibfield  {journal} {\bibinfo  {journal} {Science Advances}\
  }\textbf {\bibinfo {volume} {4}},\ \bibinfo {pages} {aat8685} (\bibinfo
  {year} {2018})}\BibitemShut {NoStop}%
\bibitem [{\citenamefont {Chen}\ \emph {et~al.}(2018)\citenamefont {Chen},
  \citenamefont {Po}, \citenamefont {Neaton},\ and\ \citenamefont
  {Vishwanath}}]{filling-np}%
  \BibitemOpen
  \bibfield  {author} {\bibinfo {author} {\bibfnamefont {Ru}~\bibnamefont
  {Chen}}, \bibinfo {author} {\bibfnamefont {Hoi~Chun}\ \bibnamefont {Po}},
  \bibinfo {author} {\bibfnamefont {Jeffrey~B.}\ \bibnamefont {Neaton}}, \ and\
  \bibinfo {author} {\bibfnamefont {Ashvin}\ \bibnamefont {Vishwanath}},\
  }\bibfield  {title} {\enquote {\bibinfo {title} {Topological materials
  discovery using electron filling constraints},}\ }\href@noop {} {\bibfield
  {journal} {\bibinfo  {journal} {Nature Physics}\ }\textbf {\bibinfo {volume}
  {14}},\ \bibinfo {pages} {55--61} (\bibinfo {year} {2018})}\BibitemShut
  {NoStop}%
\bibitem [{\citenamefont {Wang}\ \emph {et~al.}(2020)\citenamefont {Wang},
  \citenamefont {Tang}, \citenamefont {Po}, \citenamefont {Vishwanath},\ and\
  \citenamefont {Wan}}]{filling-prb}%
  \BibitemOpen
  \bibfield  {author} {\bibinfo {author} {\bibfnamefont {Di}~\bibnamefont
  {Wang}}, \bibinfo {author} {\bibfnamefont {Feng}\ \bibnamefont {Tang}},
  \bibinfo {author} {\bibfnamefont {Hoi~Chun}\ \bibnamefont {Po}}, \bibinfo
  {author} {\bibfnamefont {Ashvin}\ \bibnamefont {Vishwanath}}, \ and\ \bibinfo
  {author} {\bibfnamefont {Xiangang}\ \bibnamefont {Wan}},\ }\bibfield  {title}
  {\enquote {\bibinfo {title} {X{F}e$_4${G}e$_2$({X}={Y}, {L}u) and
  {M}n$_3${P}t: {F}illing-enforced magnetic topological metals},}\ }\href@noop
  {} {\bibfield  {journal} {\bibinfo  {journal} {Phys. Rev. B}\ }\textbf
  {\bibinfo {volume} {101}},\ \bibinfo {pages} {115122} (\bibinfo {year}
  {2020})}\BibitemShut {NoStop}%
\bibitem [{\citenamefont {K$\ddot{\mathrm{o}}$zii}\ \emph
  {et~al.}(2016)\citenamefont {K$\ddot{\mathrm{o}}$zii}, \citenamefont
  {Venderbos},\ and\ \citenamefont {Fu}}]{sc-1}%
  \BibitemOpen
  \bibfield  {author} {\bibinfo {author} {\bibfnamefont {Vladyslav}\
  \bibnamefont {K$\ddot{\mathrm{o}}$zii}}, \bibinfo {author} {\bibfnamefont
  {Jorn W.~F.}\ \bibnamefont {Venderbos}}, \ and\ \bibinfo {author}
  {\bibfnamefont {Liang}\ \bibnamefont {Fu}},\ }\bibfield  {title} {\enquote
  {\bibinfo {title} {Three-dimensional majorana fermions in chiral
  superconductors},}\ }\href@noop {} {\bibfield  {journal} {\bibinfo  {journal}
  {Science Advances}\ }\textbf {\bibinfo {volume} {2}},\ \bibinfo {pages}
  {e1601835} (\bibinfo {year} {2016})}\BibitemShut {NoStop}%
\bibitem [{\citenamefont {Ono}\ and\ \citenamefont {Shiozaki}(2021)}]{sc-2}%
  \BibitemOpen
  \bibfield  {author} {\bibinfo {author} {\bibfnamefont {Seishiro}\
  \bibnamefont {Ono}}\ and\ \bibinfo {author} {\bibfnamefont {Ken}\
  \bibnamefont {Shiozaki}},\ }\bibfield  {title} {\enquote {\bibinfo {title}
  {Symmetry-based approach to nodal structures: Unification of compatibility
  relations and gapless point classifications},}\ }\href@noop {} {\bibfield
  {journal} {\bibinfo  {journal} {arXiv: 2102.07676}\ } (\bibinfo {year}
  {2021})}\BibitemShut {NoStop}%
\bibitem [{\citenamefont {Fruchart}\ \emph {et~al.}(2020)\citenamefont
  {Fruchart}, \citenamefont {Zhou},\ and\ \citenamefont {Vitelli}}]{dual}%
  \BibitemOpen
  \bibfield  {author} {\bibinfo {author} {\bibfnamefont {Michel}\ \bibnamefont
  {Fruchart}}, \bibinfo {author} {\bibfnamefont {Yujie}\ \bibnamefont {Zhou}},
  \ and\ \bibinfo {author} {\bibfnamefont {Vincenzo}\ \bibnamefont {Vitelli}},\
  }\bibfield  {title} {\enquote {\bibinfo {title} {Dualities and non-abelian
  mechanics},}\ }\href@noop {} {\bibfield  {journal} {\bibinfo  {journal}
  {Nature}\ }\textbf {\bibinfo {volume} {577}},\ \bibinfo {pages} {636--640}
  (\bibinfo {year} {2020})}\BibitemShut {NoStop}%
\end{thebibliography}%
\end{document}